\documentclass[12pt,preprint]{aastex}

\usepackage{graphicx}
\usepackage{natbib}
\usepackage{amsmath}
\usepackage{longtable}
\usepackage{array}
\usepackage{footnote}
\usepackage{lipsum}
\usepackage{booktabs}
\usepackage{epstopdf}
\usepackage{pdfpages}
\usepackage{rotating}
\usepackage{amssymb}
\usepackage{url}
\usepackage{lscape}
\usepackage{multirow}

\newcommand{\+}{$^+$}
\newcommand{\2}{$_2$}

\newcommand{\cc}{cm$^{-3}$}
\newcommand{\kms}{km s$^{-1}$}
\newcommand{\sig}{$\sigma $}

\slugcomment{}

\shorttitle{HEXOS Spectral Survey Towards Orion South}
\shortauthors{Tahani et al.}

\begin{document}


\title{Analysis of the Herschel/HEXOS Spectral Survey Towards Orion South: A massive protostellar envelope with strong external irradiation}


\author{K. Tahani} \and
\author{R. Plume}
\affil{Department of Physics \& Astronomy, University of Calgary, Calgary, AB, Canada T2N 1N4 }
\email{ktahani@ucalgary.ca}

\author{E. A. Bergin}
\affil{Department of Astronomy, University of Michigan, 500 Church Street, Ann Arbor, MI 48109, USA}

\and

\author{V.~ Tolls,\altaffilmark{1}, T.~G.~Phillips,\altaffilmark{2}, E. Caux\altaffilmark{3, 4}, S. Cabrit\altaffilmark{5},  J. R. Goicoechea\altaffilmark{6}, P. F. Goldsmith\altaffilmark{7}, D. Johnstone\altaffilmark{8}, D. C. Lis\altaffilmark{5, 2},
L. Pagani\altaffilmark{5}, K. M. Menten\altaffilmark{9}, H. S. P. M\"uller\altaffilmark{10}, V. Ossenkopf-Okada\altaffilmark{10}, J. C. Pearson\altaffilmark{7},
F. F. S. van der Tak\altaffilmark{11, 12}
}

\altaffiltext	{	1	}{	Harvard-Smithsonian Center for Astrophysics, 60 Garden Street, Cambridge, MA 02138, USA}
\altaffiltext	{	2	}{	California Institute of Technology, Cahill Center for Astronomy and Astrophysics 301-17, Pasadena, CA 91125, USA}
\altaffiltext	{	3	}{	Universit\'e de Toulouse, UPS-OMP, IRAP, 31028 Toulouse, France}
\altaffiltext	{	4	}{	 CNRS, IRAP, 9 Av. Colonel Roche, BP 44346, 31028 Toulouse Cedex 4, France}
\altaffiltext	{	5	}{	LERMA, Observatoire de Paris, PSL Research University, CNRS, Sorbonne Universit\'es, UPMC Univ. Paris 06, F-75014, Paris, France}
\altaffiltext	{	6	}{	Instituto de Ciencia de Materiales de Madrid (ICMM-CSIC). Sor Juana Ines de la Cruz 3, E-28049 Cantoblanco, Madrid, Spain}
\altaffiltext	{	7	}{	Jet Propulsion Laboratory, Caltech, Pasadena, CA 91109, USA}
\altaffiltext	{	8	}{	National Research Council Canada, Herzberg Institute of Astrophysics, 5071 West Saanich Road, Victoria, BC V9E 2E7, Canada}
\altaffiltext	{	9	}{	Max-Planck-Institut f\"ur Radioastronomie, Auf dem H\"ugel 69, 53121 Bonn, Germany }
\altaffiltext	{	10	}{	I. Physikalisches Institut, Universit\"at zu K\"oln, Z\"ulpicher Str. 77, 50937 K\"oln, Germany}
\altaffiltext	{	11	}{	SRON Netherlands Institute for Space Research, PO Box 800, 9700 AV, Groningen, The Netherlands}
\altaffiltext	{	12	}{	Kapteyn Astronomical Institute, Groningen, The Netherlands}

\begin{abstract}

We present results from a comprehensive submillimeter spectral survey toward the source Orion South, based on data obtained with the HIFI instrument aboard the \textit{Herschel Space Observatory}, covering the frequency range 480 to 1900 GHz.  We detect 685 spectral lines with S/N $>$ 3$\sigma$,  originating from 52 different molecular and atomic species. We model  each of the detected species assuming conditions of Local Thermodynamic Equilibrium.   This analysis provides an estimate of the physical conditions of Orion South (column density, temperature, source size,  \& V$_{LSR}$).  We find evidence for three different cloud components: a cool (T$_{ex} \sim 20-40$ K), spatially extended ($> 60''$), and quiescent ($\Delta V_{FWHM} \sim 4$ km s $^{-1}$) component; a warmer (T$_{ex} \sim 80-100$ K), less spatially extended ($\sim 30''$), and dynamic ($\Delta V_{FWHM} \sim 8$ km s $^{-1}$) component, which is likely affected by embedded outflows; and a kinematically distinct region  (T$_{ex}$ $>$ 100 K; V$_{LSR}$ $\sim$  8 km s $^{-1}$), dominated by emission from species which trace ultraviolet irradiation, likely at the surface of the cloud.  We find little evidence for the existence of a chemically distinct ``hot core'' component, likely due to the small filling factor of the hot core or hot cores within the \textit{Herschel} beam. 
We find that the chemical composition of the gas in the cooler, quiescent component of Orion South more closely resembles that of the quiescent ridge in Orion-KL.  
The gas in the warmer, dynamic component, however, more closely resembles that of the Compact Ridge and Plateau regions of Orion-KL, suggesting that higher temperatures and shocks also have an influence on the overall chemistry of Orion South.

\end{abstract}


\section{INTRODUCTION}
\label{intro}

To date, about 200 different molecular species have been detected in the interstellar medium \citep{2011STMP..241...27M}\footnote{Also see: \url{http://www.astro.uni-koeln.de/cdms/molecules} and \url{http://www.astrochymist.org/astrochymist_ism.html}}.  However, our understanding of the total molecular inventory of individual sources is poor, since few sources have been systematically surveyed in any frequency band due to the large amount of observing time required to perform unbiased spectral surveys \citep[e.g.][]{Blake:1987ApJ...315..621B, Schilke:1997ApJS..108..301S, 1998ApJS..117..427N, 2001ApJS..132..281S, 2005ApJS..156..127C, Furlan:2006ApJS..165..568F, 2010A&A...517A..96T, 2014ApJ...789....8N}.  Therefore, we do not truly understand the origin of the chemical complexity observed in interstellar space. Understanding this complexity is important to comprehend details of the formation of stars, planets and life.   

Regardless of how complex chemistry arises in interstellar space, the chemical composition (and subsequent chemical evolution) can, in turn, affect the physical conditions (and subsequent dynamical evolution) of a star forming region \citep[e.g see][]{2009ARA&A..47..427H, Garrod:2006A&A...457..927G, Garrod:2008ApJ...682..283G}. For example, the overall molecular (and to some degree atomic) content can play an important role in regulating the gas pressure by changing the temperature of the gas via the process of heating and cooling through line-absorption and emission; \citep{Ceccarelli:1996ApJ...471..400C, 1978ApJ...222..881G}. In addition, molecular ions can affect the strength of coupling between the gas and the magnetic fields (which is related to magnetic turbulent support, e.g. \citealp{Williams:1998p4835}). Thus, there is a complex feedback between the physical and chemical conditions in an interstellar gas cloud that either helps drive the star formation process, or hinders it, and which may help determine the masses of the newly formed stars. 

In order to understand the origin of chemical complexity in interstellar space and how this chemistry evolves and affects the process of  star formation in the Universe (as well as the formation of planets and pre-biotic chemical species), we require unbiased and complete surveys of spectral lines that span a broad range of wavelengths.  These types of datasets are needed so that we can sample a wide variety of molecular and atomic species, as well as obtain multiple emission lines from each of the species, in order to extract the physical conditions in the gas. Fortunately, with the advent of sensitive, high-resolution spectrometers for millimeter/submillimeter wavelengths, especially those developed for space-based observatories, it is now possible to obtain such surveys and to begin to address these issues \citep[e.g.][]{Crockett:2014ApJ...787..112C, 2012A&A...546A..87Z, Kama:2013A&A...556A..57K, 2015A&A...574A..71K}.

The key project \textit{Herschel}  observations of EXtraOrdinary Sources (HEXOS) \citep{Bergin:2010p4658} was designed to address issues related to the chemical composition of massive star forming regions. HEXOS has obtained spectral line surveys of the Orion-KL, Orion South (hereafter Orion-S), and Orion Bar (Nagy et al. in prep.) regions within the Orion A Molecular Cloud, at high frequencies that are not easily accessible from ground based observatories (480$-$1900 GHz). Both Orion-KL and Orion-S are relatively nearby \citep[420 pc;][]{Menten:2007A&A...474..515M} massive star forming regions close to the Orion Nebula. The nearby Trapezium OB stars are the source of high energy photons, which produce Photon Dominated Regions (PDRs) throughout the region. The UV flux (6$<$E$<$13.6 eV) in the vicinity of Orion-S is  $\chi = 1.1 \times 10^5 \chi_0$ \citep{Herrmann:1997ApJ...481..343H, 2015ApJ...812...75G}, where $\chi_0 = 2.7 \times 10^{-4}$ ergs s$^{-1}$ cm$^{-2}$ sr$^{-1}$ \citep{1996ApJ...468..269D}. Observations of \cite{2015ApJ...812...75G} and \cite{2010AJ....140..985O} suggest that the HII region lies mostly in front of the molecular material, but may wrap behind, at least part of, the Orion-S molecular cloud. That at least some of the Orion-S molecular gas is located in front of ionized material has been convincingly demonstrated by Very Large Array absorption measurements of the H$_2$CO 6 cm $1_{10}-1_{11}$ transition \citep{1993ApJ...409..282M}.

Despite the fact that the far-infrared luminosity of Orion-S  \citep[$8.5 \times 10^3 \ L_\odot$;][]{Mezger:1990A&A...228...95M} is more than an order of magnitude below that of KL, a number of energetic outflows associated with Orion-S suggest ongoing star formation. For example, CO J=2$-$1 SMA observations   \citep[][]{Zapata:2005p5918} revealed a highly collimated bipolar outflow extending $\sim 30''$ over the velocity range $\sim -80$ to $\sim -26$ km s$^{-1}$ and $\sim 22$ to $\sim 82$ km s$^{-1}$ oriented NW-SE. The sub-millimeter continuum source with a deconvolved size  $\leq 0.6''$ and an integrated flux of 116.2 $\pm$ 9.0 mJy at 1.3 mm is well centered on the bipolar outflow axis, $\alpha_{2000}=05^h35^m13.550^s$, $\delta_{2000} = -05^\circ23'59.14''$. In addition, another quite extended, collimated,
low-velocity (5 km s $^{-1}$) CO outflow has been observed, oriented NE-SW \citep{Schmid-Burgk:1990ApJ...362L..25S}, and a low-velocity (10 km s $^{-1}$) bipolar SiO (5$-$4) outflow with a length $\sim 30''$ (oriented NE-SW) has been reported by \cite{Ziurys:1990ApJ...356L..25Z}. Four other SiO outflows are also listed by \cite{Zapata:2006ApJ...653..398Z}. 

Despite the presence of star formation activity, as indicated by the IR luminosity and molecular outflows, BIMA observations of a few selected species by \cite{McMullin:1993p6318} suggest that the chemistry of Orion-S resembles that of the Orion-KL quiescent ridge and has fewer, narrower and weaker lines than KL. These observations may imply that Orion-S is a more quiescent and younger star forming region, in which the star formation activity has not had time to significantly alter the dynamics and chemistry of the region. This idea is also consistent with dynamical ages from outflow observations in Orion-S \citep[i.e.][]{{Schmid-Burgk:1990ApJ...362L..25S},{2000AJ....119.2919B}, {Zapata:2005p5918}}. Assuming no projection effects, the maximum corresponding dynamical age for the largest outflow is found to be less than 45000 years which is still remarkably young \citep{Schmid-Burgk:1990ApJ...362L..25S}. The dynamical age for all the other outflows can be shown to be less than 5000 years  \citep{{2000AJ....119.2919B}, {Zapata:2005p5918}}. A more detailed comparison between Orion-S and Orion-KL is, therefore, of great interest, since the two regions presumably formed under similar conditions, but could have very different chemical abundances possibly based on differences in their ages, densities, temperatures, radiation fields, etc.

In this paper we present  a comprehensive study of the \textit{Herschel}/HIFI spectral survey toward Orion-S. The observations presented here were obtained as part of HEXOS and span over 1.2 THz of  frequencies, mostly not accessible from the ground.
In \S 2 we present our observations and data reduction methods, including the removal of off-position contamination,  line identification, and Gaussian fitting of the spectral features. Our results (including LTE modeling of each individual species) together with a chemical comparison of Orion-S are presented in \S 3. The conclusions are provided in \S 4.

\section{OBSERVATIONS \& DATA REDUCTION}
\label{obs}

The data presented in this paper were taken with the  Heterodyne Instrument for the Far-Infrared (HIFI) \citep{deGraauw:2010p4577}, one of three instruments aboard the \textit{Herschel Space Observatory} \citep{2010A&A...518L...1P}. HIFI operated over the frequency range 480$-$1900 GHz (with two gaps: one at 1280$-$1430 GHz, due to the switch between SIS and HEB detectors \citep{Roelfsema:2012gb}, and one at 1540$-$1570 GHz, which was an observational time saving strategy since this frequency range was expected to have few transitions). HIFI was separated into 14 different bands (1a, 1b, ..., 7b). Each receiver band had independent channels for horizontal (H) and vertical (V) polarizations, each with its dedicated Wide Band Spectrometer (WBS) having a native spectral resolution of 1.1 MHz \citep{Roelfsema:2012gb}.  Bands 1$-$5 were observed with a LO redundancy of 6, whereas Bands 6  and 7 used a redundancy of 2. Redundancy refers to the number of observations of each sky frequency with different LO settings. For example, redundancy of 6 means that each frequency in the band was observed with 6 different LO settings. This redundancy was necessary in order to distinguish lines originating from the upper and the lower sidebands \citep{2002A&A...395..357C}.  A redundancy of 2 was sufficient for bands 6  and 7 due to the relatively lower density of transitions at these high frequencies. The central position of Orion-S was $\alpha_{2000} = 5^h35^m13.44^s$, $\delta_{2000} = -5^\circ24'08.1''$.  All observations were taken in Dual Beam Switch (DBS) mode using the Fast Chop option ($>$0.5 Hz chop frequency).

We used the hifiPipeline task in HIPE 9.0 for all data reduction. The hifiPipeline task is a pre-compiled script in HIPE used to process level 0 data to any higher level (e.g. 0.5, 1.0, etc.). See \cite{2010ASPC..434..139O} for a description of the various data products. Spurious spectral features were removed and  fully calibrated,  double side band (DSB) spectra  were deconvolved into single sideband (SSB) spectra (e.g. level 2.5).  Additional details on data reduction and observational parameters can be found in \cite{Bergin:2010p4658} and \cite{Crockett:2010A&A...521L..21C}. 
\textbf{After processing by HIPE 9.0, the H and V polarizations were co-added (except for Band 4a which, due to processing errors specific to this band, had much noisier V polarization data that were excluded) and then the spectra were Hanning smoothed by two to sixteen channels (see Table \ref{tab:noise}) to improve the signal to noise ratio.}  Results are provided in Table \ref{tab:noise}, which shows typical values for the 1$\sigma$ rms noise, system temperature, velocity and frequency resolutions after smoothing, and the number of channels smoothed for each band. The 1$\sigma$ rms noise is calculated from line-free regions of the spectrum immediately adjacent to the lines. A noise range is provided since the noise is not uniform across the bands. 

\begin{figure}[b]
\centering
\includegraphics[width=\textwidth]{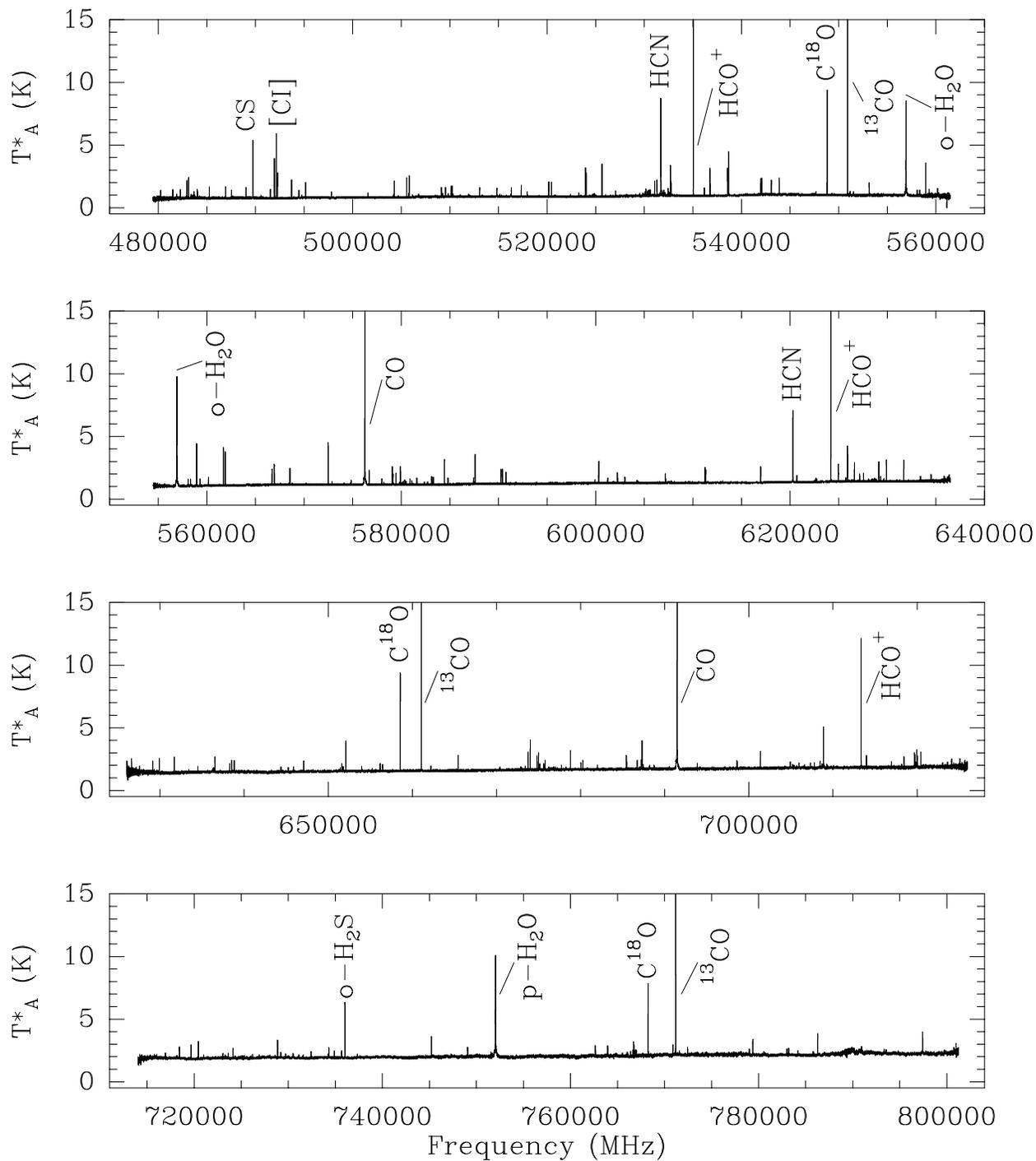}
\caption{HEXOS/HIFI spectral scans of (from top to bottom) band 1a, band 1b, band 2a and band 2b after Hanning smoothing. Resolutions, noise levels, and smoothing factors for each band are listed in Table \ref{tab:noise}. Baselines are not subtracted and  some of the strongest lines are labelled.}
\label{fig:1.1}
\end{figure}

\begin{figure}[b]
\centering
\includegraphics[width=\textwidth]{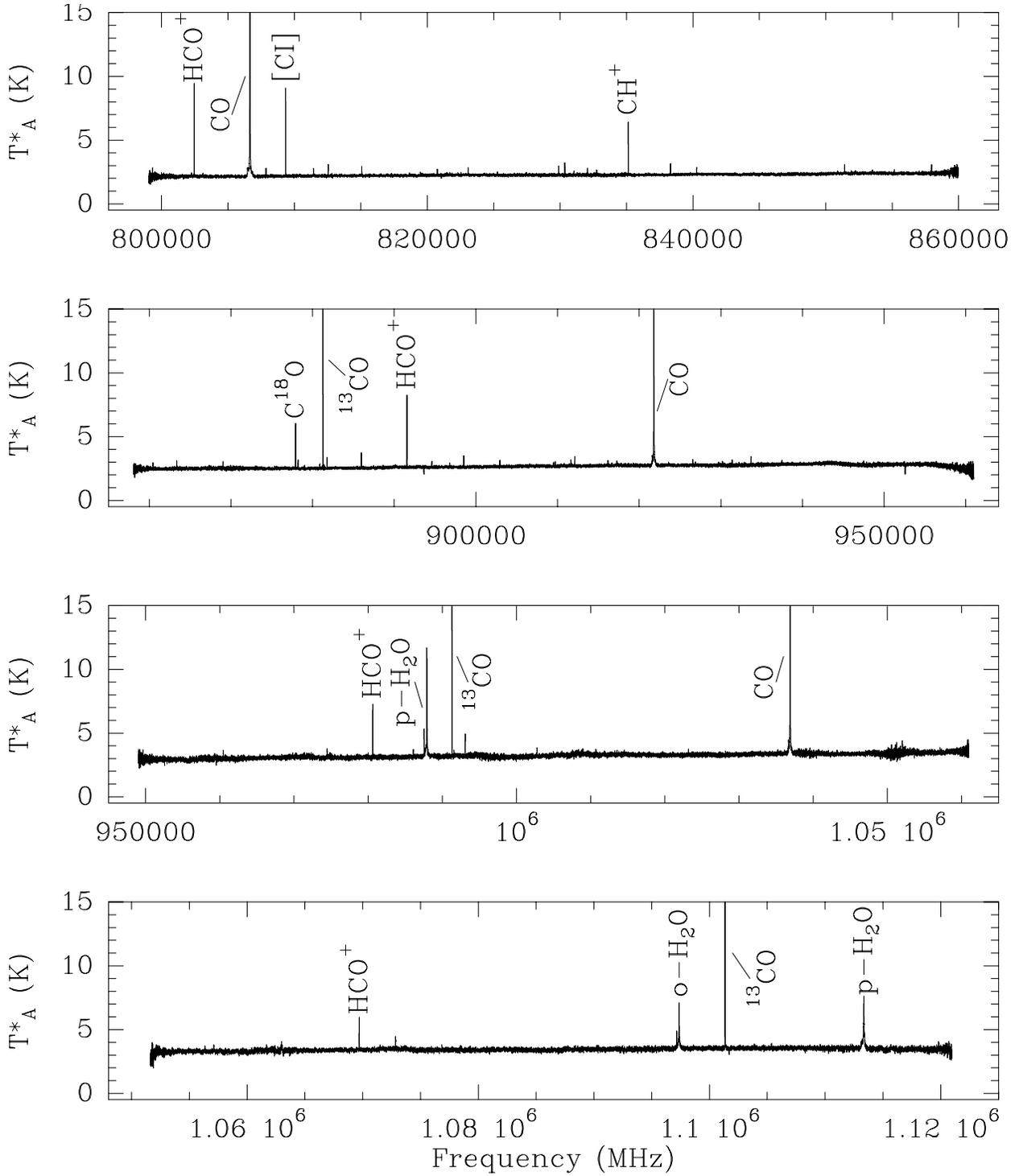}
\caption{Same as Figure 1 except for band 3a, band 3b, band 4a and band 4b.}
\label{fig:1.2}
\end{figure}

\begin{figure}[b]
\centering
\includegraphics[width=\textwidth]{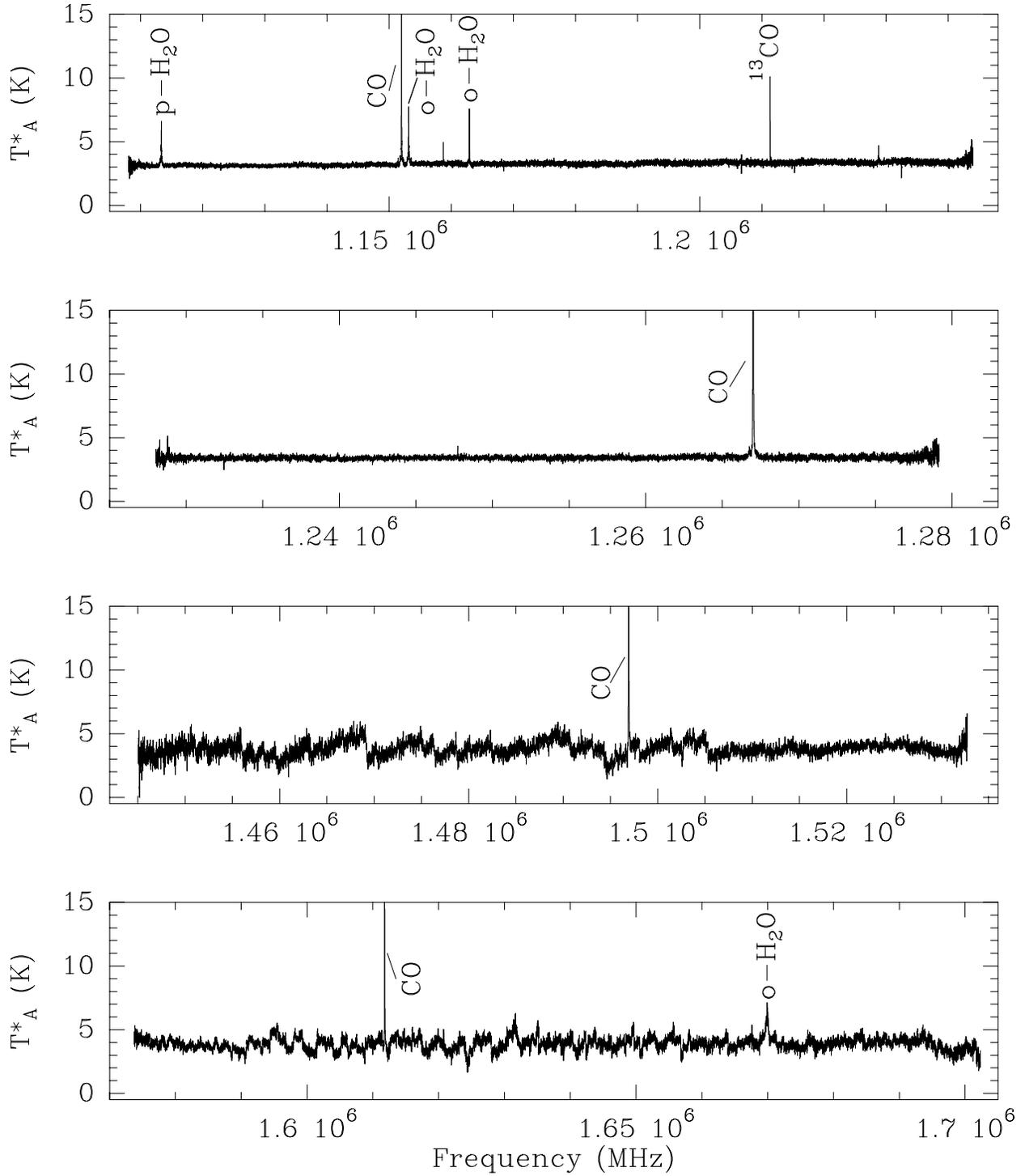}
\caption{Same as Figure 1 except for band 5a, band 5b, band 6a and band 6b. The higher noise level in band 6 is due to the HEB mixers which produce higher noise in comparison with the SIS mixers used in the first five bands.}
\label{fig:1.3}
\end{figure}

\begin{figure}[b]
\centering
\includegraphics[width=\textwidth]{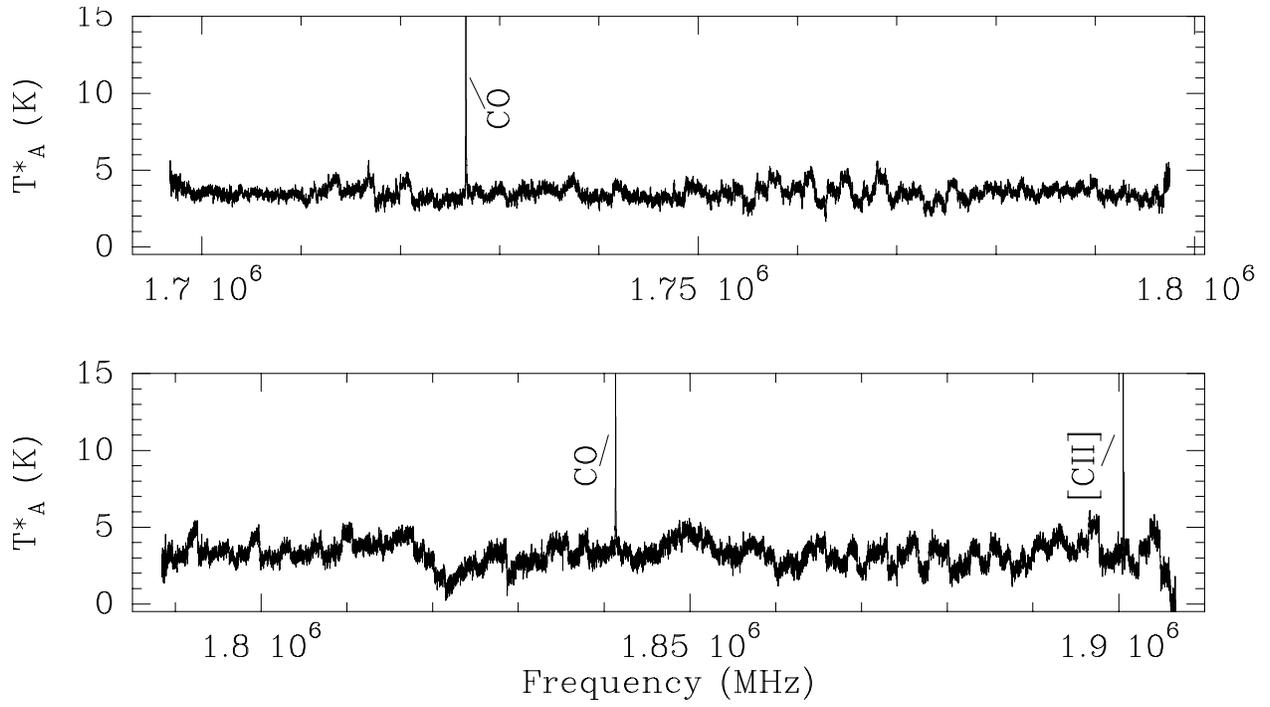}
\caption{Same as Figure 1 except for band 7a, and band 7b. The higher noise level in band 7 is due to the HEB mixers which produce higher noise in comparison with the SIS mixers used in the first five bands.}
\label{fig:1.4}
\end{figure}

All data in this paper are presented on the T$_{A}$ temperature scale and, for subsequent analysis, were converted to T$_{mb}$ using the main beam efficiencies in \cite{Mueller:2014}\footnote{\url{http://herschel.esac.esa.int/twiki/bin/view/Public/HifiCalibrationWeb#HIFI_performance_and_calibration.}}.  The final data after deconvolution and spectral smoothing (bands 1a$-$7b) are shown in Figures \ref{fig:1.1}-\ref{fig:1.4}, in which the  T$_A$ range is in Kelvin and frequency in MHz. The strongest lines are labeled in each band, and in order to  make the residual noise recognizable and comparable from one band to another the intensity is fixed to 15 K for all bands. \textbf{Note that certain broad features, like the one near 790 GHz, are most likely due to excess noise, since individual observations show quite a few noise spikes in these spectral ranges. }

\subsection{Removal of Off-Position Contamination}
\label{contamination}

As described in the previous section, the HEXOS Orion-S spectral survey was observed in DBS mode.\textbf{ Since the observations used the chopper in this mode, the reference positions were fixed to $\sim$3 arc minutes from the target position, with an angle set by observatory constraint.} In a crowded field like the Orion Molecular Cloud region, it is very likely that the reference position is not free of emission or absorption for some or all molecular lines detected. Typically, emission (absorption) in the reference position appears as an absorption (emission) like feature in the final spectrum. Since the case of absorption in the reference beam is rare, due to a low continuum flux, we will consider only emission. Figure \ref{Orion_KL_dust} an image of the dust emission at a wavelength of  $250~\mu m$ obtained with the \textit{Herschel}/SPIRE instrument. The positions of the Orion-S observations and the two reference observations for each of the 14 HIFI bands are overlaid. The diameters of the circles shown represent the FWHM of the individual beams for the center frequencies of each HIFI band. It is apparent that a few of the reference observations (on the east side) were located near the Orion Bar region, making reference beam line contamination very likely. In addition, we even captured emission in the lower-J $^{12}$CO transitions in the opposite chopping direction.

The identification of potentially contaminated lines was first performed after the deconvolution by checking the line profile of the detected lines. Once lines were identified, we performed additional tests by subtracting the Level 0.5 nod2 reference spectrum from the associated nod1 reference spectrum of the scans that cover the frequency ranges of these lines. If the resulting spectrum showed only noise we assumed no emission in the reference spectra (we never experienced the case that emission in both reference beams cancelled out perfectly, which would hide this problem). If the resulting spectrum showed an emission line, there was emission in the nod1 reference beam; and if the spectrum showed an absorption feature, there was emission in the nod2 reference beam.

In order to remove the emission in the reference beams, we used the \textit{Herschel}/HIPE hifiPipeline task to create the Level 0.5 product, in which the reference observations are still separated from the target observations. A custom HIPE script then extracted all affected reference scans and the neighboring scans taken before and after the affected scans that have slightly changed LO-settings such that the contaminating emission is at sufficiently different intermediate frequencies (IF). The repair is based on the assumption that the bandpass of all observations changes only very little with small changes in the LO-setting. The primary change is in the amplitude, while the shape of the bandpass changes negligibly. Thus, we could use the neighboring reference scans to repair the contaminated reference flux.

The first step of the repair was to determine the IF-frequency interval [$f_{l}$, $f_h$], covering the reference beam emission and two smaller, abutting intervals [$f_{l} - \Delta_l$, $f_l$] and [$f_{h}$, $f_h + \Delta_h$], indicated by the green areas in Fig \ref{hexos_os_contsm}, for scaling. The next step included extracting and averaging the flux of the neighboring scans over the frequency range [$f_{l} - \Delta_l$, $f_h + \Delta_h$] (Ref 1 and Ref 2 in Figure  \ref{hexos_os_contsm}). Then, to properly scale the averaged flux to replace the contaminated flux, we determined the ratios of the original reference flux to the averaged flux over the two green intervals, interpolated the corresponding values over the interval containing the contaminated flux (the white area in Figure \ref{hexos_os_contsm}), and calculated the new reference flux by multiplying the averaged flux with the just determined ratio over the entire frequency range [$f_{l} - \Delta_l$, $f_h + \Delta_h$]. This new reference flux (New Ref) now replaced the original reference flux (Orig Ref). From here on, we continued to use the \textit{Herschel}/HIPE hifiPipeline task to create the Level 1.0 and higher products.

The lines that needed repairs are: B1a ([CI]), B2b (H$_2$S, C$^{18}$O, $^{13}$CO, C$^{17}$O), B3a (CO, [CI], CH$^+$), B3b (C$^{18}$O, $^{13}$CO, C$^{17}$O, CO), B4b (H$_2$S, C$^{18}$O, $^{13}$CO), B5a (C$^{17}$O, CO, C$^{18}$O, $^{13}$CO), B7a (CO), and B7b (CO, [CII]). Figure \ref{hexos_os_final} shows an example of how the repair recovered the true line profile of the [CII] 158 $\mu m$ line.

\subsection{Line Identification}
\label{sec:id}

We used CASSIS\footnote{CASSIS was developed by IRAP-UPS/CNRS.  See: \url{http://cassis.irap.omp.eu}}
, a Java based software package designed to analyze astrophysical spectroscopic data to perform the line identification and modeling. Our line identification procedure involved two main steps. First we visually identified the strongest (well above 5$\sigma$ signal-to-noise) and best known emission lines in the spectrum (e.g. from CO, CS, HCO$^+$, HCN, H$_2$O, etc.) and some of their isotopologues utilizing the JPL\footnote{http://spec.jpl.nasa.gov/ftp/pub/catalog/doc/catintro.pdf} \citep{Pearson:2005IAUS..235P.270P} and CDMS\footnote{http://www.astro.uni-koeln.de/cdms/catalog} \citep{Muller:2005JMoSt.742..215M} spectral line databases.  These databases include tabulated values of the central frequency error for each transition. Although in some cases, the difference between the listed centroid  frequency of a particular transition from these two catalogues is larger than their given error bars, the observed line width of the transition usually compensates for this ambiguity and makes the identification robust.  Many of the strongest identified species are shown in Figures \ref{fig:1.1}-\ref{fig:1.4}.  In these cases, line blending (e.g. the appearance of more than one transition/species at a single frequency) is not considered a problem since the emission from the well-known species will invariably overwhelm the weak emission from a less well-known and, presumably, lower abundance blended line.   

\textbf{Once the strongest emission lines were accounted for, we examined} all other spectral lines in our data that had a signal-to-noise ratio above 3$\sigma$ (in peak intensity).  We first performed the line identification via visual inspection of each spectral feature in each HIFI band and compared the transition's frequency to those listed in the databases. From the possible database entries we investigated all species with transitions that fell within a Doppler velocity range of 5.5 to 8.5 km s$^{-1}$ (i.e. within $\pm$1.5 km s$^{-1}$ of the assumed central velocity of Orion-S). Within this velocity range we examined a smaller sub-sample of possible spectral lines with upper state excitation energies (E$_{\rm up}$) less than 1500 K. If a single database entry from this sub-sample matched the observed spectral feature, we considered this to be a tentative identification. In order to confirm or reject this tentative identification, we then searched for other predicted transitions of the selected species in all of the HIFI bands.  If we saw other spectral features in the data that matched the predicted frequencies, we accepted the initial line identification as likely correct.  If we did not, then the initial line identification was still considered to be only tentative, since we realize that the absence of other predicted transitions may be due to special excitation conditions.  Therefore, in both cases, in order to finally confirm or reject our initial line identifications,  Local Thermodynamic Equilibrium (LTE) modeling was performed (described in detail in Section 3), which allowed us to determine if all the observed spectral features  from the tentatively detected species could be theoretically reproduced under uniform excitation conditions.  

Our LTE models explored excitation conditions with T$_{ex} \le$ 1500 K (where T$_{ex}$ is the excitation temperature - equal to the kinetic temperature in LTE),  E$_{\rm up}$ $\le$ 1500 K, and total species column density $\le$  10$^{17}$ cm$^{-2}$. If the LTE model produced emission at the frequency of the spectral feature then  the line identification was considered confirmed. If not, the species was assumed to have been incorrectly identified and a new identification for that spectral feature was sought.  Note, at this stage we are simply trying to produce some visible model emission at the frequency of the spectral feature and not trying to fit or replicate the observed spectral line profile. This will be performed in a subsequent stage described in Section 3.1. If a spectral feature could not be reproduced by an LTE model of any species, or if there was no database entry at the frequency of the observed spectral feature, that feature was listed as an unidentified line (32 lines in total). Visual inspection of the original,  DSB spectra indicates that all of these features are ``ghosts" (i.e. artifacts of the deconvolution routine). A list of all identified species is given in Table  \ref{tab:IdentifiedSpecies}.
A frequency ordered list of all spectral features above $3\sigma$ in intensity (as well as their peak intensity) is given in Table \ref{tab:Frequencies}. Ghost lines are identified as ``ghost". In total we identified 52 different species (including isotopologues) which are responsible for 685 transitions (including the blended lines) in the HIFI spectra. It is, of course, possible that additional species and transitions exist in Orion-S, but at intensities too weak to be detected. This will be addressed in Section 3.3.2. 

In some cases, our LTE modeling resulted in a particular spectral feature being reasonably explained by a superposition of lines from more than one species/transition (i.e. blended lines). In order to account for the possible effects of line blending, we performed modeling of all transitions of the two species in all HIFI bands.  After obtaining good fits to the unblended transitions, we were able to determine how much each species contributed to the blended feature. An example of a blended line is shown in Figure \ref{fig:Blended}, in which the data are shown by the black histogram, the solid red line indicates a C$^{17}$O transition, the solid blue line indicates an H\2CO transition and the solid green line is the superposition of these two model results. Note that the solid green line in Figure \ref{fig:Blended} does not represent a multicomponent Gaussian fit but rather the LTE modeling required to reproduce the line profiles (see Section \ref{ltemodeling}). Given the relatively few spectral lines in Orion-S, significant blending was only a problem in 6 of 685 lines detected.  These blended lines are indicated by a ``b" superscript in Tables \ref{tab:Frequencies}  and \ref{tab:Gaussian}. In Table \ref{tab:Frequencies} both species are listed. Blended lines were excluded from the modeling analysis presented in Section \ref{sec:result}. 

\subsection{ Line Profiles}
\label{sec:gauss}

Although all lines above the 3$\sigma$ S/N level in intensity were identified, Gaussian fitting and subsequent modeling was only performed on  lines that were above the 5$\sigma$ noise level (where the noise is calculated from line-free regions of the spectrum immediately adjacent to the line).  Gaussian fits were obtained using the Levenberg-Marquardt algorithm as implemented in the CASSIS software package and was done independently from LTE modeling, the latter of which will be described in Section 3.1. A linear or a second order baseline was fit to the data prior to Gaussian fitting, but was not removed so we could include the continuum in subsequent modeling (important for absorption lines). 

\begin{figure}
\centering
\includegraphics[scale=0.7]{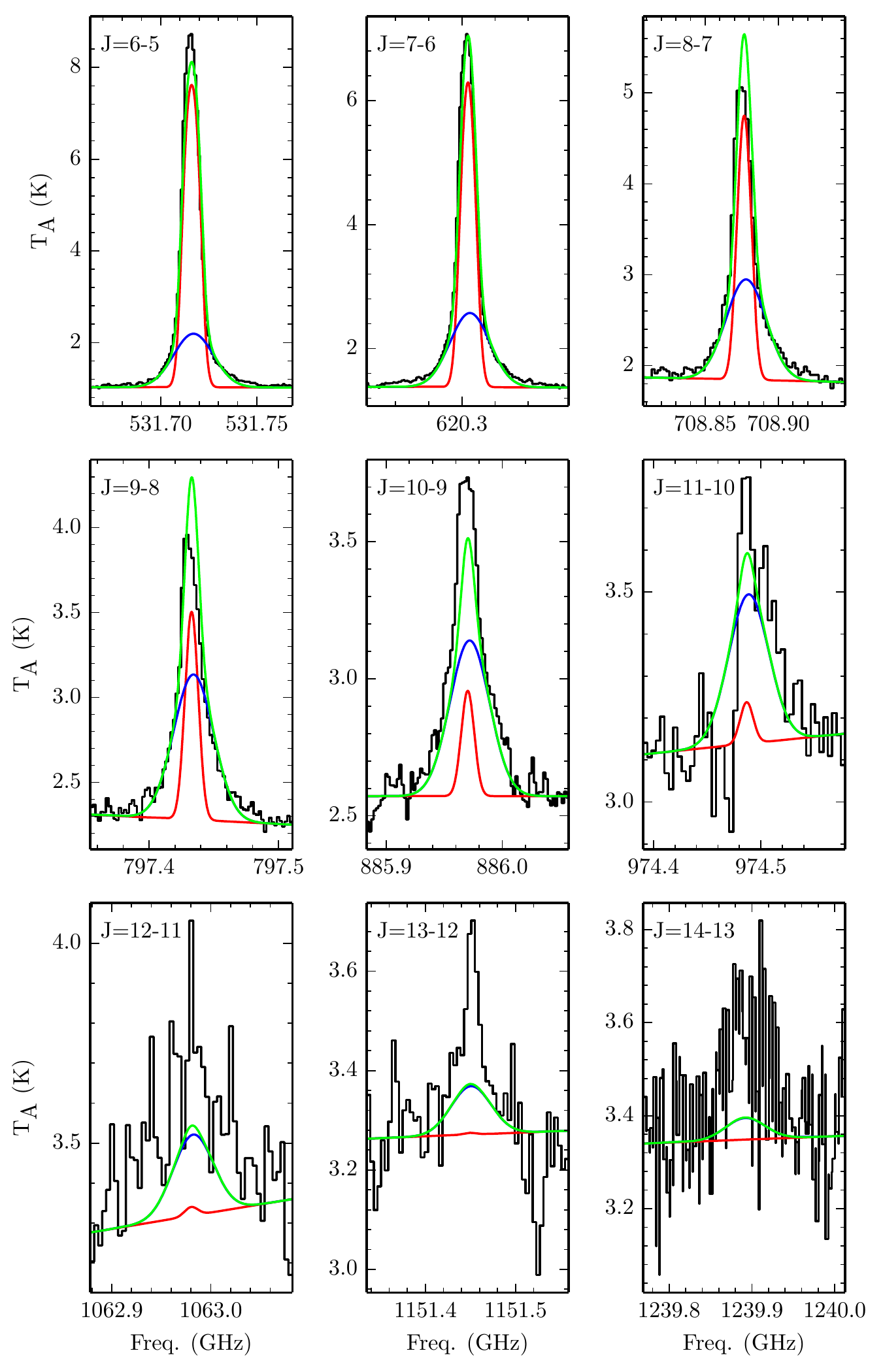}
\caption{Two component LTE modeling for HCN. Black histogram shows the data.  Models of the narrow and the broad components are shown by the red and blue lines respectively. The superposition of the two components is shown in green. LTE model parameters are provided in Table \ref{tab:LTE2}.} 
\label{HCN}
\end{figure}

In most cases, a single component Gaussian fit to a specific species could reasonably reproduce the observed lines. However, in some cases a two component Gaussian fit was needed, one component being narrow ($\Delta V_{FWHM}$ = 3$-$5 km s$^{-1}$) and the other broad ($\Delta V_{FWHM}$ = 7$-$14 km s$^{-1}$); the latter could be the effect of a hot core, an outflow, or a shock. Figure \ref{C13O} provides an example of a species that needed only a single component fit ($^{13}$CO), whereas Figure \ref{HCN} shows the spectra for HCN, a typical example of a species requiring a two component fit. In addition, $^{12}$C$^{16}$O line profiles are the only ones among 52 identified species, which clearly had a non-Gaussian shape, probably due to self absorption. A few other transitions are seen in absorption rather than emission and are listed in Table \ref{tab:Frequencies} with negative intensities. 

Table ~\ref{tab:Gaussian} shows the results of the Gaussian fitting for each species. The reported result for each individual line profile is the best Gaussian fit. We should note that, usually, the signal to noise ratio is lower, as we move to higher frequencies. In case of a two component fit, the narrower component is referred as ``main" and the broad component is referred to as ``wing". The first column in Table ~\ref{tab:Gaussian} is the transition quantum number (an explanation of the quantum numbers is provided on the CDMS  and JPL websites). The centroid frequency of the fitted Gaussian profile is listed in the second column. T$_A$ and $V_{LSR}$ are respectively the observed antenna temperature and centroid velocity of the corresponding Gaussian fit. $\Delta V_{FWHM}$ is the ``Full Width Half Maximum", and $\int{T_A}dV$ (km s$^{-1}$) is the integrated line intensity. 

\begin{figure}
\centering
\includegraphics[scale=1.2]{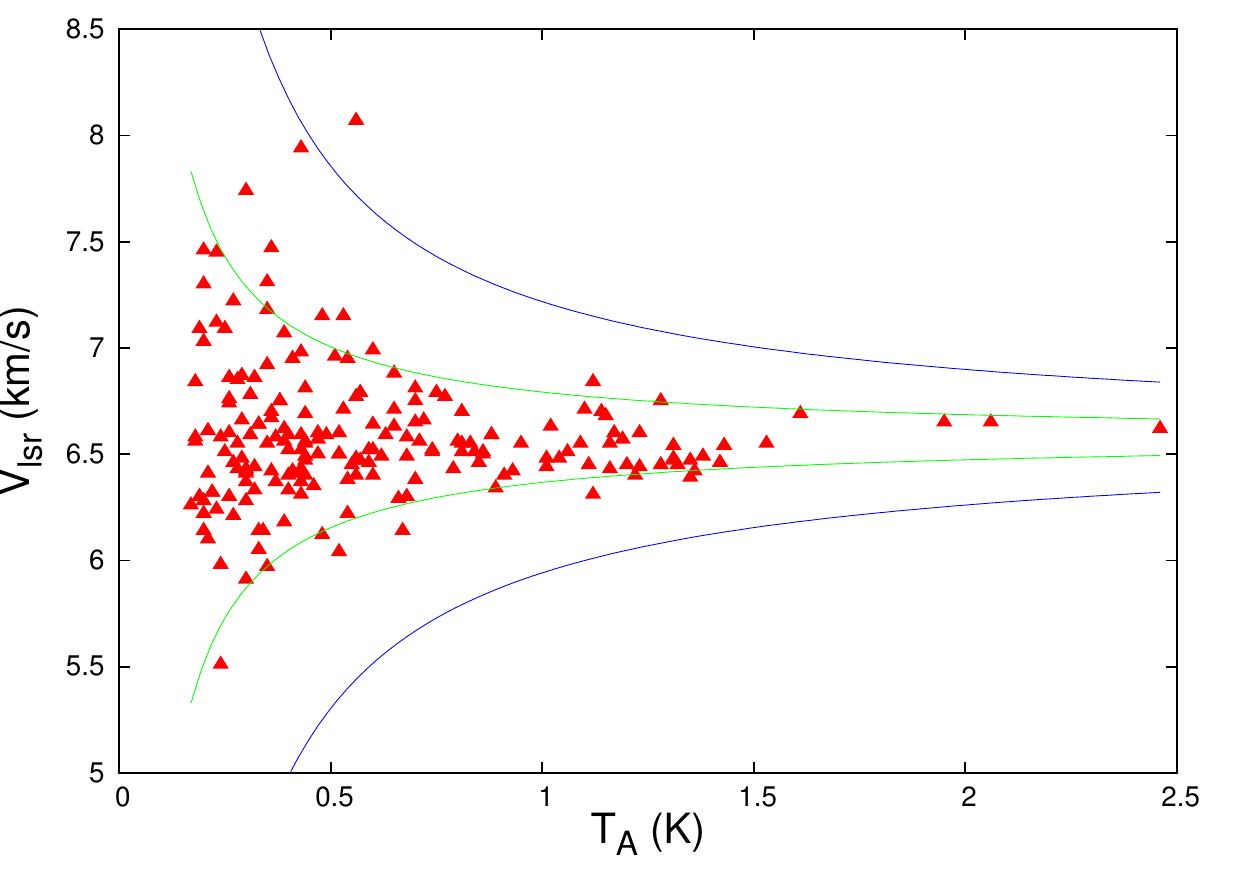}
\caption{Plot of the measured V$_{LSR}$ vs T$_A$ derived from independent Gaussian fits to each of the A-CH$_3$OH and E-CH$_3$OH transitions listed in Table \ref{tab:Gaussian}. The green and blue lines are the 1$\sigma$ and 3$\sigma$ (respectively) theoretical error envelope for the V$_{LSR}$ determined from Gaussian fitting of noisy lines predicted by \cite{2004A&A...428..327P}. Curves are calculated assuming $<\Delta V_{FWHM}> \sim 5$ km s$^{-1}$ and T$_{rms} \sim 0.1$ K (typical values for methanol).}
\label{fig:VvsT}
\end{figure}

\begin{figure}
\centering
\includegraphics[scale=1.2]{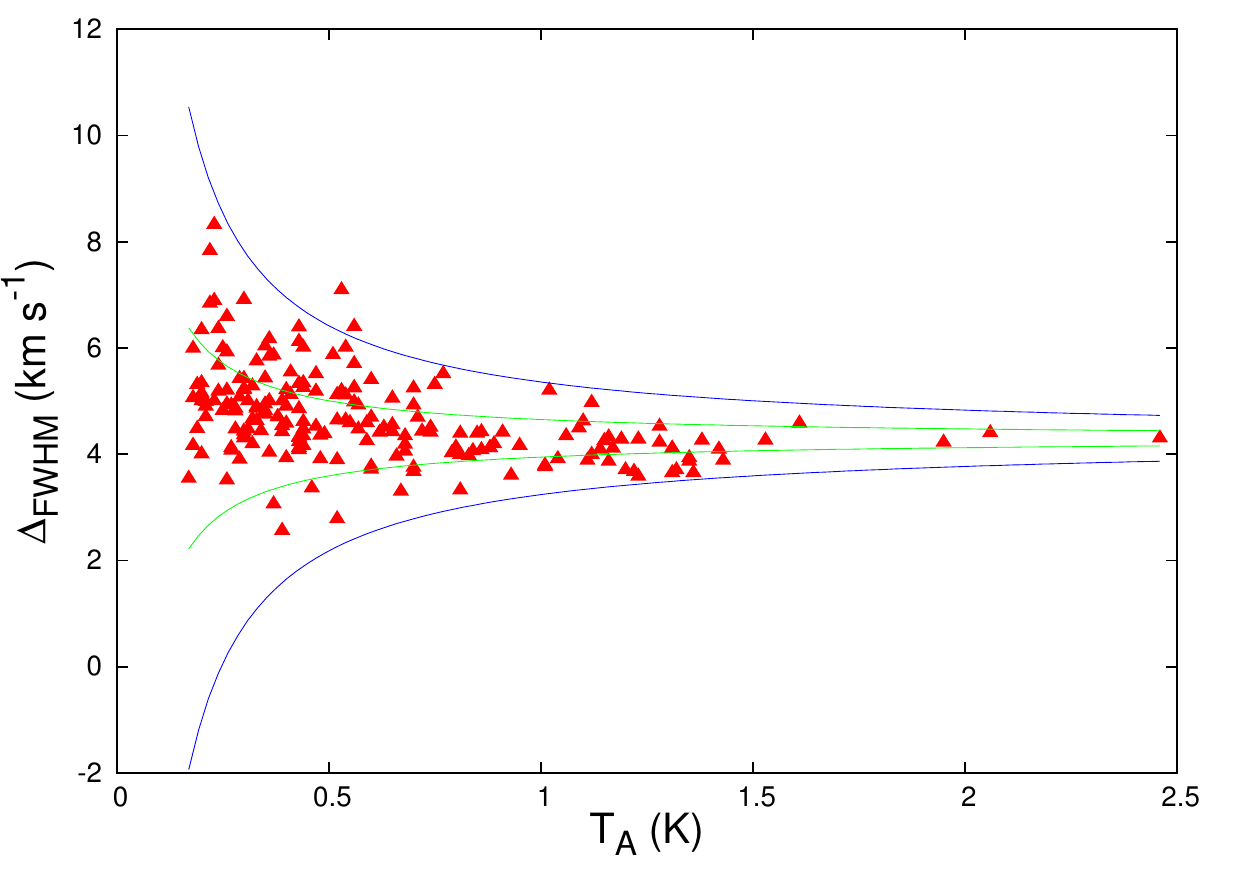}
\caption{Plot of the measured $\Delta$V$_{FWHM}$ vs T$_A$ derived from independent Gaussian fits to each of the A-CH$_3$OH and E-CH$_3$OH transitions listed in Table \ref{tab:Gaussian}. The green and  blue lines are the 1$\sigma$ and 3$\sigma$ (respectively) theoretical error envelope for the $\Delta_{FWHM}$ determined from Gaussian fitting of noisy lines predicted by \cite{2004A&A...428..327P}. Curves are calculated assuming $<\Delta V_{FWHM}> \sim 5$ km s$^{-1}$ and T$_{rms} \sim 0.1$ K (typical values for methanol)}
\label{fig:dVvsT}
\end{figure}

Note that Gaussian fitting was performed on each line separately (i.e. we did not utilize a single set of Gaussian parameters to fit all transitions simultaneously). This implies that each transition of a given species can have slightly different V$_{LSR}$ and $\Delta V_{FWHM}$.  This effect is best demonstrated using methanol  as an example, since it has the largest number of transitions.  The mean values of V$_{LSR}$ and $\Delta V_{FWHM}$ for methanol A \& E combined is 6.6$\pm$0.3 \kms\ and 4.7$\pm$0.9 \kms\ respectively, where the errors are the 1$\sigma$ standard deviations about the mean.  The calculated scatter about the mean, however, is an intensity dependent parameter. Figures \ref{fig:VvsT} and \ref{fig:dVvsT} plot V$_{LSR}$ and $\Delta V_{FWHM}$ vs T$_A$ ($>$ 5$\sigma$) for all the fitted methanol A \& E transitions in Table ~\ref{tab:Gaussian}, and clearly show that the scatter in both fitted parameters decreases with increasing T$_A$. For transitions with T$_A <$ 0.5 K,  $<$V$_{LSR}> = 6.59\pm 0.39$ \kms\ and $<\Delta V_{FWHM}> = 4.99\pm 0.92$ \kms.  Whereas for transitions with T$_A >$ 0.5 K,  $<$V$_{LSR}> = 6.57\pm 0.23$ \kms\  and $<\Delta V_{FWHM}> = 4.39\pm 0.67$ \kms.  This suggests that most of the observed scatter in these parameters is not due to the emission itself but it is due to our Gaussian fitting procedure, which clearly is subject to larger errors for weaker lines. This behaviour is predicted by \cite{2004A&A...428..327P} who show that the error in  V$_{LSR}$ and $\Delta V_{FWHM}$ from Gaussian fitting increases with decreasing signal to noise. The green and the blue lines on Figures \ref{fig:VvsT} and \ref{fig:dVvsT} denote, respectively, the 1$\sigma$ and 3$\sigma$ theoretical error envelope. These were calculated from equation A.1 in \cite{2004A&A...428..327P} assuming $<\Delta V_{FWHM}> \sim 5$ km s$^{-1}$ and T$_{rms} \sim 0.1$ K (typical values for methanol) and  illustrate this effect quite clearly.

\begin{figure}
\centering
\includegraphics[scale=1.2]{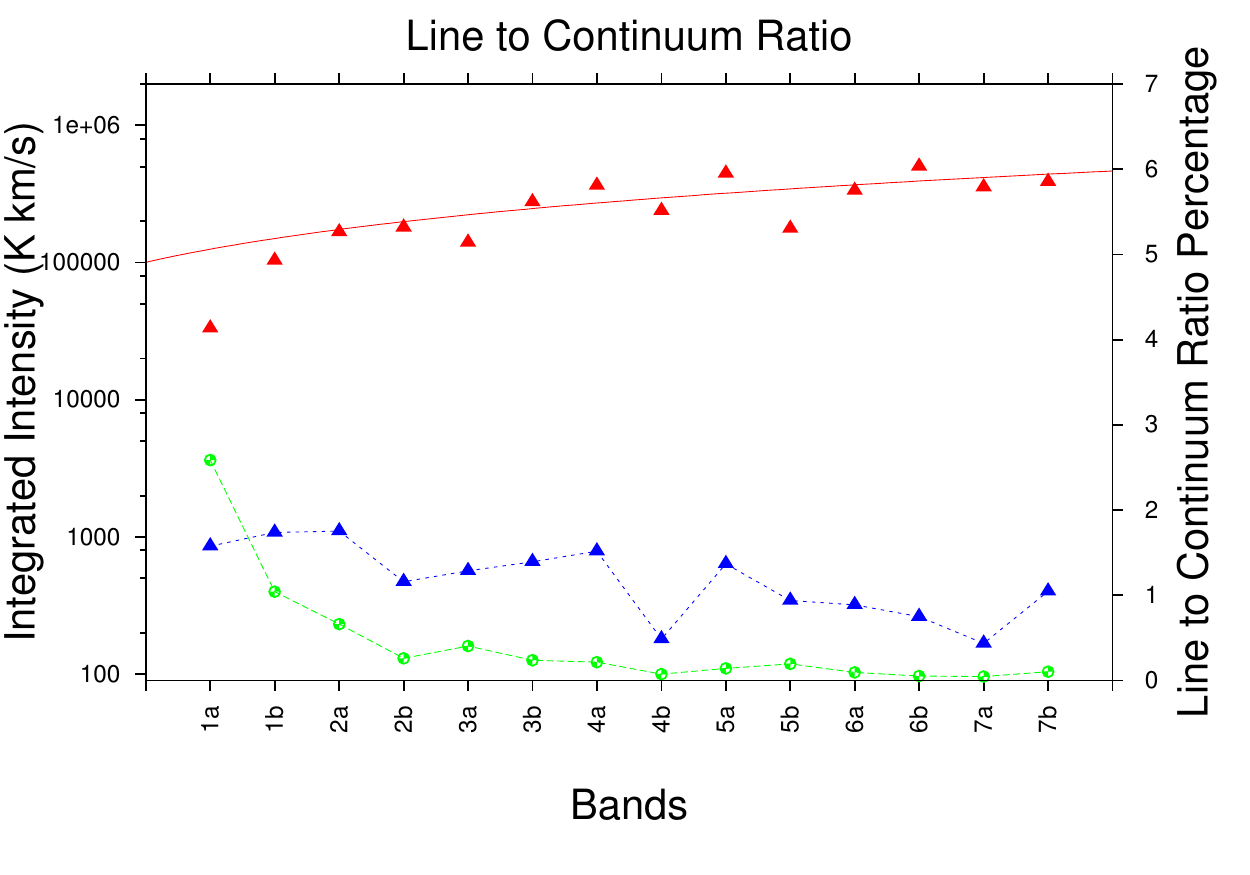}
\caption{Plot of the continuum emission integrated over each band  (red triangles),  integrated line emission in each band (blue triangles), and  the line to continuum ratio percentage (green circles). The red line is a power law fit to the continuum emission using a modified black body function (see text for details). 
}
\label{fig:ltoc}
\end{figure}

HIFI data obtained in beam switching mode generally provide quite a good measure of the continuum. Therefore, in each band, we have integrated the emission over the entire frequency range to obtain the line+continuum emission. Summation over the integrated intensities, listed in Table \ref{tab:Gaussian}, of all the transitions in each band provides the corresponding total line emission.
Comparing the two provides the line to continuum ratio, which is interesting for the interpretation of broadband continuum images of star forming regions.  The advantage of our  data is that the line and continuum emission are measured in the same beam with the same instrument and, therefore, there are no complications that arise from cross-calibration between different instruments or beam sizes.   
Figure \ref{fig:ltoc} plots the band integrated continuum emission (red triangles), the integrated line emission in each band (blue triangles) and, on the right y-axis, the percentage line to continuum ratio (green circles).  The figure shows that the line to continuum ratio is $\sim 3-1$\% in bands 1a$-$2a and drops to less than $\sim 0.5$\% in the higher bands.  These are both smaller than the $\sim 10$\% seen in Orion-S in the 300 GHz band 
\citep{groesbeck:1995} suggesting that the line to continuum ratio generally decreases with increasing frequency. 
The dramatic drop in the line to continuum ratio between Band 1a to 2b is due to two factors.  Over this frequency range, the continuum emission rises by a factor of a few while, at the same time, the number of spectral lines and their corresponding intensity drops by a factor of a few. The red line is a power law fit to the red triangles using a modified black body in which the Planck function is multiplied by $\kappa_o (\nu/\nu_o)^\beta$, where $\kappa_o$ is the dust mass opacity coefficient.  The best fitted value for $\beta$ is 1.0. This value is consistent with the behaviour of dust in other studies of star forming regions \citep[e.g.][]{2009ApJ...696.2234S}.

\section{RESULTS \& DISCUSSION}
\label{sec:result}

\subsection{LTE Modeling}
\label{ltemodeling}

LTE modeling assumes that the gas is in Local Thermodynamic Equilibrium, meaning that the density is sufficiently high that collisions dominate the excitation. The LTE modeling capability implemented in CASSIS has 5 input variables, $\boldsymbol{N_t ,}$ $\boldsymbol{T_{ex} ,}$ $\boldsymbol{V_{LSR} ,}$ $\boldsymbol{\Delta V_{FWHM} ,}$ $\boldsymbol{\Omega}$, where $N_t$ is the total column density, T$_{ex}$ is temperature, and $\Omega$ is the size of the emitting region (which couples to the variable HIFI beam sizes to take into account beam dilution effects).  Note that by definition, under LTE conditions, the excitation temperature that determines the relative populations of the upper and the lower level of a spectral line, the rotation temperature that describes the populations of all the rotational levels of one species and the gas kinetic temperature are all identical. Each combination of these variables produces a Gaussian model spectrum for each transition of the selected species. Note that, unlike the Gaussian fitting procedure which fits the V$_{LSR}$ and $\Delta V_{FWHM}$ to each line separately, for the LTE modeling we obtain a single average value of V$_{LSR}$ and $\Delta V_{FWHM}$ for all transitions of a given species.

In order to find the set of parameters which produce synthetic spectra that best fit the observed spectral line profiles, we used a Markov Chain Monte Carlo (MCMC) method implemented in CASSIS \citep[e.g.][]{Guan:2007math......3021G}. The MCMC method randomly picks a seed in the five dimensional parameter space (that  we call the $X_0$ state). Then it randomly chooses one of the nearest neighbors (called the $X_1$ state), as specified by a variable step size, which is calculated for each iteration. The $\chi^2$ of the new state is calculated and if $p = \chi^2(X_0) / \chi^2(X_1) >1$ then the new state is accepted. If $p =\chi^2(X_0) / \chi^2(X_1) <1$ this new state might still be accepted with a certain acceptance probability. If the new state is rejected the $X_0$ state will remain, and another random nearby state will be picked as the $X_1$ state. Having a finite probability to accept a new position even if the $\chi^2$ is worse, ensures we do not converge directly to a local minimum, but instead forces better sampling of the full parameter space. The code runs with several initial random states and, usually, when the variance among different clusters of states is smaller than the variance of each cluster, it is assumed to have converged to the correct solution \citep[][]{Hastings, Roberts}.  When the code approaches convergence, it calculates a number of models and $\chi^2$ values in a tight cluster surrounding the ``best'' solution.  This allows us to calculate a median value for each fitted parameter and its statistical standard deviation, which are listed in Table \ref{tab:LTE}.

Despite the fact that we identify all transitions above 3$\sigma$, for our modeling we only utilize transitions above 5$\sigma$, while neglecting the blended lines. However, when exploring the validity of our models we also investigate frequency regions where potential transitions of the selected species exist in the molecular line databases, but were not  detected above 3$\sigma$. This ensures that our models do not produce synthetic spectra where no transitions are actually observed. At the beginning of the procedure we usually let all five parameters vary. However, frequently we were able to find good solutions for the $V_{LSR}$, and FWHM after the first convergence of the code. Therefore, on subsequent runs, we fixed the $V_{LSR}$ and FWHM and allowed the other three parameters to vary. This significantly speeds up the computational time of subsequent runs. Once we obtain a good fit, we run the code five to ten more times  to ensure that different runs converge to the same solution within the error bars. In some cases, after running the code five to ten times, the scatter of the converged solutions  is larger than the standard deviation of any of the individual solutions. In this case, in Tables \ref{tab:LTE}  and \ref{tab:LTE2} we report the average of the multiple runs (i.e. we take the median value of each run and compute the average between runs) and the standard deviation of the solutions about this new average.  In all cases we let the source size ($\Omega$) vary up to 90$''$ twice that of the largest HIFI beam ($\sim 45''$). However, if the source size is larger than the largest beam it is essentially unconstrained (although there is some sensitivity to source sizes that are larger than the beam since the beam is Gaussian in shape and not a tophat profile). In such cases, the source is simply considered to be extended in nature. Table \ref{tab:LTE} provides the results of our MCMC $\chi^2$ fitting of the spectral lines listed in Table \ref{tab:Gaussian} (i.e. those with S/N $> 5 \sigma$).  Column 1 is the species, column 2 lists the median total column density of the species and the standard deviation, column 3 is the excitation temperature, column 4 the FWHM line width, column 5 is the source size ($\Omega$), and column 6 is the median LSR velocity.  In cases where the error is not listed, the parameter was fixed in the MCMC fitting routine. Table \ref{tab:LTE} shows that the column density uncertainties range from 10$-$50\%.  \textbf{To ensure that our assumption of LTE is valid,  we ran the same MCMC fitting procedure using the non-LTE (RADEX) models implemented in CASSIS for a handful of species  (N$_2$H$^+$, SO$_2$, $^{13}$CO, and DCO$^+$).  In all cases, the non-LTE column densities are consistent and within the reported error bars of the LTE models, as shown in Table \ref{tab:LTE}.} Given that the errors in Table \ref{tab:LTE} are the statistical uncertainties on the LTE solutions, to account for the possibility that LTE is not always a good approximation, we suggest that uncertainties on the high side of this range are probably appropriate.

Figure \ref{C13O} provides an example of one component modeling for $^{13}$CO. The apparent shift in centroid velocity between the data (black histogram) and the LTE model (red Gaussian curve) is seen in a number of species and can also be seen by comparing the tabulated V$_{LSR}$ listed in Tables \ref{tab:Gaussian} and \ref{tab:LTE}. These apparent shifts of a few tenths of a km s$^{-1}$ are caused by  the fact that the spectral lines are not perfectly Gaussian in shape; in some cases possibly due to optical depth effects. In addition, the MCMC routine optimizes a number of free parameters to obtain the best overall physical model which fits all spectral lines simultaneously; as opposed to the Gaussian fitting routine, which simply fits a mathematical Gaussian profile to each spectral line separately. Thus, the median V$_{LSR}$ and $\Delta$V$_{FWHM}$ determined from the LTE modeling may not perfectly match the actual V$_{LSR}$ and $\Delta$V$_{FWHM}$ of any individual transition.

Although one component modeling usually results in a remarkably good fit to the observations (e.g C$^{18}$O, CH, CCH, etc), there are some cases in which a second (broad) component   is necessary to properly reproduce the observations.  This is independent of the  broad line wings seen in the transitions of some species that required a two component Gaussian fitting as mentioned in  Section \ref{sec:gauss}.  In some cases, even species that were well fit by a single Gaussian component required two component LTE modeling, one with a narrow line width, and the other with a broad line.  This is because, in these cases, there is no single combination of model parameters (notably T$_{ex}$ and N$_{tot}$) that could reproduce the intensities of all transitions simultaneously. The two component LTE modeling implementation in CASSIS uses a two slab model; from the perspective of the observer, component 1 is the front slab and component 2 is the slab located behind it.  The code allows  component 1 to absorb emission from component 2. The results of the two component MCMC LTE modeling are listed in Table \ref{tab:LTE2}. Figure \ref{HCN} provides an example of a species for which two component modeling \textbf{was required. To see} if this could be the result of non-LTE effects, we also attempted to model these species  using the RADEX code \citep{2007A&A...468..627V} as implemented in CASSIS. For the RADEX modeling we used the identical parameter range as for the LTE modeling. RADEX, however, invokes one additional free parameter, namely the H\2 volume density which we allowed to range from 10$^2$ to 10$^{10}$ cm$^{-3}$. In all cases, the  RADEX modeling was unable to produce a good fit to all transitions unless a second physical component was included.

Note that $^{12}$CO is not presented in either Table  \ref{tab:LTE}  or Table  \ref{tab:LTE2}  due to the presence of self-absorption from foreground material which complicated the modeling procedure.  We do, however, model the broad shock/outflow component separately for the analysis presented in Section \ref{comparison}.

\subsection{Comments on Individual Species}

The detected species listed in Table \ref{tab:IdentifiedSpecies} can be related to a variety of physical processes that exist in the ISM such as:  shocks, UV irradiation by nearby OB stars, and hot core chemistry. In this section we discuss \textbf{some specific molecules in the context of these} physical processes.

\subsubsection{Tracers of UV Irradiation }

Given the high UV flux in the Orion-S region  \citep[$\chi = 1.1 \times 10^5 \chi_0$; ][]{Herrmann:1997ApJ...481..343H}, it is not surprising that we detect  a wide variety of UV tracers.

\begin{figure}
\centering
\includegraphics[scale=1.2]{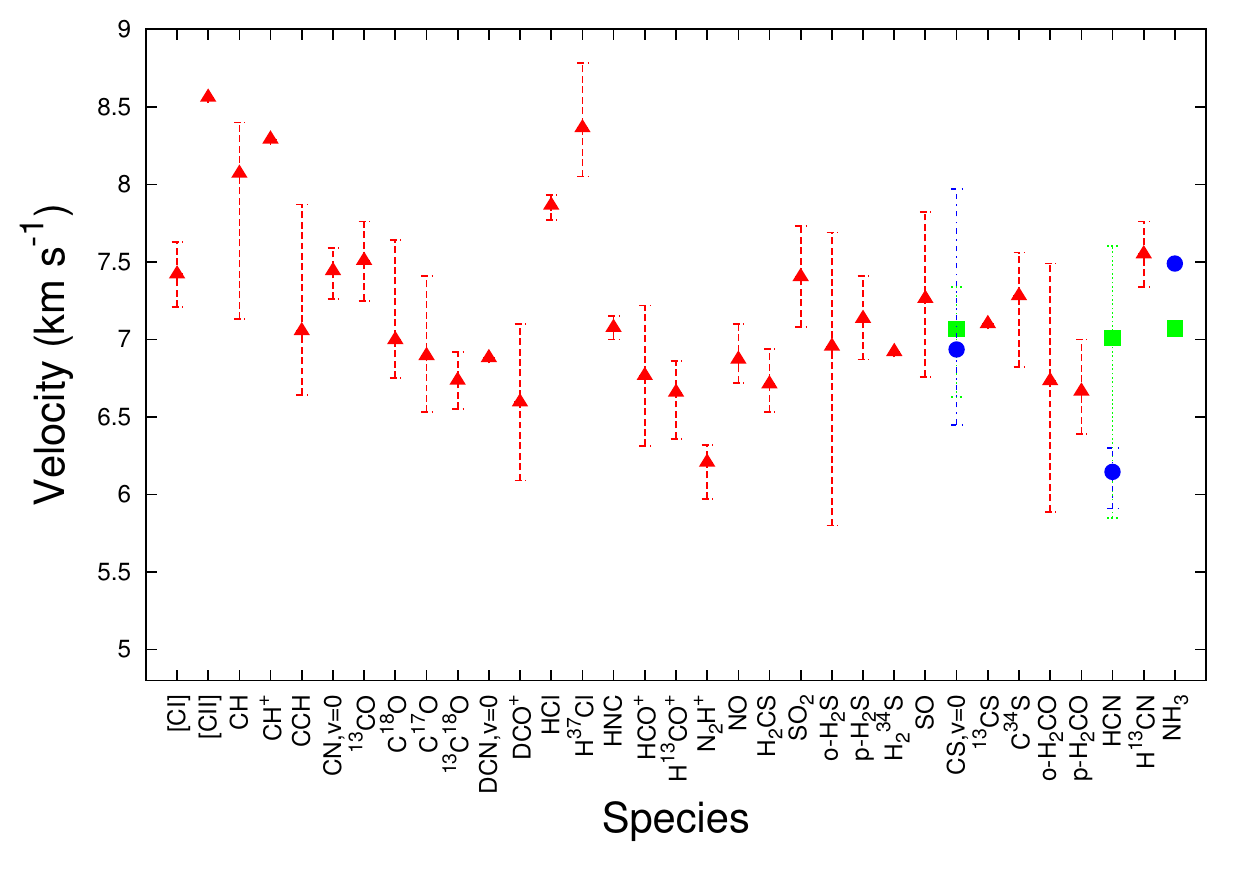}
\caption{Mean V$_{LSR}$ for each species derived from the Gaussian fits (Table \ref{tab:Gaussian}). Red triangles indicate the mean V$_{LSR}$ for species fit by a single Gaussian component. In cases requiring two component Gaussian fits, the narrow component is indicated by blue circle and the broad component is indicated by a green square. Error bars reflect the range in fitted V$_{LSR}$ values provided in Table \ref{tab:Gaussian}.} 
\label{velocity}
\end{figure}

{\bf [CI] \& [CII]:} [CI] and [CII] are the fine structure lines of neutral atomic and singly ionized carbon. Both have been seen over large regions of the ISM.  [CI] is known to trace PDRs at the UV illuminated surfaces of GMCs \citep{Papadopoulos:2004, 1999ApJ...512..768P, Tielens:1985}, and [CII] is a tracer of the interface between the diffuse warm ionized medium and the outermost surface of GMCs \citep{Velusamy:2012}.  [CII] is also thought to be a tracer of CO ``dark gas'' \citep{2010A&A...521L..17L}.  We have detected all [CI] and [CII] transitions accessible to HIFI, i.e. both of the [CI] ground-state fine-structure transitions: $^3P_1$ $\rightarrow$ $^3P_0$ and $^3P_2\rightarrow$ $^3P_1$ (V$_{LSR} \sim 7.5$ km s $^{-1}$), and the single [CII] transition: $^2P_{3/2} \rightarrow$ $^2P_{1/2}$ (V$_{LSR} \sim 8.6$ km s $^{-1}$), toward Orion-S.   With only one transition of [CII], we modeled the column density assuming that the excitation temperature in the PDR was between 200$-$500 K.  The velocity of [CII] is considerably different from the velocity of the dense, quiescent cloud component of Orion-S as traced by C$^{18}$O, CS, DCO$^+$, HCO$^+$, etc. (e.g. $\sim$ 7 km s $^{-1}$; see Figure \ref{velocity}).  This suggests that [CII] is tracing a kinematically distinct component of Orion-S; most probably photoevaporating material moving away from the molecular clump surfaces \citep[e.g.][]{2015ApJ...812...75G}.  [CI], however, does not have a velocity that is dramatically different from the quiescent cloud component and is, in fact, similar to that of C$^{18}$O (Figure \ref{velocity}).  This is probably due to the fact that neutral atomic carbon exists slightly deeper in the cloud (A$_V > 3-4$) where it is still mixed with molecular material \citep[see e.g.][]{{hollenbach:1997}, {2012A&A...542L..17M}}. 

{\bf CH$^+$, CH \& CCH:} CH$^+$, CH \& CCH are often associated with PDRs, with the former two also being tracers of ``CO-dark molecular gas'' \citep{2013A&A...550A..96N, Gerin:2010A&A...521L..16G}. In addition, CH$^+$ and SH$^+$ can also form via turbulent chemistry in the diffuse ISM \citep{2012A&A...540A..87G}. Transitions above 5\sig\ detected toward Orion-S for these species are listed in Table \ref{tab:Gaussian}. Enough transitions of CH and CCH were detected above 5$\sigma$ that we could model the emission from these species (Table \ref{tab:LTE}), both of which were well fit by 1 component models.  For CH$^+$, we provide a range of column densities for a range of excitation temperatures between 30$-$200 K.  From the Gaussian fits in Table \ref{tab:Gaussian}, both CH and CH$^+$ have similar kinematics (V$_{LSR} > 8.0$ km s $^{-1}$), whereas the CCH has V$_{LSR} \sim 7.2$ km s$^{-1}$ (see Figure \ref{velocity}).  This suggests that CH and CH$^+$ trace the same region as the [CII] emission i.e. the UV illuminated surface of the cloud, although possibly a deeper and denser region  of the PDR as suggested by \cite{2001ApJ...558L.105P}. CCH, which has a velocity closer to that of the quiescent gas, likely arises from deeper layers in the cloud \citep{2015A&A...578A.124N}.  The formation pathways for these species may help clarify these velocity differences. For example, CH$^+$ forms by an endothermic reaction: C$^+$ + H$_2$ $\rightarrow$ CH$^+$ + H \citep{1996MNRAS.279L..41F}. The formation of CH follows after a hydrogen abstraction reaction with CH$^+$ to form CH$_2^+$ and a subsequent dissociative recombination.  Since these two species are closely linked to the C$^+$ abundance, through one or two steps in the reaction network, it makes sense that they would be linked physically and, therefore, kinematically. The formation of CCH, however, involves additional steps in the reaction chain, starting with the formation of C$_2$H$_2^+$ followed by dissociative recombination to form CCH \citep[e.g.][]{1980ApJ...239..844W}. Since this requires additional reactions involving molecular material, this species is probably more closely linked to the denser molecular gas.

{\bf SH$^+$ \& CO$^+$:} SH$^+$ \& CO$^+$ are also species thought to trace regions with enhanced UV fields \citep{2013A&A...550A..96N}.  We detect two weak ($< 5 \sigma$) hyperfine components of SH$^+$ in Orion-S, which are too weak to be fitted or modeled but which, interestingly, are seen in emission rather than the usual absorption line profiles seen in the diffuse ISM \citep{2012A&A...540A..87G}.  Although we do not report the Gaussian fit parameters of SH$^+$ or CO$^+$ due to the weakness of the transitions, inspection of their lines suggests  V$_{LSR}$ of  $\sim$ 8.5 km s $^{-1}$ which is virtually identical to the [CII] velocity. Although SH$^+$ can form via turbulent chemistry in the diffuse ISM, given the strength of the UV field in Orion-S, it is likely that the main formation pathway is S$^+$ + H$_2$ $\rightarrow$ SH$^+$ + H. Therefore, like [CII], SH$^+$ probably also originates in the PDR at the surface of the cloud. The same is true of CO$^+$ which has a similar V$_{LSR}$ (8.5 km s $^{-1}$) as SH$^+$ and [CII], and like CH$^+$ forms directly from C$^+$ via the reaction OH + C$^+$ $\rightarrow$ CO$^+$ + H.

{\bf CN \& HCN:} CN \& HCN have both been detected in Orion-S. While both molecules are good tracers of warm dense gas, the CN/HCN abundance ratio is suggested to be an indirect measure of the UV field \citep[e.g.][]{Fuente:1993A&A...276..473F}; i.e.  if the ratio is significantly larger than 1, then the UV field is thought to be enhanced.
We have identified the  N = 5$-$4, N = 6$-$5 and N = 7$-$6 transitions of CN above the 5$\sigma$ noise level toward Orion-S and have modeled \textbf{these transitions using a narrow component} (see Table \ref{tab:LTE}).  The transitions of HCN (J = 6$-$5 to 13$-$12), however, exhibit the characteristic broad line wings that required two component Gaussian fitting and LTE modeling (see Table \ref{tab:LTE2}).
Figure \ref{HCN} shows the LTE model fit to our HCN observations. Since the fits are constrained by the rms noise in each spectrum, the higher frequency transitions (which tend to have much larger noise) appear to be less well fit than the lower frequency/lower noise transitions.  They are, however, still acceptable fits to the data within the given noise levels. Comparing the CN column density with the narrow component of HCN, we obtain a CN/HCN abundance ratio of 1.2$\pm$0.6 indicating a moderately enhanced UV field.  Given the high critical densities of these transitions ($\ge 10^8$ \cc ) it is unlikely that they originate at the UV illuminated cloud surface.  Instead, both their critical densities and the CN/HCN \textbf{abundance ratio of 1.2 suggest that they arise} deeper in the cloud (around A$_V > 5$; \citealp{Fuente:1993A&A...276..473F}) and, therefore, are more closely associated with the dense molecular gas.  This is also borne out by their velocities, which are similar to the dense, quiescent cloud component (Figure \ref{velocity}).

\subsubsection{Complex Organic Molecules and precursors}
\label{sec:complex}

Complex organic molecules are often associated with hot core chemistry.  Unlike Orion-KL in which a plethora of complex organic molecules were detected \citep{Crockett:2014ApJ...787..112C}, in Orion-S we only detect a handful of molecules that might be considered complex.

{\bf CH$_3$OH:} Methanol is an asymmetric top molecule, whose internal rotation results in two distinct symmetry species A${-}$CH$_3$OH  and E${-}$CH$_3$OH.  In total we observed 359 methanol transitions above 3$\sigma$ toward Orion-S, 170 A${-}$CH$_3$OH  and 189 E${-}$CH$_3$OH. 198 of the lines were above the 5\sig\ noise level, 111 A${-}$CH$_3$OH  and 87 E${-}$CH$_3$OH. 
While methanol is known to be a good temperature probe \citep[e.g.][]{Beuther:2005ApJ...632..355B, Wang:2011p4682}, detailed modeling of methanol is beyond the scope of this paper and will be the subject of future work.

{\bf H\2CO:} Formaldehyde is another commonly used tracer of gas temperature \citep[e.g.][]{Mangum:1993ApJS...89..123M}, in which the \textbf{two hydrogen atom spins separate the molecule} into distinct ortho and para species. Transitions of H\2CO above the 5\sig\ noise level are listed in Table \ref{tab:Gaussian}. We needed two component LTE modeling for both the ortho and para H\2CO molecules since one component models could not simultaneously reproduce all observed transitions.  The modeling (Table \ref{tab:LTE2}) results in low temperatures for the narrow component (T$_{ex} \sim 45-50$~K) and higher temperatures (T$_{ex} \sim 150-165$~K) for the broad component.  The large linewidths and the fact that the estimated source sizes are quite large ($> 45''$) may indicate that the high temperature H$_2$CO emission arises from shocks in the outflows rather than from a ``hot core'' region. Ortho to para ratios in the narrow and broad components are 0.8$\pm$0.1 and 0.6$\pm$0.1 respectively. These low values are also consistent with our results for H\2S (see below). The spin temperatures associated with the ortho and para species are 7$\pm$1 K and 6$\pm$1 K respectively, \textbf{suggesting that if formaldehyde formed under LTE conditions} that the formation temperature was very low.

{\bf CH$_3$OCH$_3$:} Dimethyl ether is a complex molecule detected toward Orion-S. Since this molecule has no transition above 5$\sigma$, we only report it as a detection in Table \ref{tab:IdentifiedSpecies}.

The lack of complex organic molecules suggests that if hot cores exist in Orion-S they are still in their infancy and have not had time to either expand dynamically or develop chemically.  This is not surprising given the very small size of the embedded submillimeter continuum sources detected in the region \citep{Zapata:2005p5918}. In addition, if these submillimeter continuum sources are indeed hot cores they are approximately 10 times smaller than the Orion-KL hot core. Therefore, the beam dilution in Orion-S would be a 100 times worse. Thus, any transitions arising from the Orion-KL hot core that have an intensity less than a few K in the survey of \cite{Crockett:2014ApJ...787..112C} would be undetectable in our survey if they originate from the considerably smaller region in Orion-S. Alternatively, it is possible that Orion-S is not a massive star forming region at all and, therefore, there are no hot cores in this region. Observations with higher spatial resolution or at lower frequencies (with associated lower excitation temperatures) would help address this issue by  revealing the presence of more complex organics. 

\subsubsection{Pure Shock Tracers}

{\bf SiO:} SiO abundances can be enhanced by more than two order of magnitudes in hot and shocked regions  \citep[e.g.][]{Iglesias:1978ApJ...226..851I, 1992A&A...254..315M} and SiO emission is often used as a tracer of molecular outflows since the SiO emission traces the outflow material itself, rather than the dense protostellar core \citep{Martin:1992A&A...254..315M}. This is believed to be due to Si-bearing dust grains being shattered by the outflow, followed by a rapid gas-phase reaction with free oxygen to produce SiO \citep[e.g.][]{1997A&A...321..293S, 2008A&A...482..809G, 2008A&A...490..695G}. A number of outflows have already been identified in Orion-S  by \cite{Zapata:2006ApJ...653..398Z}. In our data, although we could not identify any SiO emission above the 5\sig\ level, we have identified three SiO transitions above the 3$\sigma$ level (J=12$-$11, 13$-$12, 14$-$13, v=0) at the velocity of 6.2 km s$^{-1}$ (similar to the quiescent gas). With additional spectral smoothing (to a velocity resolution of $\sim$ 4 km s$^{-1}$) it is clear that these transitions are real, with S/N $> 5\sigma$. These transitions are quite broad ($\Delta$V $\sim$ 20$-$30 km s$^{-1}$), which is reasonable considering the observed characteristics of the SiO outflows as seen by \cite{Zapata:2006ApJ...653..398Z}. Given the existence of such high-J transitions with excitation energies above $\sim$ 150 K, this indicates the presence of at least a small amount of hot shocked SiO in Orion-S.

\subsubsection{Tracers of Quiescent Gas}

{\bf CO,$^{13}$CO, C$^{18}$O, C$^{17}$O, \& $^{13}$C$^{18}$O:} For all carbon monoxide isotopologues, excluding $^{12}$C$^{16}$O itself, one component Gaussian fitting and LTE modeling match the observations remarkably well. $^{12}$C$^{16}$O itself, however, exhibits a broad line wing (clearly tracing an outflow/shock) and, due to the presence of  self-absorption, was not modeled. The existence of an outflow is not visible in any of the $^{12}$C$^{16}$O isotopologue transitions due to their lower abundances. The LTE modeling of $^{13}$CO is shown in Figure \ref{C13O}. For all $^{12}$C$^{16}$O isotopologues (except $^{13}$C$^{18}$O)  we see transitions from J=5$-$4 to 11$-$10 above the 3$\sigma$ level. For $^{13}$C$^{18}$O the highest transition we detect above 3$\sigma$ is J=7$-$6. The higher J transitions are buried in the larger noise of the higher frequency HIFI bands.  For $^{12}$C$^{16}$O, however, we detect lines up to J = 16$-$15.  For the isotopologues, the typical V$_{LSR}$ is approximately 7 km s $^{-1}$, indicating that these species trace the quiescent gas in the cloud.  The V$_{LSR}$ of the main isotopologue, however, is often a bit higher than this, probably due to the Gaussian fits being skewed by the presence of self-absorption in the spectra, or due to the fact that with its high opacity $^{12}$C$^{16}$O may be tracing a different physical region of the cloud. Interestingly, in Table \ref{tab:LTE}, a correlation can be seen between the CO isotopologues and the derived excitation temperature; with the more optically thick species, which trace the cloud surface (e.g. $^{13}$CO)  having a higher temperature than the optically thin ones which preferentially trace the interior (e.g. $^{13}$C$^{18}$O).  This suggests that the external UV field is responsible for much of the heating in Orion-S \citep[see also][]{1990ApJ...356L..63T}. This is different than the usual case of isolated star formation, in which the gas is predominantly heated internally by the process of gravitational collapse and the formation of an embedded protostar.

{\bf Deuterium-Bearing Molecules:} Deuterated species are  subject of considerable interest in the ISM, since the D/H ratio in molecular clouds can be considerably enhanced over the cosmic value of $\sim$ 10$^{-5}$. In Orion-S we detect only a few deuterated species: DCN, DCO$^+$, and HDO, which all have velocities similar to that of the quiescent gas. Enhanced deuteration can occur because fractionation reactions involving deuterium are favoured in low-temperature environments associated with pre-stellar cores, the resultant deuterated molecules can freeze onto grains, and then be released back into the gas phase when star formation activity begins to heat the natal gas \citep[e.g.][]{2007prpl.conf...47C}.  Thus, deuterated species such as  DCN and HDO can trace the chemical history of the gas. In Orion-S we found the DCN/HCN column density ratio to be 0.02$\pm$0.01, suggesting considerable enhancement in cold gas. The DCO$^+$/HCO$^+$ column density ratio is 0.03$\pm$0.02.
While the DCO$^+$ abundance  can be enhanced in cold gas via H$_2$D$^+$, \cite{2009A&A...508..737P} have shown that deuteration can also occur in the gas phase of warm regions like the Orion Bar via the CH\2D$^+$ ion. 
Although we detected  three HDO transitions,  it was the only species for which we were not able to find any models that converged to a good solution. Therefore, there is no way to give even a rough estimate for the D/H ratio in water.

{\bf N$_2$H$^+$:} While N$_2$H$^+$, J = 1$-$0, is often associated with cold, dense gas, we detect N$_2$H$^+$  transitions from J = 6$-$5 to 10$-$9.  LTE modeling indicates excitation temperatures of $\sim 47$ K, suggesting that even the dense gas in Orion-S is quite warm. Previous observations of CH$_3$C\2H in Orion-S \citep{1994ApJ...431..674B} confirm this  idea.  
The V$_{LSR}$ of N$_2$H$^+$  (Figure \ref{velocity}) also suggests that it originates from the quiescent gas. The upper limit for the column density of N$_2$D$^+$ with the same excitation condition as found for N$_2$H$^+$ is 5$\times 10^{11}$ cm$^{-2}$. This provides a rough estimate of the D/H ratio of $<$ 0.03.

\subsubsection{Tracers of Both Shocked and Quiescent Gas}

As previously mentioned, there are a number of species, for which we had to invoke two component LTE modeling in order to fit the observed transitions (see Table \ref{tab:LTE2}).  
Narrow spectral components are usually associated with quiescent gas, whereas broader spectral components trace more dynamic gas  that is often associated with shocks.  This suggests that species listed in Table \ref{tab:LTE2} can simultaneously exist in both quiescent and shocked gas components.  This is not surprising, since  \cite{Bachiller:1997} show that, in the bipolar outflow L1157, while some species are clearly quiescent gas tracers, many species exist in both components.  Of these latter species, their abundances in the shocked gas are often an order of magnitude or more higher than their abundances in the quiescent gas.  We will explore the issue of abundances further in Section \ref{comparison}.  Here, however, we will briefly discuss  some of the species listed in Table \ref{tab:LTE2} as possible tracers of both shocked and quiescent gas.

{\bf H$_2$O: } While H$_2$O is not listed in Table \ref{tab:LTE2}, it is an important molecule in the ISM and has been the subject of a number of important studies using the \textit{Herschel Space Observatory} in both shocked and unshocked gas.   Both the ortho and para forms of H$_2$O were detected in Orion-S, as well as one transition of o-H$_2^{18}$O.  The H$_2$O transitions required two component Gaussian fitting due to the presence of a broad line wing in the spectra (Table \ref{tab:Gaussian}).  
The modeling of water is a complex affair and is beyond the scope of this paper.  However, \cite{2014A&A...572L..10C} modeled the ortho and para H$_2^{18}$O in Orion-S and found LTE column densities of 2$\times 10^{11}$ cm$^{-2}$ and 2$\times 10^{12}$ cm$^{-2}$ respectively, which suggests an ortho to para ratio of 0.1, indicating that it is unlikely that water formed under LTE conditions. Their non-LTE analysis of the data, however, brings the ratio up to a factor of 2. 
\cite{2014A&A...572L..10C} also show that the ortho to para ratio is $\sim 0.3$ in the nearby Orion Bar. Both values are well below the usual value of 3, which indicates non-LTE formation mechanism for water in both Orion-S and the Bar, possibly due to photodesorption from dust grains.

{\bf H\2S \& H$_2^{34}$S:} H\2S is an asymmetric rotor which has ortho and para spin modifications. It  is considered to be a tracer of high temperature  grain surface  chemistry \citep[e.g.][]{Watson:1982ASSL...93..357W}. Similar to SO,  despite the fact that we fit the H\2S transitions with a single Gaussian in Table \ref{tab:Gaussian}, H\2S also  required a second physical component in order to obtain a good $\chi ^2$ fit from the LTE modeling process.  As Table \ref{tab:LTE2} shows, we modeled the H\2S emission from the ortho and para spin modifications separately. In both cases, the narrow component has a linewidth of $\sim$3.6 km s $^{-1}$, a low excitation temperature (24 K), and is fairly extended (emission extending beyond the \textit{Herschel} beam) whereas the broad component has a larger linewidth ($\sim$ 8 km s $^{-1}$), is warmer ($\sim$ 80 K), and yet is still fairly extended ($> 35''$). The ortho to para ratio in the narrow component is 1.1$\pm$0.3 and in the broad component is 0.9$\pm$0.1 indicating a spin temperature of 9$\pm$2 K. These values are consistent with those determined for H\2O in Orion-S by \cite{2014A&A...572L..10C} and for Formaldehyde (above).  Similarly, this low ortho to para ratio suggests either a very low formation temperature for H\2S or that non-LTE effects had an important role in its formation. 
To model H$_2^{34}$S, we coupled its single ortho transition with those of the common isotopologue. The isotopic ratio is another possible free parameter in CASSIS, which assumes that the other parameters of both isotopologues are identical. The isotopic ratio converged to 31$\pm$9. Note, however, that there is  only one weak transition  of H$_2^{34}$S that we used to determine this ratio. For comparison, typical values for the $^{32}$S/$^{34}$S ratio in galactic molecular clouds are $\sim$ 19$\pm$8 \citep{1998A&A...337..246L} which is consistent with the solar value of 23 \citep{{1989GeCoA..53..197A}}.

{\bf CS:} CS is a well-known tracer of dense gas due to its high critical density \citep{Plume:1992ApJS...78..505P}. We observe many transitions of CS, from J = 10$-$9  to J = 19$-$18.   
Like CO, CS \textbf{requires two components to be successfully fit} by a Gaussian profile and modeled; one narrow ($\Delta V_{FWHM} \ \sim$ 4.1 km s $^{-1}$), extended ($\Omega = 67''$), and cool (T = 37 K), and the other broader ($\Delta V_{FWHM} \ \sim$ 10 km s $^{-1}$), moderately extended ($\Omega = 35''$), and warm (T = 108 K). The broad component is not seen in the spectra of the CS isotopologues but was needed to successfully model the transitions. $^{13}$CS and C$^{34}$S were modeled in a fashion similar to that of H$_2^{34}$S (i.e. the transitions of the isotopologues modeled simultaneously with those of the common isotope, and found to have isotopic ratio of 46$\pm$17 and 14$\pm$5 respectively).  The Gaussian fitted  (Table \ref{tab:LTE}) and modeled velocity (Table \ref{tab:LTE2}) of both components is $\sim 7$ km s$^{-1}$,  suggesting that both components originate in the same material (Figure \ref{velocity}).

{\bf HCN:} The transitions of HCN (J = 6$-$5 to 13$-$12) exhibit the characteristic broad line wing that required both two component Gaussian fitting and LTE modeling (see Table \ref{tab:LTE2}).  As is usual for other species,  the broad component is hotter (67 K vs 34 K),  broader (13.4 \kms\ vs 4.4 \kms), and less spatially extended (41$''$ vs 64$''$) than the narrow component.  While the broad component's V$_{LSR}$ is lower than that of narrow component (6.6 \kms\ vs 7.2 \kms) both are consistent with the systemic velocity of Orion-S (Figure \ref{velocity}).

{\bf HCO\+:} HCO\+ is another well-known tracer of both dense molecular gas and outflows. We detected the J = 6$-$5 to J = 13$-$12 transitions of HCO$^+$ in Orion-S, which span a wide range of physical conditions (E$_{\rm up}$= 90 K, n$_{crit} \sim 3.2 \times 10^7$ cm$^{-3}$ to E$_{\rm up}$=389 K , n$_{crit} \sim 6.0 \times 10^8$ cm$^{-3}$). HCO\+ has some of the strongest lines seen in our survey of Orion-S.  Despite the fact that we fit the HCO\+ transitions with a single Gaussian in Table \ref{tab:Gaussian}, HCO\+  required a second physical component in order to obtain a good $\chi ^2$ fit from the LTE modeling process (see Table \ref{tab:LTE2}).  Both components have a velocity of $\sim$ 7 \kms\ and are fairly warm ($\sim$ 69 K), suggesting a common origin.

{\bf NH$_3$:}  We detect two transitions above the 5$\sigma$ level in Orion-S: one in emission  ($1_{0,0} - 0_{0,1}$) and one in absorption ($2_{1,1} - 1_{1,0}$).  In fact, we detect 3 additional transitions of NH$_3$ (all in absorption) but since they were just below the 5$\sigma$ level, we do not report them in Table \ref{tab:Gaussian}. Modeling these transitions simultaneously requires two components: a cold (T $\sim 20$K), quiescent ($\Delta V \sim 4.2$ km s $^{-1}$) layer of gas in front of a warmer (T $\sim 36$K), broader ($\Delta V = 10$ km s $^{-1}$) component.
The V$_{LSR}$ values of both NH$_3$ components are consistent with the systemic velocity of the cloud (Figure \ref{velocity}). The presence of absorption lines provides additional evidence for the existence of two components in Orion-S (one warm and one cool). The fact that the low energy transition is seen in emission, whereas the higher energy transition is seen in absorption may be related to the beam size and the strength of the continuum. At 572 GHz the continuum is weaker than it is at 1215 GHz (see Figure \ref{fig:ltoc}) and may be too beam-diluted to see absorption. However, at higher frequencies, the beam couples better to the source and absorption may become more prevalent. This suggests that NH$_3$ may not come from a high density, hot region which is consistent with our conclusion that there are no hot cores in Orion-S. 

{\bf SO\2:}  SO\2 is also often a tracer of shocks, since it can freeze onto grain mantles at early evolutionary times when the gas is cold and dense, and later be returned to the gas phase by shocks \citep[e.g.][]{Millar:1990A&A...231..466M, 2013A&A...556A.143E}. Toward Orion-S we detected a number of SO$_2$ transitions above the 5$\sigma$  level (see Table \ref{tab:Gaussian}). Given the fact that all the observed SO\2 lines are relatively weak, one component modeling matches the observations remarkably well. LTE modeling for this species (Table \ref{tab:LTE}) shows that the SO$_2$ is fairly warm (T$_{ex}$ $\sim 150$ K), broad ($\Delta V$ $\sim$ 6.7 km s$^{-1}$), and extended ($\sim$ 65$''$).  The velocity of SO\2 is also similar to the velocity of the quiescent gas (Figure \ref{velocity}).

 {\bf SO:}  
 In contrast with SO\2, and despite the fact that we fit the SO transitions with a single Gaussian in Table \ref{tab:Gaussian}, SO  required a second physical component in order to obtain a good $\chi ^2$ fit from the LTE modeling process (see Table \ref{tab:LTE2}). The narrow component has a line width of 3.8 \kms, a moderately low temperature (34 K), and is extended beyond the Herschel beam. Despite the fact that the broad component ($\Delta V_{FWHM}$ = 11 \kms) is much warmer (122 K), it is still quite extended in size (35'').

The large linewidths (7$-$13 \kms), high temperatures (70$-$150 K), and extended size ($> 30''$) determined for the second physical components of these species suggest that 
 the embedded outflows seen in Orion-S \citep[][]{Zapata:2005p5918, Schmid-Burgk:1990ApJ...362L..25S, Ziurys:1990ApJ...356L..25Z} have affected a large volume of the region both thermally and dynamically. Whether or not these shocks have affected the chemistry of the gas will be examined in Section 3.3.

\subsection{Chemical Comparison with Orion-KL}

One of the main goals of this project is to explore the chemical differences and similarities between Orion-S and Orion-KL.  As mentioned in Section \ref{intro}, a  detailed comparison between the chemical abundances in Orion-S and Orion-KL is useful, since both regions presumably formed under similar conditions, but could have developed very different chemical abundances based on differences in their ages, densities, temperatures, radiation fields, etc.    As part of the HEXOS survey, and in a direct analogue to our study,  \cite{Crockett:2014ApJ...787..112C} have observed the same frequency range in Orion-KL using the same instrument.  Therefore, we have a perfectly matched database, with which to compare our results.  In this section we will compare the chemistries of these two regions.
All of the  Orion-KL data  are taken from the HEXOS survey of this region as listed in \cite{Crockett:2014ApJ...787..112C} using the column densities derived from their XCLASS LTE modeling of the data.

\subsubsection{Chemical Abundances in Orion-S}

Our LTE modeling produces column densities but, to obtain chemical abundances, we must scale each column density by the H$_2$ column density, which is not known directly. Therefore, we use C$^{18}$O as a proxy for the H$_2$ column density. The  C$^{18}$O transitions we observe in Orion-S  are optically thin and have only a single narrow component (Table \ref{tab:Gaussian}) that is well fit by  a column density of $3.5 \times 10^{16}$ cm$^{-2}$ (Table \ref{tab:LTE}). To convert this to an H$_2$ column density we use a C$^{18}$O:H$_2$ conversion factor of 1.7$\times10^{-7}$ \citep{1997ApJ...491..615G} to obtain N${_{H_2}}$ =   2.1$\times$10$^{23}$ cm$^{-2}$. 

Other species have been modeled using both a narrow and broad component (Table \ref{tab:LTE2}), the latter possibly indicative of gas affected by shock. To compute the chemical abundances in the broad component of species listed in Table \ref{tab:LTE2}, we need an estimate of the C$^{18}$O column density specifically present in this broad component.  Since the broad component is too optically thin to be detected in C$^{18}$O or even $^{13}$CO, we rely on $^{12}$CO instead for this purpose. In this case, we attempt to model the broad (i.e. line wing) component of our observed $^{12}$CO lines by fitting three Gaussian components to each transition.  The first two Gaussians  fit the ``main'' component of the asymmetric CO profile (see Table \ref{tab:Gaussian}) and the third fits the outflow ($\Delta$V $\sim 18$ \kms ).  \textbf{This third component} is then modeled via our LTE procedure to
estimate the physical parameters of the $^{12}$CO outflow (T$_{ex} \sim 200$ K; N(CO) $ \sim 7.1 \times 10^{16}$ cm$^{-2}$; $\Omega > 30''$).  Dividing this column density by the canonical $^{12}$CO:C$^{18}$O abundance ratio of 500:1 provides a C$^{18}$O column density of $1.4 \times 10^{14}$ cm$^{-2}$ for the broad component in Orion-S; a factor of $\sim 250$ smaller than the C$^{18}$O column density in the narrow component as measured by directly modeling our C$^{18}$O observations. Using the same C$^{18}$O:H\2 scaling relationship as above we obtain N${_{H_2}}$ =   8.2$\times$10$^{20}$ cm$^{-2}$  

\begin{figure}
\centering
\includegraphics[scale=1.2]{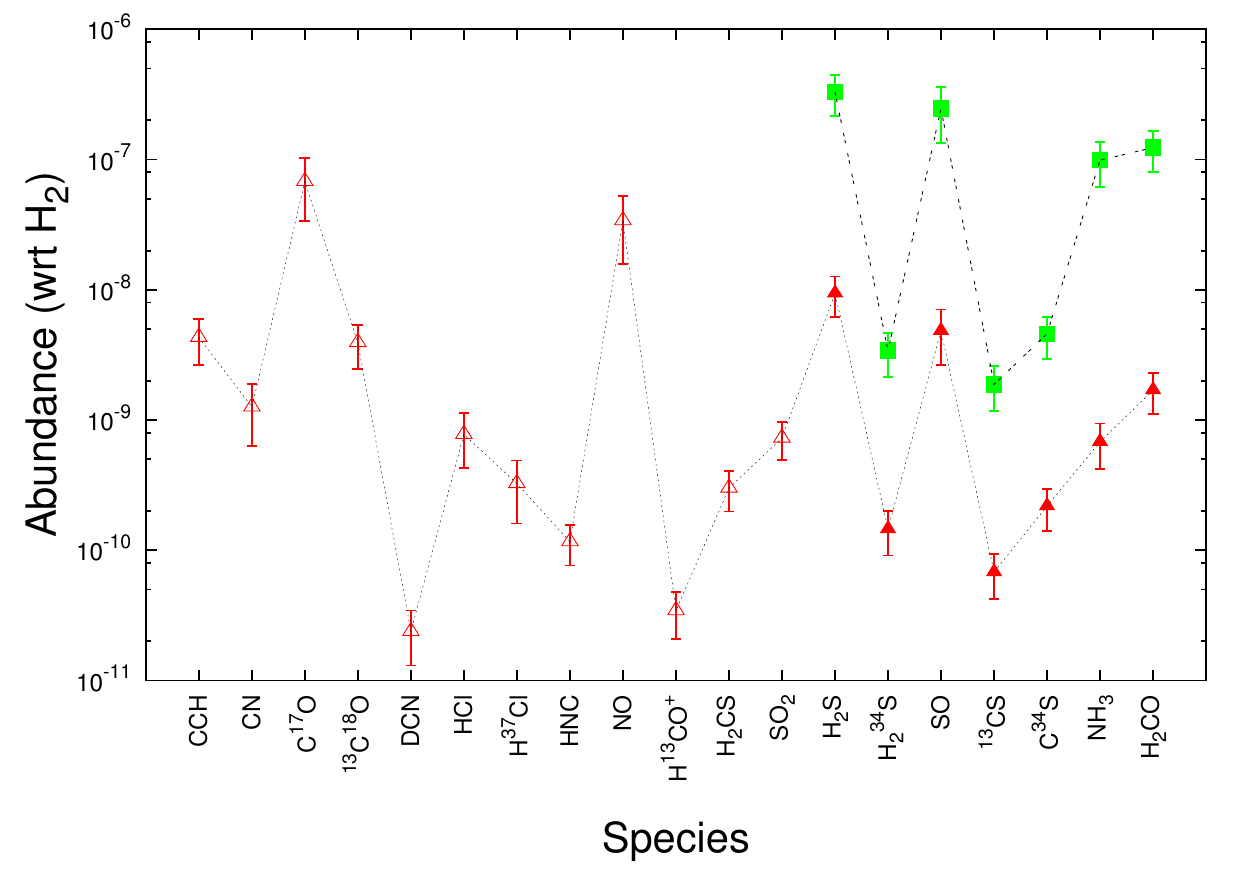}
\caption{The abundance (with respect to H$_2$) of species listed in Tables \ref{tab:LTE} \& \ref{tab:LTE2}. Open red triangles indicate the abundance ratio for species fitted by a single component LTE model in Orion-S. In cases requiring two component LTE models for the Orion-S data, the abundance ratio for the narrow component is indicated by solid red triangles and that for the broad component is indicated by solid green squares. Therefore, the dotted line connects species/components that likely trace quiescent gas, whereas the dashed line connects species/components that may trace shocked gas. }
\label{fig:xos}
\end{figure}

Therefore, dividing the modeled column densities of  the broad components of the species in Table \ref{tab:LTE2} by $8.2 \times 10^{20}$ cm$^{-2}$ gives the abundance (with respect to H$_2$) of all species in Orion-S that are possibly affected by shocks. Dividing the modeled column densities of the rest of the species in Table \ref{tab:LTE}, as well as the narrow component of the species in Table \ref{tab:LTE2}, by $2.1 \times 10^{23}$ cm$^{-2}$ gives the abundance (with respect to H$_2$) of all species in Orion-S that likely originate in quiescent gas. The results are provided in Table \ref{tab:wrtH2} and illustrated in Figure \ref{fig:xos}, which  clearly show that in the broad component (green squares) the abundances are enhanced by a factor of 10$-$100 with respect to the narrow component (red triangles). Abundance enhancements of this magnitude indicate classic shock behavior \citep{Bachiller:1997}. This suggests that shock chemistry is playing an important role in Orion-S. 


\subsubsection{Species Common to Both Orion-KL and Orion-S}
\label{comparison}

 \cite{Crockett:2014ApJ...787..112C} detected $\sim$ 13,000 lines from 39 different molecules (79 species if one includes all the isotopologues).  This is considerably more than the 685 lines from 52 species (including isotopologues) that we have detected in Orion-S. In addition, the lines in Orion-KL are typically an order of magnitude stronger than those seen in Orion-S.  A more interesting comparison, however, is to examine the abundances of species common to the two sources. 
 
\cite{Crockett:2014ApJ...787..112C} produce column densities for many of the detected species in Orion-KL. To compare with the chemical abundances in Orion-S (Section 3.3.1), we must also scale by the H$_2$ column density in Orion-KL. For Orion-KL we use the C$^{18}$O column densities derived by \cite{Plume:2012ApJ...744...28P} as a proxy, which breaks down the results for each of the four known kinematic components:  the Hot Core (V$_{LSR}$ $\sim 4-6$  and  $\Delta V \sim 7-12$ km s$^{-1}$), the Plateau (V$_{LSR}$ $\sim 7-11$  and  $\Delta V _{LSR} \geq 20$ km s$^{-1}$), the Compact Ridge (V$_{LSR}$ $\sim 7-9$  and  $\Delta V \sim 3-6$ km s$^{-1}$), and the Extended Ridge (V$_{LSR}$ $\sim 8-10$  and $\Delta V \sim 2-4$ km s$^{-1}$) \citep{Blake:1987ApJ...315..621B}. The H$_2$ column density can then be produced using the same C$^{18}$O:H$_2$ conversion factor of 1.7$\times$10$^{-7}$. 

Producing abundances in this way does depend on the assumptions regarding the  C$^{18}$O:H$_2$ abundance ratio.  However, by dividing the abundance of a given species in Orion-KL by the abundance of the same species in Orion-S, we eliminate the C$^{18}$O:H$_2$ abundance ratio altogether and are essentially normalizing to the C$^{18}$O column density in each region. This does, of course, assume that C$^{18}$O abundances are the same in both sources, which may be reasonable based upon similarities between the observed C$^{18}$O:C$^{17}$O ratios (e.g. 2.5 in Orion-S, 3.0 in the Hot Core, 6.5 in the Compact Ridge, 3.5 in the Plateau, and 2.3 in the Extended Ridge). These ratios are also consistent with those found by \cite{2004ApJ...610..320L}. 

Therefore, we are essentially creating the following ratio:
$$
\frac{X_{KL}}{X_S}  = \frac{\left ( \frac{N_i}{N_{C^{18}O}} \right )_{KL}}{\left ( \frac{N_i}{N_{C^{18}O}} \right )_S} ~~~~(1)
$$
where N$_i$ is the column density of species $i$ and N$_{C^{18}O}$ is the column density of C$^{18}$O.  The subscripts KL  and S refer to this ratio in Orion-KL and Orion-S respectively.  Given the four distinct kinematic components of Orion-KL, we create this ratio for the Hot Core (HC), the Plateau (P), the Compact Ridge (CR), and the Extended Ridge (ER) separately and, again, use different values for the C$^{18}$O abundance in Orion-S depending on whether the species in question has a narrow or broad spectral line profile.

\begin{figure}
\centering
\includegraphics[scale=1.2]{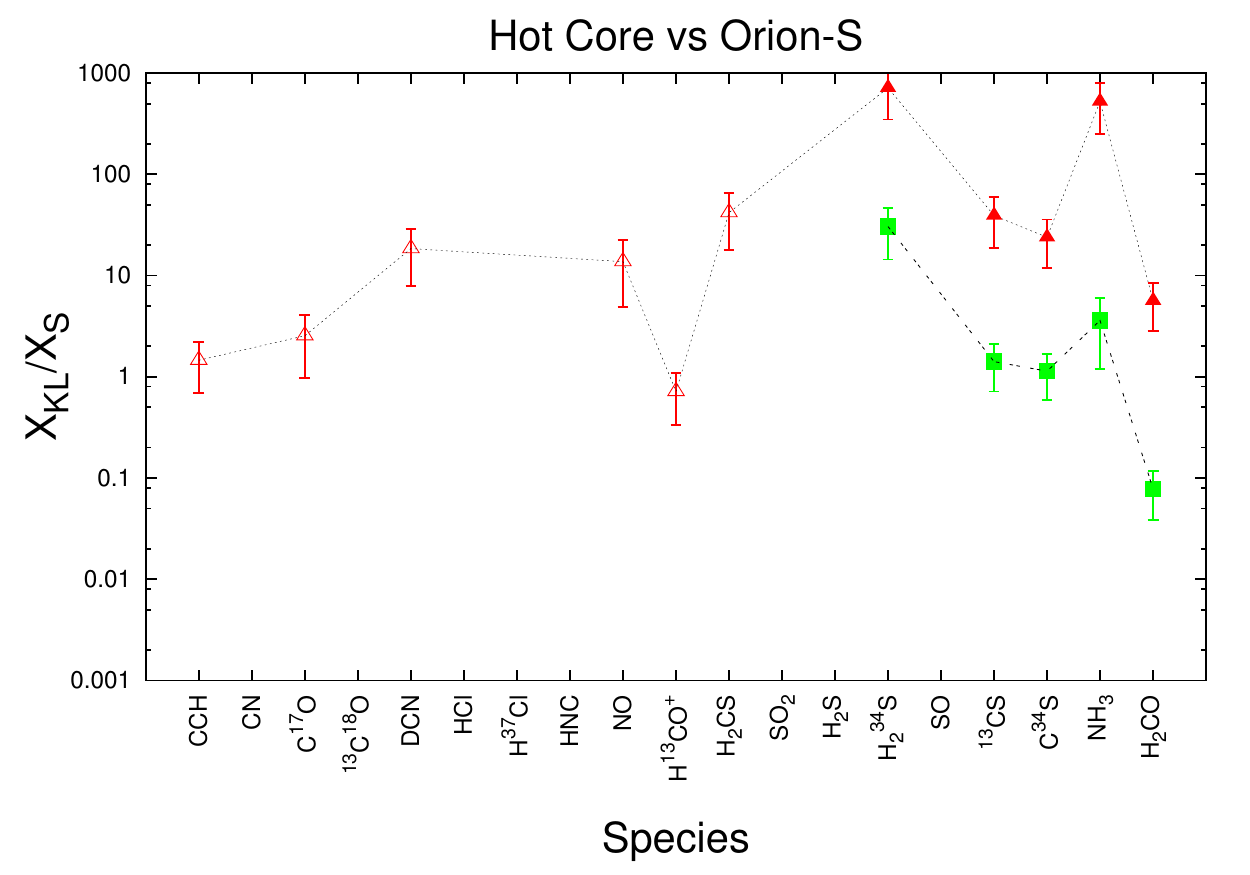}
\caption{
Comparison of the abundances of species detected in the Orion-KL Hot Core to those in Orion-S as given by equation 1 in Section \ref{comparison}. Symbols are the same as described in Figure 15.
}
\label{HC}
\end{figure}

Figure \ref{HC} shows the comparison between Orion-S and the Orion-KL Hot Core. Note that, by common use, the term ``hot core'' refers to a dense, warm region surrounding a central high mass protostellar object that dominates its energetics \citep{2000prpl.conf..299K}. It has been argued that the eponymous hot core in the Orion-KL region does not fulfil this criterion \citet{2011A&A...529A..24Z}.  Rather,  these authors suggest that this region is rather powered by the aftermath of the explosion caused by a stellar merger event \citep{2005AJ....129.2281B}. Regardless of this, in the present paper we are comparing the chemical abundances of Orion-S with those in what is traditionally referred to as the ``hot core'' component of Orion KL. The x-axis indicates the species and the y-axis shows the ratio as calculated from Equation 1.  Open red triangles indicate molecules for which one component LTE models in Orion-S were sufficient.  In cases where we required two components to model the Orion-S data, the solid red triangles indicate the ratio for the narrow component and the solid green squares indicates the ratio for the broad component. The dotted line connects species/components that likely trace quiescent gas, whereas the dashed line connects species that have a broad component. Not all species detected and modeled in Orion-S are represented in this figure.  This is due to the fact that \cite{Crockett:2014ApJ...787..112C} did not model all their detected species (notably the atomic species), nor did they provide column densities for species in which the lines were optically thick (e.g. CO, $^{13}$CO, HCO$^+$, CS) in Orion-KL.  Note also that not every species listed in Figure \ref{HC} has a symbol associated with it (e.g. CN, HCl, SO, etc.).  This is because emission from these species  were not attributed to the HC, but to one or more of the other kinematic components in Orion-KL. Error bars are calculated from the statistical uncertainties  determined from our LTE modeling of Orion-S ($\sim$ 10$-$50\%; see Table \ref{tab:LTE} and \ref{tab:LTE2}) and with the assumption of 10\%  error bars of the reported column densities in Orion-KL \cite{Crockett:2014ApJ...787..112C} which includes the effects of  calibration errors, pointing errors, etc.  However, to account for the possibility that LTE is not a good approximation in either Orion-S or Orion-KL we add an additional 40\% error to the column densities. This value is based on a comparison of LTE versus non-LTE column density calculations for the Orion-KL Extended Ridge \citep{Crockett:2014ApJ...787..112C}.

Inspection of Figure \ref{HC} clearly shows that the abundances of species in the Orion-KL HC are significantly higher than those in the narrow component of Orion-S (dotted line in Figure \ref{HC}).  Except for CCH, C$^{17}$O, and  H$^{13}$CO$^+$, the abundances in the HC are $\gtrsim 10$ times larger than those in Orion-S.  Examining the abundance ratios in the narrow component, we obtain $\left <\frac{X_{KL}}{X_S} \right > = 135$ (SD = 260) where SD is the standard deviation about the mean. The large standard deviation simply reflects the enormous scatter in the ratios (note that the y-axis in Figure \ref{HC} is on the log scale).
Although still not a good match, the disagreement is smaller for species that have a broad component, (dashed line in Figure \ref{HC}).
In this case we obtain $\left <\frac{X_{KL}}{X_S} \right > = 6$ (SD = 12).  Given the lack of complex molecules noted in Section \ref{sec:complex} and the poor match in the abundances between the Orion-KL HC and  Orion-S, this suggests that the gas detected in this study of  Orion-S does not originate in a hot core.

\begin{figure}
\centering
\includegraphics[scale=1.2]{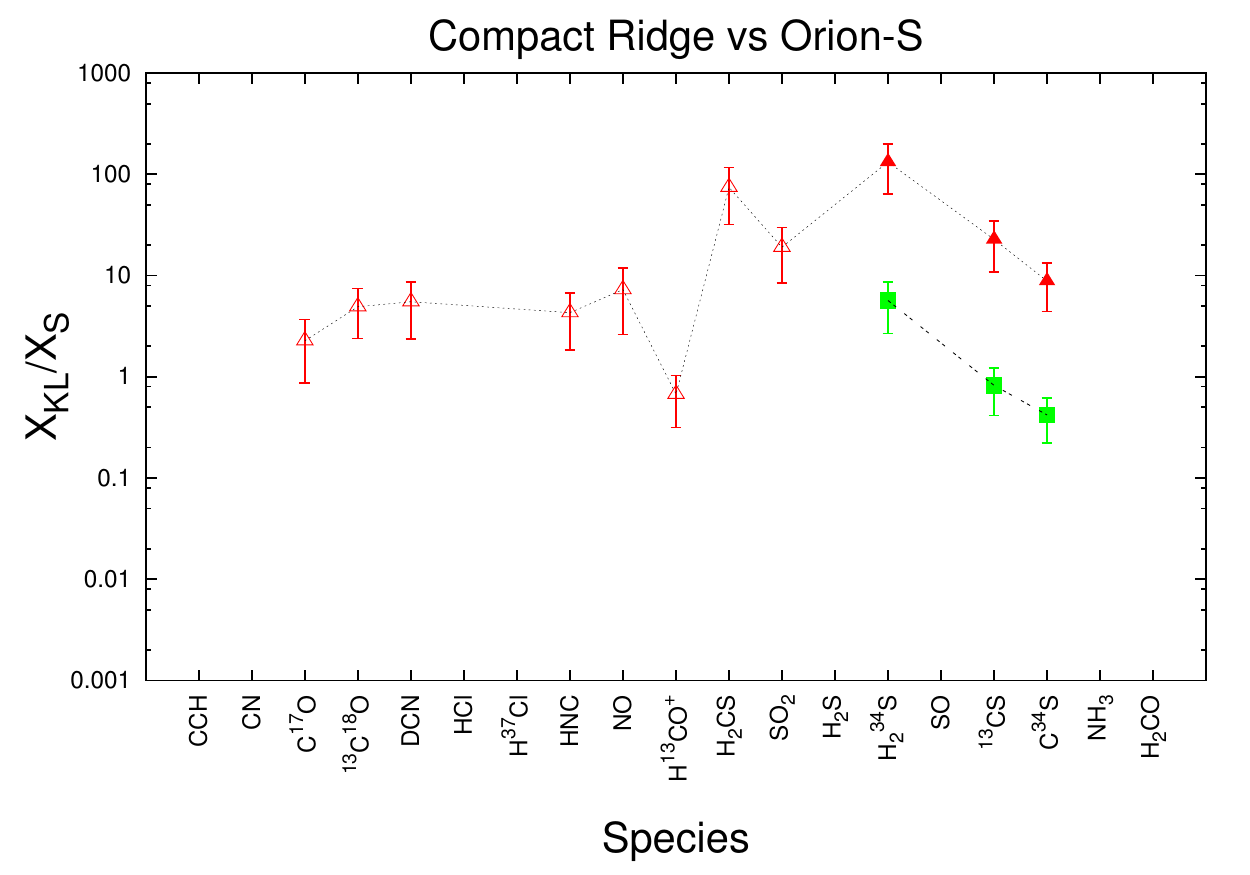}
\caption{ Same as for Figure \ref{HC} except for the Orion-KL Compact Ridge}
 \label{CR}
\end{figure}

Figure \ref{CR} shows that the abundances of species in the Orion-KL CR are also higher than those in the narrow component of Orion-S  ($\left <\frac{X_{KL}}{X_S} \right > = 23$;  SD = 45) but the agreement is better than it is for the HC.   Again, the match is better to the broad component ($\left <\frac{X_{KL}}{X_S} \right > = 1$;  SD = 2) of the two component fits in Orion-S (dashed line) than it is to the narrow component.   

\begin{figure}
\centering
\includegraphics[scale=1.2]{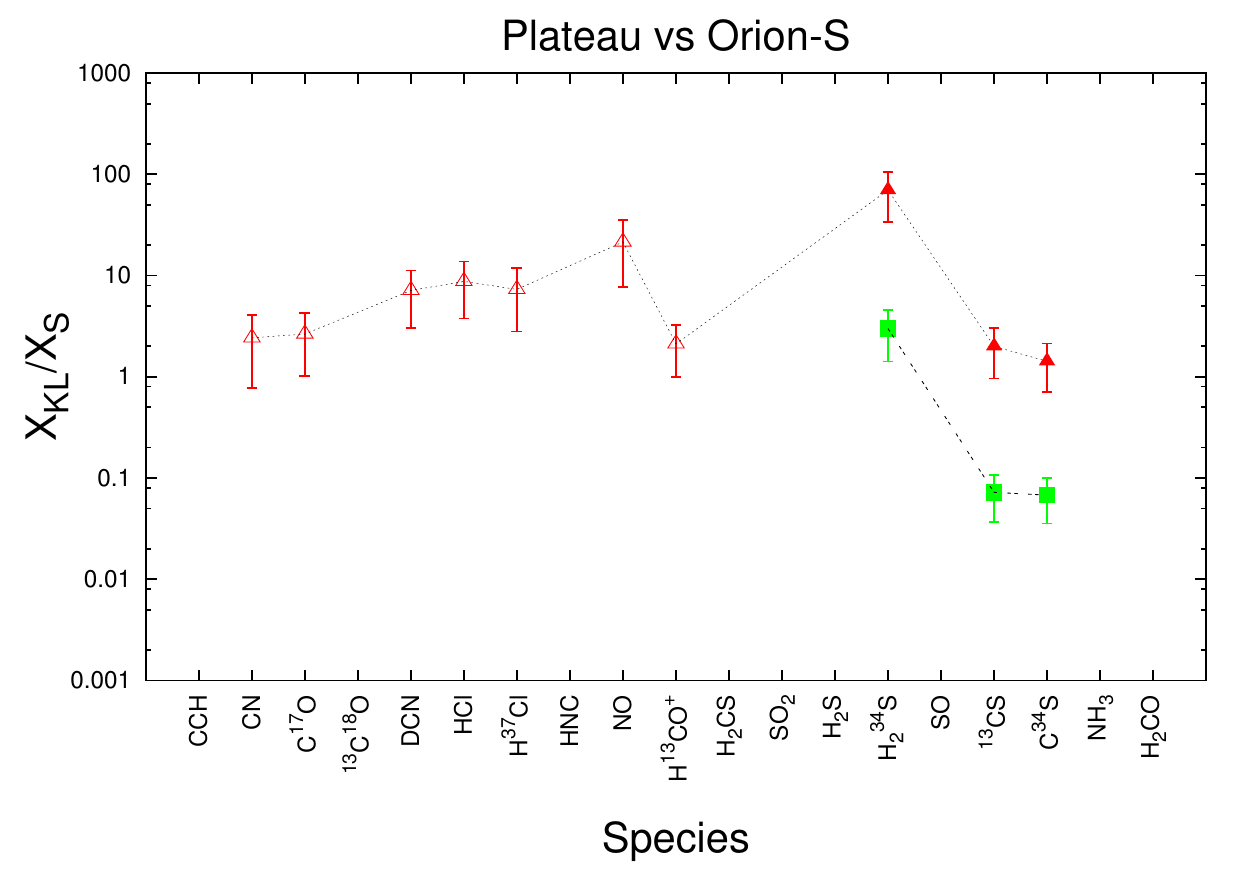}
\caption{ Same as for Figure \ref{HC} except for the Orion-KL Plateau.}
\label{P}
\end{figure}

Figure \ref{P} shows the comparison between Orion-S and the Orion-KL Plateau region.  The match between abundances here is clearly better than it is for the HC or the CR   with $\left <\frac{X_{KL}}{X_S} \right > = 14$ (SD = 22) for the narrow component (dotted line) and $ 1$ (SD = 2) for the broad component (dashed line).  
Note that although SO and SO$_2$ are often associated with shocked gas they do not appear in Figure \ref{P}. This is because these molecules were optically thick and  \cite{Crockett:2014ApJ...787..112C} did not provide column densities.  

\begin{figure}
\centering
\includegraphics[scale=1.2]{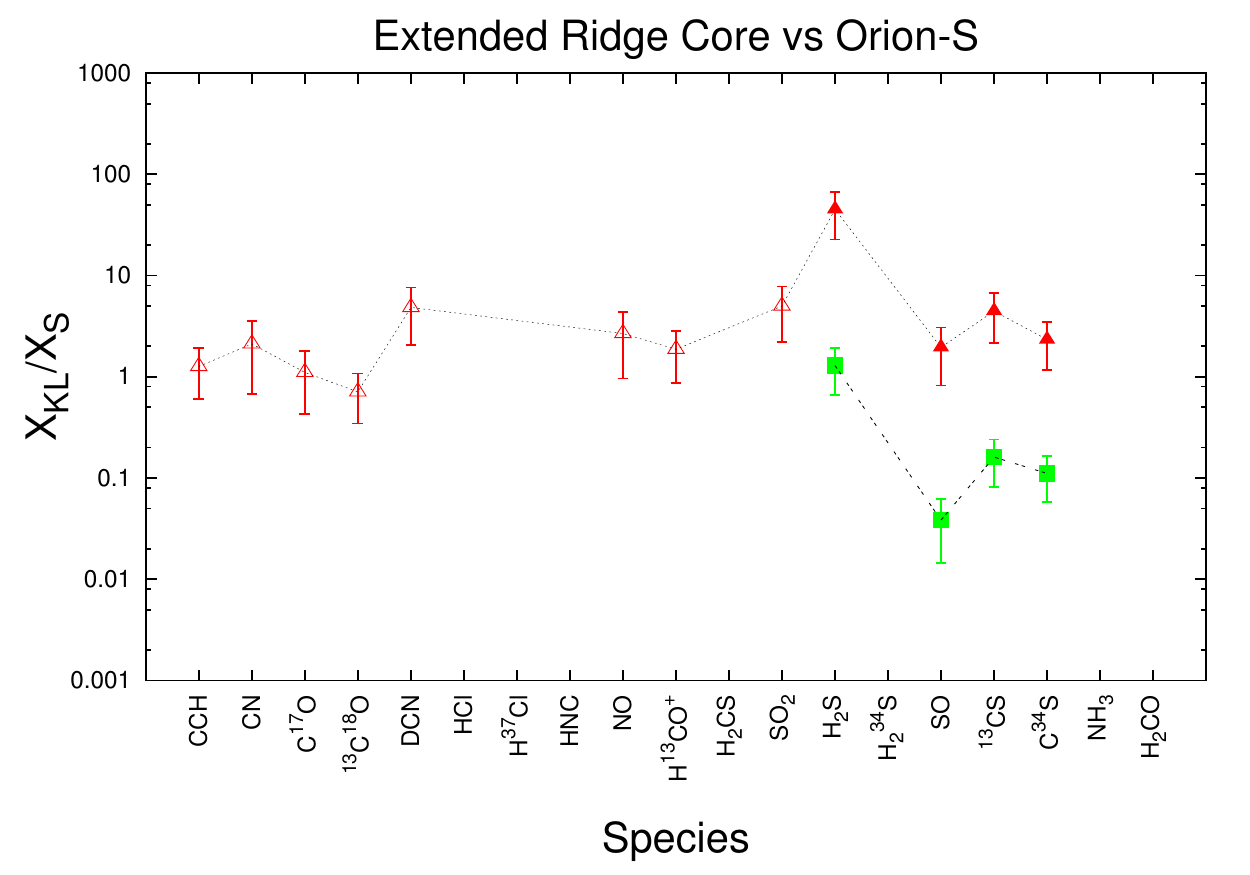}
\caption{ Same as for Figure \ref{HC} except for the Orion-KL Extended Ridge.}
\label{ER}
\end{figure}

The best agreement with molecular abundances in the narrow component of Orion-S is with the Extended Ridge of Orion-KL (Figure \ref{ER}) where  we obtain  $\left <\frac{X_{KL}}{X_S} \right > = 7$ (SD = 14).  For the broad component (dashed line) we obtain $\left <\frac{X_{KL}}{X_S} \right > = 0.3$ (SD = 0.5).  

Given that the best match to the abundances in the narrow component of Orion-S is the ER of Orion-KL, it seems as though these species/components do indeed trace quiescent gas.  In particular, it probably is the same gas out of which both star forming regions have been formed.  The broad component of Orion-S, however, seems better matched to the CR and Plateau of Orion-KL. Figures \ref{HC} to \ref{ER} only have a few broad component points and, therefore, it is difficult to make any strong statistical arguments based on these data alone. However, this evidence along with the chemical abundance analysis presented in Section 3.3.1 provide fairly strong support for the idea that shocks have also had an influence on the chemistry of Orion-S.  

\subsubsection{Species Detected in Orion-KL but not in Orion-S}

It is well known that Orion-KL has an incredibly rich molecular chemistry \citep[e.g.][]{Schilke:1997ApJS..108..301S, 2005ApJS..156..127C, 2007A&A...476..791O, 2007A&A...476..807P, 2006A&A...454L..47L, 2005IAUS..235P.203T}.  However, it is possible that the species that were detected in Orion-KL, but not in Orion-S, exist in the latter source, but at levels too weak to be detected.  Some of these species might be observable with ALMA at lower frequencies. In this Section, we explore this possibility by providing upper limits for the abundances of all the species detected in Orion-KL by  \cite{Crockett:2014ApJ...787..112C} but not detected in Orion-S.  

Modeling was accomplished by fixing T$_{ex}$, $\Delta V_{FWHM}$, $\Omega$, and V$_{LSR}$ and finding the column density that produced transitions whose intensities were $< 3 \sigma$ across all the HEXOS bands.  
Since we were able to model all of our species using, at most, two components, we determine two different column density upper limits:  one assuming that the undetected emission arises from the narrow component and assuming that it originates in the broad component.  For the narrow component, we used fixed values of T$_{ex} = 40$K, $\Delta V_{FWHM} = 4$ km s $^{-1}$, $\Omega = 60''$, and V$_{LSR} = 7.1$, which were  found to be typical for the narrow component (see Tables \ref{tab:LTE}  and  \ref{tab:LTE2}).  \textbf{For the broad component, we used} fixed values of T$_{ex} = 80$K, $\Delta V_{FWHM} = 8$ km s $^{-1}$, $\Omega = 40''$, and V$_{LSR} = 7.1$, which were  found to be typical for the broad component (see Table \ref{tab:LTE2}). Results are listed in Table \ref{tab:OrionKL}.  Column 1 is the species name, column 2 the upper limit total column density assuming the gas arises in the narrow component, column 3 the upper limit total column density assuming the gas originates in the broad component, and column 4 lists the maximum upper state energy for the model (i.e. no transitions with E $>$ E$_{\rm up}$ were modeled).  Different values of E$_{\rm up}$ were used for different molecules to keep the number of modeled transitions to a reasonable value. 

There are also a number of species that we detected (S/N $> 3\sigma$), but did not model, since their S/N was $< 5 \sigma$.  Using the same assumptions as for the undetected species we provide upper limits for the abundances for these species in Table \ref{tab:upperS}. The column density limits listed in Table \ref{tab:upperS} are approximately an order of magnitude or more smaller than the column densities of the same species detected in Orion-KL \citep{Crockett:2014ApJ...787..112C}.

A possible question is whether the species listed in Table \ref{tab:OrionKL} could actually exist in Orion-S even though they are undetected. We inspect the excitation conditions of these species, examining E$_{\rm up}$ of all possible transitions, to determine whether they are detectable given the reported noise of the HIFI bands. A portion of these species have E$_{\rm up}$ much greater than 500 K (e.g. H$_2$O v$_2$, H$^{13}$CN v$_2$ = 1, HC$_3$N, HC$_3$N v=0, HCN v$_2$ = 1, HCN v$_2$ = 2, OCS, etc.). Based on the analysis done in this paper, Orion-S can barely excite species with E$_{\rm up}$ $>$ 500 K. Therefore, transitions of these species would not be observed above the noise, even if they were present. In addition, there are some complex organic species with transition of E$_{\rm up}$  $<$ 100 K which we also did not detect  (e.g.  CH$_2$NH, CH$_2$DOH, CH$_2$NH, CH$_3$OCHO, etc.). This is not surprising given the absence of hot core chemistry in Orion-S.

\section{SUMMARY AND CONCLUSION}

We have presented results from a comprehensive spectral survey toward Orion South, taken with the HIFI instrument aboard the \textit{Herschel Space Observatory} covering the frequency range 480 to 1900 GHz with a resolution of 1.1 MHz.  We detected 685 spectral lines with S/N $>$ 3\sig\  originating from 52 different molecular and atomic species. Using the CASSIS spectral line analysis software package, we modeled each of the detected species assuming conditions of Local Thermodynamic Equilibrium.   Based on this modeling, we  found evidence for three different cloud components: a cool (T$_{ex} \sim 20-40$ K), spatially extended ($> 60''$), and quiescent ($\Delta V_{FWHM} \sim 4$ km s $^{-1}$) component; a warmer (T$_{ex} \sim 80-100$ K), less spatially extended ($\sim 30''$), and more dynamic ($\Delta V_{FWHM} \sim 8$ km s $^{-1}$) component, which is likely affected by embedded outflows; and a kinematically distinct region dominated by emission from species which trace UV irradiation. \textbf{Indirect evidence to support the existence of the first two components can be inferred from \cite{McMullin:1993p6318} who mapped the region in a few spectral lines (SiO, H$^{13}$CO$^+$, SO$_2$, CH$_3$OH, and HC$_3$N) with the BIMA array. Their H$^{13}$CO$^+$ and HC$_3$N data confirm the existence of a fairly extended ($\sim$1$'$) quiescent (FWHM $\sim$3 km s$^{-1}$) component, whereas the SO$_2$ and CH$_3$OH data reveal a smaller emitting region ($\sim$20$''$) of warm gas ($\sim$75 K).  While the spectra for the latter two species are too weak to determine line widths, their SiO data reveal a similarly small region (offset by only a few arc seconds from the SO$_2$ and CH$_3$OH emission peaks)  with broad line widths ($\sim$7 km s$^{-1}$).  In addition, \cite{McMullin:1993p6318} reports column densities of SO$_2$ and H$^{13}$CO$^+$ of $<$ 2$\times$10$^{-10}$ and 4$\times$10$^{-11}$ respectively which compare favourably to the values reported in Table 7. Finally, while there are no higher resolution observations to confirm the existence of the third component (i.e. the UV irradiated region), since CO$^+$ is only ever detected in PDRs, its presence in our data strongly suggests that such a component must exist. }

We also presented a comprehensive chemical abundance comparison between Orion-KL and Orion-S; two star forming regions that potentially formed from the same natal molecular gas but are at different evolutionary stages.  Based on a paucity of complex molecules in Orion-S, we found little chemical evidence for the existence of a significant ``hot core'' component. This is likely due to the fact that either the hot cores associated with the embedded star formation have either not had sufficient time to develop chemically, or that they are simply too small for their line emission to be detected in the large Herschel beam, or that Orion-S is not a massive star forming region and hot cores  massive enough to produce their characteristic rich spectra simply do not exist. The presence of a number of UV tracers such as [CII], [CI], CH, CH$^+$, SH$^+$, CO$^+$, and the fact that transitions of these species have velocities that are 1-1.5 km s$^{-1}$, higher than those of the quiescent gas, suggest that these species arise from a kinematically distinct PDR; most likely the UV illuminated surface of the cloud.
 The best match to the chemical abundances in the cooler, quiescent gas in Orion-S is with the quiescent extended ridge of Orion-KL, indicating that most of the gas in Orion-S is still quiescent as well, and relatively unaffected by  higher temperature or UV driven chemistry.  The best agreement  with the warmer, broad component of Orion-S is with the Orion-KL Plateau  and Compact Ridge regions, suggesting that shocks  have had an influence on the overall chemistry in Orion-S.

\acknowledgements
\section*{Acknowledgements}

 HIFI was designed and built by a consortium of institutes and university departments from across 
Europe, Canada and the United States under the leadership of SRON Netherlands Institute for Space
Research, Groningen, The Netherlands and with major contributions from Germany, France and the US. 
Consortium members are: Canada: CSA, U.Waterloo; France: CESR, LAB, LERMA,  IRAM; Germany: 
KOSMA, MPIfR, MPS; Ireland, NUI Maynooth; Italy: ASI, IFSI-INAF, Osservatorio Astrofisico di Arcetri- 
INAF; Netherlands: SRON, TUD; Poland: CAMK, CBK; Spain: Observatorio Astron—mico Nacional (IGN), 
Centro de Astrobiolog'a (CSIC-INTA). Sweden:  Chalmers University of Technology - MC2, RSS \& GARD; 
Onsala Space Observatory; Swedish National Space Board, Stockholm University - Stockholm Observatory; 
Switzerland: ETH Zurich, FHNW; USA: Caltech, JPL, NHSC.
We also need to acknowledge the support by the Deutsche Forschungsgemeinschaft (DFG) via the collaborative research grant SFB 956, project C1 \& C3, as well as the ERC and the Spanish MINECO for funding support under grants ERC-2013-Syg-610256 and AYA2012-32032.
Support for this work was provided, in part, by a National Sciences and Engineering Research Council of Canada (NSERC) grant to R. Plume and K. Tahani and by NASA through an award issued by JPL/Caltech. This work was carried out in part at the Jet Propulsion Laboratory, which is operated for NASA by the California Institute of Technology.

{\it Facilities:} \facility{Herschel}.

\bibliographystyle{apj.bst}
\bibliography{orions}

\clearpage


\begin{table}[t]
\caption{Data Smoothing and Noise Characteristics}
\begin{center}
\begin{tabular}{lcccccc}
\hline \hline
 & Smoothing & Freq. Res. & Vel. Res. & T$_{sys}$ & 1$\sigma$ rms \\
 & Channels        & (MHz)        & (km s$^{-1}$) & (K) &  (K)  \\                      
\hline
band 1  &	 	2	&	2.2	&	1.0$-$1.4	& $\sim100$&	$\sim0.02-0.05$\\
band 2	&	 	4	&	4.4	&	1.6$-$2.1	&$\sim150$&	$\sim0.03 - 0.07$\\
band 3	&	 	4	&	4.4	&	1.4$-$1.7	&$\sim200$&	$\sim0.04-0.09$\\
band 4	&	 	8	&	8.8	&	2.4$-$2.8	&$\sim400$&	$\sim0.1-0.2$\\
band 5	&	 	8	&	8.8	&	2.1$-$2.4	&$\sim1000$&	$\sim0.1-0.3$\\
band 6	&	 	16	&	17.6	&	3.1$-$3.7	&$\sim1300$& $\sim0.3-2.0$\\
band 7	&	 	16	&	17.6	&	2.8$-$3.1	&$\sim1300$& $\sim0.4-4.0$\\
\hline \hline
\end{tabular}
\end{center}
\label{tab:noise}
\end{table}

\begin{table}																								
\caption{Identified species in Orion-S} \label{tab:IdentifiedSpecies}																								
\begin{tabular}{c c c c}																								
\hline																								
atoms 		&			di-atomic molecules   &	multi-atomic molecules	& Ionized species\\			
\hline																				
\hline																			
C-atom	&	CO	&	CCH	&	C$^+$\\
	&	$^{13}$CO	&	DCN	&	CH$^+$\\
	&	$^{13}$C$^{18}$O	&	HNC	&	CO$^+$\\
	&	C$^{18}$O	&	HCN	&	DCO$^+$\\
	&	C$^{17}$O	&	H$^{13}$CN	&	H$^{13}$CO$^+$\\
	&	CS	&	HC$^{15}$N	&	HC$^{18}$O$^+$\\
	&	$^{13}$CS	&	HDO$\;^{\rm a}$	&	HCO$^+$\\
	&	C$^{34}$S	&	o${-}$H$_2$O$\;$	&	HCS$^+$\\
	&	CH	&	p${-}$H$_2$O$\;$	&	N$_2$H$^+$\\
	&	CN	&	o${-}$H$_2^{18}$O$\;^{\rm a}$	&	SH$^+$\\
	&	HCl$\;^{\rm a}$ 	&	p${-}$H$_2^{18}$O$\;^{\rm a}$	&	\\
	&	H$^{37}$Cl$\;^{\rm a}$	&	o${-}$H$_2$S$\;$	&	\\
	&	HF$\;^{\rm a}$	&	p${-}$H$_2$S$\;$	&	\\
	&	NO$\;^{\rm a}$	&	H$_2^{33}$S	&	\\
	&	SiO 	&	H$_2^{34}$S	&	\\
	&	SO	&	H$_2$CS	&	\\
	&		&	o${-}$H$_2$CO$\;$	&	\\
	&		&	p${-}$H$_2$CO$\;$	&	\\
	&		&	o${-}$NH$_3$	&	\\
	&		&	p${-}$NH$_3$	&	\\
	&		&	A${-}$CH$_3$OH$\;$	&	\\
	&		&	E${-}$CH$_3$OH$\;$	&	\\
	&		&	CH$_3$OCH$_3$	&	\\
	&		&	NH$_2$	&	\\
	&		&	SO$_2$	&	\\

\hline																								
\end{tabular}																								
\footnotesize{\hspace{6cm} $^{\rm a}$ JPL database used for these species. For all other species CDMS database is used.}\\																								
\footnotesize{}																								
\end{table}																								

\begin{table}[t]								
\caption{All observed lines above the $3 \sigma$ noise level in order of increasing frequency.} \label{tab:Frequencies}								
\begin{minipage}[t]{0.5\linewidth}								
	\begin{tabular}{lcr}							
	\hline \hline							
	\textbf{Band 1a} \\							
	\hline \hline							
	Frequency	&	T$_A$	&	Species\\ 			
	   (GHz)	&	(K)	&	\\			
	\hline							
	480.2699	&	0.7	&	A${-}$CH$_3$OH	\\
	481.5056	&	0.7	&	A${-}$CH$_3$OH	\\
	481.9167	&	0.3	&	C$^{34}$S	\\
	482.2179	&	0.1	&	A${-}$CH$_3$OH	\\
	482.2833	&	0.8	&	E${-}$CH$_3$OH	\\
	482.9598	&	1.4	&	E${-}$CH$_3$OH	\\
	483.0808	&	0.2	&	A${-}$CH$_3$OH	\\
	483.1418	&	1.7	&	E${-}$CH$_3$OH	\\
	483.3898	&	0.3	&	E${-}$CH$_3$OH	\\
	483.3983	&	0.1	&	A${-}$CH$_3$OH	\\
	483.4623	&	0.2	&	A${-}$CH$_3$OH	\\
	483.4728	&	0.2	&	E${-}$CH$_3$OH	\\
	483.5393	&	0.2	&	A${-}$CH$_3$OH	\\
	483.5533	&	0.3	&	A${-}$CH$_3$OH	\\
	483.5668	&	0.3	&	E${-}$CH$_3$OH	\\
	483.5818	&	0.1	&	A${-}$CH$_3$OH	\\
	483.6868	&	0.6	&	E${-}$CH$_3$OH	\\
	483.7633	&	0.3	&	A${-}$CH$_3$OH	\\
	484.0058	&	0.8	&	A${-}$CH$_3$OH	\\
	484.0238	&	0.7	&	E${-}$CH$_3$OH	\\
	484.0718	&	0.4	&	E${-}$CH$_3$OH	\\
	484.2703	&	0.2	&	SO$_2$	\\
	485.2638	&	0.9	&	A${-}$CH$_3$OH	\\
	486.9419	&	0.9	&	A${-}$CH$_3$OH	\\
	487.5319	&	0.6	&	A${-}$CH$_3$OH	\\
	487.6639	&	0.2	&	H$_2$CS	\\
	489.0379	&	0.9	&	A${-}$CH$_3$OH	\\
	489.7509	&	4.5	&	CS	\\
	490.5970	&	0.2	&	HDO	\\
	491.5520	&	0.7	&	A${-}$CH$_3$OH	\\
	491.9335	&	0.2	&	SO$_2$	\\
	\hline \hline							
	\end{tabular}							
\end{minipage}								
\begin{minipage}[t]{0.5\linewidth}								
	\begin{tabular}{lcr} 							
								
	\hline \hline							
	Frequency	&	T$_A$	&	Species\\ 			
	   (GHz)	&	(K)	&	\\			
	\hline							
	491.9690	&	3.1	&	o${-}$H$_2$CO	\\
	492.1615	&	4.7	&	[CI]	\\
	492.2795	&	2.0	&	A${-}$CH$_3$OH	\\
	492.7841	&	0.1	&	CH$_3$OCH$_3	$	\\
	493.7000	&	1.4	&	A${-}$CH$_3$OH	\\
	493.7350	&	1.4	&	A${-}$CH$_3$OH	\\
	494.4820	&	0.6	&	A${-}$CH$_3$OH	\\
	494.7781	&	0.2	&	SO$_2$	\\
	495.1741	&	1.2	&	E${-}$CH$_3$OH	\\
	496.9226	&	0.1	&	E${-}$CH$_3$OH	\\
	497.8296	&	0.4	&	A${-}$CH$_3$OH	\\
	501.5897	&	0.4	&	A${-}$CH$_3$OH	\\
	503.0142	&	0.1	&	H$_2^{34}$S	\\
	504.2008	&	0.2	&	DCO$^+$	\\
	504.2948	&	1.3	&	E${-}$CH$_3$OH	\\
	504.6783	&	0.1	&	SO	\\
	505.5658	&	1.5	&	o${-}$H$_2$S	\\
	505.7633	&	0.3	&	A${-}$CH$_3$OH	\\
	505.8343	&	1.7	&	p${-}$H$_2$CO	\\
	506.1548	&	0.2	&	E${-}$CH$_3$OH	\\
	506.7728	&	0.2	&	H$_2$CS	\\
	506.8273	&	0.2	&	DCN	\\
	508.5369	&	0.1	&	$^{13}$CS	\\
	508.7069	&	0.1	&	SO$_2$	\\
	509.0924	&	0.1	&	A${-}$CH$_3$OH	\\
	509.1469	&	0.8	&	p${-}$H$_2$CO	\\
	509.2939	&	0.3	&	HDO	\\
	509.5654$^b$	&	0.8	&	E${-}$CH$_3$OH	\\
		&		&		+ o${-}$H$_2$CO		\\
	509.8314	&	0.3	&	p${-}$H$_2$CO	\\
	510.1559	&	0.9	&	o${-}$H$_2$CO	\\
	510.2389	&	0.9	&	o${-}$H$_2$CO	\\

	\hline \hline							
	\end{tabular}							
\end{minipage}								
\end {table}								
								
\clearpage								
								
\begin{table}								
\begin{minipage}[t]{0.5\linewidth}								
	\begin{tabular}{lcr} 							
	\hline \hline							
	Frequency	&	T$_A$	&	Species\\ 			
	   (GHz)	&	(K)	&	\\			
	\hline

	510.3459	&	0.3	&	A${-}$CH$_3$OH	\\
	510.9114	&	0.1	&	HC$^{18}$O$^+$	\\
	511.0904	&	0.2	&	SO$_2$	\\
	511.5024	&	0.1	&	SO$_2$	\\
	511.7166	&	0.1	&	CH$_3$OCH$_3	$	\\
	511.9455	&	0.2	&	HCS$^+$	\\
	513.0775	&	0.7	&	p${-}$H$_2$CO	\\
	513.3610	&	0.1	&	H$_2$CS	\\
	513.3698	&	0.1	&	CH$_3$OCH$_3	$	\\
	514.8535	&	0.7	&	SO	\\
	515.1700	&	0.2	&	A${-}$CH$_3$OH	\\
	515.3335	&	0.2	&	A${-}$CH$_3$OH	\\
	515.8230	&	0.1	&	Ghost	\\
	516.2616	&	0.2	&	HC$^{15}$N	\\
	516.3361$^b$	&	0.7	&	SO	\\
		&		&		+ H$_2$CS		\\
	517.3551	&	0.9	&	SO	\\
	517.9706	&	0.4	&	H$^{13}$CN	\\
	520.1801	&	1.2	&	E${-}$CH$_3$OH	\\
	520.4612	&	1.1	&	H$^{13}$CO$^+$	\\
	520.7297	&	0.1	&	A${-}$CH$_3$OH	\\
	520.8817	&	0.1	&	SiO	\\
	522.4057	&	0.1	&	H$_2$CS	\\
	523.1002	&	0.0	&	E${-}$CH$_3$OH	\\
	523.2752	&	0.3	&	E${-}$CH$_3$OH	\\
	523.4827	&	0.1	&	$^{13}$C$^{18}$O	\\
	523.9727	&	2.2	&	CCH	\\
	524.0352	&	1.8	&	CCH	\\
	524.2682	&	0.1	&	E${-}$CH$_3$OH	\\
	524.5847	&	0.1	&	E${-}$CH$_3$OH	\\
	524.6673	&	0.1	&	E${-}$CH$_3$OH	\\
	524.7418	&	0.2	&	E${-}$CH$_3$OH	\\
	\hline \hline							
	\end{tabular}							
\end{minipage}								
\begin{minipage}[t]{0.5\linewidth}								
	\begin{tabular}{lcr} 							
	\hline \hline							
	Frequency	&	T$_A$	&	Species\\ 			
	   (GHz)	&	(K)	&	\\			
	\hline							
								
	524.8053	&	0.2	&	E${-}$CH$_3$OH	\\
	524.8628	&	0.2	&	E${-}$CH$_3$OH	\\
	524.9098	&	0.2	&	E${-}$CH$_3$OH	\\
	524.9483	&	0.1	&	E${-}$CH$_3$OH	\\
	525.0548	&	0.1	&	E${-}$CH$_3$OH	\\
	525.6663	&	2.5	&	o${-}$H$_2$CO	\\
	526.0363	&	0.1	&	SH$^+$	\\
	526.0453	&	0.1	&	SH$^+	$\\
	526.5233	&	0.1	&	A${-}$CH$_3$OH	\\
	526.5483	&	0.1	&	E${-}$CH$_3$OH	\\
	526.7878	&	0.1	&	E${-}$CH$_3$OH	\\
	527.0548	&	0.4	&	A${-}$CH$_3$OH	\\
	527.1738	&	0.1	&	E${-}$CH$_3$OH	\\
	527.6608	&	0.1	&	E${-}$CH$_3$OH	\\
	528.1813	&	0.1	&	E${-}$CH$_3$OH	\\
	528.6833	&	0.1	&	E${-}$CH$_3$OH	\\
	529.1419	&	0.2	&	E${-}$CH$_3$OH	\\
	529.2904	&	0.1	&	SO$_2$	\\
	529.5409	&	0.2	&	E${-}$CH$_3$OH	\\
	529.8679	&	0.2	&	E${-}$CH$_3$OH	\\
	529.9739	&	0.2	&	SO$_2$	\\
	530.0699	&	0.2	&	A${-}$CH$_3$OH	\\
	530.1234	&	0.3	&	C$^{34}$S	\\
	530.1849	&	0.6	&	E${-}$CH$_3$OH	\\
	530.0244	&	0.1	&	E${-}$CH$_3$OH	\\
	530.3174	&	0.4	&	E${-}$CH$_3$OH	\\
	530.4559	&	0.5	&	E${-}$CH$_3$OH	\\
	530.5499	&	0.5	&	E${-}$CH$_3$OH	\\
	530.6124	&	0.5	&	E${-}$CH$_3$OH	\\
	530.6484	&	0.5	&	E${-}$CH$_3$OH	\\
	530.8244	&	0.2	&	E${-}$CH$_3$OH	\\
	531.0804	&	1.2	&	E${-}$CH$_3$OH	\\
								
	\hline \hline							
	\end{tabular}							
\end{minipage}								
\end {table}								
								
\clearpage								
								
\begin{table}								
\begin{minipage}[t]{0.5\linewidth}								
	\begin{tabular}{lcr} 							
	\hline \hline							
	Frequency	&	T$_A$	&	Species\\ 			
	   (GHz)	&	(K)	&	\\			
	\hline							
	531.3199	&	1.3	&	A${-}$CH$_3$OH	\\
	531.6384	&	0.2	&	A${-}$CH$_3$OH	\\
	531.7159	&	7.5	&	HCN	\\
	531.8709	&	0.4	&	A${-}$CH$_3$OH	\\
	531.8924	&	0.3	&	A${-}$CH$_3$OH	\\
	532.0334	&	0.4	&	E${-}$CH$_3$OH	\\
	532.0709	&	0.2	&	E${-}$CH$_3$OH	\\
	532.1349	&	0.3	&	A${-}$CH$_3$OH	\\
	532.3239	&	0.1	&	CH$_3$OCH$_3	$	\\
	532.4669	&	0.6	&	E${-}$CH$_3$OH	\\
	532.5684	&	0.4	&	E${-}$CH$_3$OH	\\
	532.7214	&	2.4	&	CH	\\
	532.7909	&	1.0	&	CH	\\
	533.3810	&	0.1	&	Ghost	\\
	535.0610	&	14.6	&	HCO$^+$	\\
	536.1925	&	0.6	&	A${-}$CH$_3$OH	\\
	536.7585	&	2.1	&	CH	\\
	536.7805	&	0.5	&	CH	\\
	536.7925	&	0.9	&	CH	\\
	538.5716	&	2.2	&	A${-}$CH$_3$OH	\\
	538.6891	&	3.4	&	CS	\\
	539.2806	&	0.1	&	E${-}$CH$_3$OH	\\
	540.4656	&	0.2	&	H$_2$CS	\\
	540.9236	&	0.1	&	E${-}$CH$_3$OH	\\
	541.7536	&	0.2	&	SO$_2$	\\
	541.8162	&	0.1	&	SO$_2$	\\
	542.0022	&	1.3	&	A${-}$CH$_3$OH	\\
	542.0832	&	1.4	&	A${-}$CH$_3$OH	\\
	543.0777	&	1.2	&	E${-}$CH$_3$OH	\\
	543.8987	&	0.1	&	HNC	\\
	545.0437$^b$	&	0.2	&	E${-}$CH$_3$OH	\\
	&		&		+ A${-}$CH$_3$OH		\\

	\hline \hline							
	\end{tabular}							
\end{minipage}								
\begin{minipage}[t]{0.5\linewidth}								
	\begin{tabular}{lcr} 							
	\hline \hline							
	Frequency	&	T$_A$	&	Species\\ 			
	   (GHz)	&	(K)	&	\\			
	\hline

	545.1032	&	0.2	&	Ghost	\\
	545.8872	&	0.2	&	Ghost	\\
	546.2488	&	0.2	&	A${-}$CH$_3$OH	\\
	547.3003	&	0.2	&	Ghost	\\
	547.6768	&	0.3	&	H$_2^{18}$O	\\
	548.8313	&	8.1	&	C$^{18}$O	\\
	549.2998	&	0.1	&	SO$_2$	\\
	549.5506	&	0.1	&	CH$_3$OCH$_3	$	\\
	549.5533	&	0.1	&	NO	\\
	550.6564	&	0.0	&	A${-}$CH$_3$OH	\\
	550.9254	&	29.4	&	$^{13}$CO	\\
	551.1874	&	0.3	&	NO	\\
	551.5159	&	0.1	&	Ghost	\\
	551.5344	&	0.3	&	NO	\\
	553.1474	&	1.0	&	E${-}$CH$_3$OH	\\
	554.0569	&	0.2	&	E${-}$CH$_3$OH	\\
	554.5784	&	0.1	&	HCS$^+$	\\
	555.6670	&	0.1	&	SO$_2$	\\
	556.9370	&	7.2	&	o${-}$H$_2$O	\\
	557.1260	&	0.1	&	H$_2$CS	\\
	558.0870	&	0.3	&	SO	\\
	558.3465	&	0.4	&	E${-}$CH$_3$OH	\\
	558.9681	&	2.5	&	N$_2$H$^+$	\\
	559.3216	&	0.4	&	SO	\\
	560.1781	&	0.5	&	SO	\\
	560.2496	&	0.2	&	E${-}$CH$_3$OH	\\
	560.2716	&	0.2	&	E{-}CH$_3$OH		\\
	560.3001	&	0.2	&	Ghost	\\
	\hline \hline

\\\\\\\\				
	\end{tabular}							
\end{minipage}								
\end{table}								
								
\clearpage								
								
\begin{table}								
\begin{minipage}[t]{0.5\linewidth}								
	\begin{tabular}{lcr} 							
	\hline \hline							
	\textbf{Band 1b} \\							
	\hline \hline							
	Frequency	&	T$_A$	&	Species\\ 			
	   (GHz)	&	(K)	&	\\			
	\hline							
	556.9370	&	8.4	&	o${-}$H$_2$O	\\
	558.0875	&	0.5	&	SO	\\
	558.3460	&	0.5	&	E${-}$CH$_3$OH	\\
	558.9681	&	3.4	&	N$_2$H$^+$	\\
	559.3206	&	0.5	&	SO	\\
	560.1781	&	0.6	&	SO	\\
	561.7136	&	3.0	&	C$^{17}$O	\\
	561.9001	&	2.7	&	o${-}$H$_2$CO	\\
	564.2522	&	0.1	&	SiO	\\
	566.7312	&	1.3	&	CN	\\
	566.9477	&	1.6	&	CN	\\
	567.2612	&	0.2	&	Ghost	\\
	567.5953	&	0.1	&	SO$_2$	\\
	568.2358	&	0.2	&	E${-}$CH$_3$OH	\\
	568.4353	&	0.2	&	CH$_3$OCH$_3	$	\\
	568.5673	&	1.3	&	E${-}$CH$_3$OH	\\
	568.7853	&	0.2	&	A${-}$CH$_3$OH	\\
	570.2434	&	0.1	&	CH$_3$OCH$_3	$	\\
	572.4974	&	3.3	&	o${-}$NH$_3$	\\
	572.8984	&	0.2	&	E${-}$CH$_3$OH	\\
	574.1399	&	0.1	&	H$_2$CS	\\
	574.8084	&	0.1	&	SO$_2$	\\
	574.8694	&	0.3	&	A${-}$CH$_3$OH	\\
	576.2050$^b$	&	0.4	&	DCO$^+$	\\
	576.2660	&	62.0	&	CO	\\
	576.3835	&	0.2	&	Ghost	\\
	576.7095	&	1.1	&	p${-}$H$_2$CO	\\
	578.0080	&	0.4	&	E${-}$CH$_3$OH	\\
	578.2165	&	0.2	&	C$^{34}$S	\\
	579.0860	&	1.4	&	A${-}$CH$_3$OH	\\
	579.1520	&	0.8	&	E${-}$CH$_3$OH	\\
\hline\hline
								
	\end{tabular}							
\end{minipage}								
\begin{minipage}[t]{0.5\linewidth}								
	\begin{tabular}{lcr}							
	\hline \hline							
	Frequency	&	T$_A$	&	Species\\ 			
	   (GHz)	&	(K)	&	\\			
	\hline							
	579.1995	&	0.1	&	DCN	\\								
	579.4605	&	0.9	&	A${-}$CH$_3$OH	\\
	579.8590	&	0.2	&	A${-}$CH$_3$OH	\\
	579.9225	&	1.4	&	A${-}$CH$_3$OH	\\
	580.0600	&	0.1	&	A${-}$CH$_3$OH	\\
	580.1760	&	0.2	&	A${-}$CH$_3$OH	\\
	580.2135	&	0.2	&	A${-}$CH$_3$OH	\\
	580.3696	&	0.3	&	E${-}$CH$_3$OH	\\
	580.4446	&	0.2	&	E${-}$CH$_3$OH	\\
	580.5026	&	0.2	&	A${-}$CH$_3$OH	\\
	580.9036	&	0.4	&	E${-}$CH$_3$OH	\\
	581.0916	&	0.3	&	E${-}$CH$_3$OH	\\
	581.6131	&	0.5	&	p${-}$H$_2$CO	\\
	582.3841	&	0.2	&	o${-}$H$_2$CO	\\
	582.7246	&	0.2	&	p${-}$H$_2$CO	\\
	583.1456	&	0.6	&	o${-}$H$_2$CO	\\
	583.3096	&	0.5	&	o${-}$H$_2$CO	\\
	584.4511	&	1.9	&	A${-}$CH$_3$OH	\\
	584.8227	&	0.5	&	A${-}$CH$_3$OH	\\
	587.4537	&	0.4	&	p${-}$H$_2$CO	\\
	587.5702	&	0.2	&	SO$_2$	\\
	587.6167	&	2.3	&	CS	\\
	589.1653	&	0.1	&	CH$_3$OCH$_3	$	\\
	589.8703	&	0.1	&	CO$^+$	\\
	590.2793	&	1.2	&	A${-}$CH$_3$OH	\\
	590.4418	&	1.1	&	A${-}$CH$_3$OH	\\
	590.7522	&	0.1	&	CH$_3$OCH$_3	$	\\
	590.7923	&	0.9	&	E${-}$CH$_3$OH	\\
	591.8218	&	0.1	&	H$_2$CS	\\
	593.9424	&	0.1	&	SO$_2$	\\
	593.9624	&	0.1	&	SO$_2$	\\
	597.2084	&	0.1	&	HCS$^+$	\\

	\hline \hline							
	\end{tabular}							
\end{minipage}								
\end{table}								
								
\clearpage								
								
\begin{table}								
\begin{minipage}[t]{0.5\linewidth}								
	\begin{tabular}{lcr} 							
	\hline \hline							
	Frequency	&	T$_A$	&	Species\\ 			
	   (GHz)	&	(K)	&	\\			
	\hline							
								
	598.5485	&	-0.1	&	Ghost	\\
	599.9280	&	0.3	&	HDO	\\
	600.3320	&	1.7	&	o${-}$H$_2$CO	\\
	600.9065	&	0.1	&	$^{13}$CS	\\
	601.2570	&	0.4	&	SO	\\
	601.8521	&	0.1	&	E${-}$CH$_3$OH	\\
	602.2346	&	0.8	&	E${-}$CH$_3$OH	\\
	602.2750	&		&	HC$^{15}$N	\\
	602.2916	&	0.4	&	SO	\\
	603.0236	&	0.5	&	SO	\\
	604.2641	&	0.2	&	H$^{13}$CN	\\
	604.3696	&	0.2	&	SO$_2$	\\
	605.8801	&	0.1	&	E${-}$CH$_3$OH	\\
	607.1762	&	0.8	&	H$^{13}$CO$^+$	\\
	607.2167	&	0.3	&	E${-}$CH$_3$OH	\\
	607.6112	&	0.1	&	SiO	\\
	607.7982	&	0.2	&	H$_2$CS	\\
	608.0984	&	0.1	&	CH$_3$OCH$_3	$	\\
	609.7087	&	0.1	&	CH$_3$OCH$_3	$	\\
	611.2688	&	1.2	&	CCH	\\
	611.3308	&	1.0	&	CCH	\\
	611.5518	&	0.1	&	SO	\\
	613.0798	&	0.1	&	SO$_2$	\\
	616.9814	&	1.2	&	E${-}$CH$_3$OH	\\
	620.3035	&	5.6	&	HCN	\\
	620.7045	&	0.6	&	o${-}$H$_2$O	\\
	622.5705	&	0.1	&	A${-}$CH$_3$OH	\\
	622.6605	&	0.3	&	A${-}$CH$_3$OH	\\
	622.7755	&	0.2	&	E${-}$CH$_3$OH	\\
	624.1801	&	0.2	&	E${-}$CH$_3$OH	\\
	624.2081	&	13.3	&	HCO$^+$	\\
	624.9616	&	1.0	&	H$^{37}$Cl		\\

	\hline \hline							
	\end{tabular}							
\end{minipage}								
\begin{minipage}[t]{0.5\linewidth}								
	\begin{tabular}{lcr} 							
	\hline \hline							
	Frequency	&	T$_A$	&	Species\\ 			
	   (GHz)	&	(K)	&	\\			
	\hline							
								
	624.9751	&	1.4	&	H$^{37}$Cl		\\
	624.9856	&	0.6	&	H$^{37}$Cl		\\
	625.7521	&	0.4	&	E${-}$CH$_3$OH	\\
	625.7601	&	0.1	&	A${-}$CH$_3$OH	\\
	625.8996	&	2.1	&	HCl		\\
	625.9166	&	2.8	&	HCl		\\
	625.9296	&	1.5	&	HCl		\\
	626.0896	&	0.2	&	SO$_2$	\\
	626.3521	&	0.1	&	C$^{34}$S	\\
	626.4996	&	0.2	&	H$_2$CS	\\
	626.5141	&	0.1	&	A${-}$CH$_3$OH	\\
	626.5566	&	0.2	&	E${-}$CH$_3$OH	\\
	626.6271	&	1.5	&	A${-}$CH$_3$OH	\\
	627.0203	&	0.1	&	CH$_3$OCH$_3	$	\\
	627.1036	&	0.1	&	Ghost	\\
	627.1721	&	0.6	&	E${-}$CH$_3$OH	\\
	627.5602	&	0.7	&	A${-}$CH$_3$OH	\\
	628.0522	&	0.2	&	A${-}$CH$_3$OH	\\
	628.1422	&	0.2	&	$^{13}$C$^{18}$O	\\
	628.3302	&	0.2	&	E${-}$CH$_3$OH	\\
	628.4482	&	0.2	&	E${-}$CH$_3$OH	\\
	628.4717	&	0.3	&	A${-}$CH$_3$OH	\\
	628.5142	&	0.2	&	A${-}$CH$_3$OH	\\
	628.5272	&	0.2	&	A${-}$CH$_3$OH	\\
	628.6627	&	0.1	&	CH$_3$OCH$_3	$	\\
	628.6982	&	0.3	&	E${-}$CH$_3$OH	\\
	628.8182	&	0.2	&	E${-}$CH$_3$OH	\\
	628.8682	&	0.2	&	A${-}$CH$_3$OH	\\
	629.1417	&	1.6	&	A${-}$CH$_3$OH	\\
	629.3232	&	0.4	&	E${-}$CH$_3$OH	\\
	629.6517	&	0.3	&	E${-}$CH$_3$OH	\\
	629.9222	&	1.7	&	A${-}$CH$_3$OH	\\

	\hline \hline							
	\end{tabular}							
\end{minipage}								
\end{table}								
								
\clearpage								
								
\begin{table}								
\begin{minipage}[t]{0.5\linewidth}								
	\begin{tabular}{lcr} 							
	\hline \hline							
	Frequency	&	T$_A$	&	Species\\ 			
	   (GHz)	&	(K)	&	\\			
	\hline

	631.2847	&	0.2	&	Ghost	\\								
	631.7038	&	1.7	&	o${-}$H$_2$CO	\\
	632.1918	&	0.2	&	SO$_2$	\\
	633.4248	&	0.4	&	A${-}$CH$_3$OH	\\
	634.5118	&	0.6	&	HNC	\\
	635.8673	&	0.3	&	A${-}$CH$_3$OH	\\
	636.3394	&	0.5	&	A${-}$CH$_3$OH	\\
	636.3979	&	0.4	&	A${-}$CH$_3$OH	\\
	636.4209	&	0.6	&	A${-}$CH$_3$OH	\\
							
	\hline \hline	
			\\\\\\\\\\\\\\\\\\\\	\\\\\\\\\\\\\\\\\\\\	
	\\\\		\\

	\end{tabular}							
\end{minipage}								
\begin{minipage}[t]{0.5\linewidth}								
	\begin{tabular}{lcr} 							
	\hline \hline							
	\textbf{Band 2a} \\							
	\hline \hline							
	Frequency	&	T$_A$	&	Species\\ 			
	   (GHz)	&	(K)	&	\\			
	\hline							
	626.5191	&	0.7	&	A${-}$CH$_3$OH	\\
	626.6271	&	1.3	&	A${-}$CH$_3$OH	\\
	627.1711	&	0.6	&	E${-}$CH$_3$OH	\\
	627.5607	&	0.5	&	A${-}$CH$_3$OH	\\
	629.1417	&	1.0	&	A${-}$CH$_3$OH	\\
	629.3242	&	0.3	&	E${-}$CH$_3$OH	\\
	629.6482	&	0.3	&	E${-}$CH$_3$OH	\\
	629.9232	&	1.2	&	A${-}$CH$_3$OH	\\
	631.7038	&	1.3	&	o${-}$H$_2$CO	\\
	633.4233	&	0.3	&	A${-}$CH$_3$OH	\\
	634.5113	&	0.5	&	HNC	\\
	636.2514	&	0.2	&	A${-}$CH$_3$OH	\\
	636.2749	&	0.3	&	A${-}$CH$_3$OH	\\
	636.3059	&	0.3	&	A${-}$CH$_3$OH	\\
	636.3349	&	0.3	&	A${-}$CH$_3$OH	\\
	636.3669	&	0.4	&	A${-}$CH$_3$OH	\\
	636.3949	&	0.4	&	A${-}$CH$_3$OH	\\
	636.4209	&	0.3	&	A${-}$CH$_3$OH	\\
	636.5199	&	0.3	&	A${-}$CH$_3$OH	\\
	636.5339	&	1.3	&	CS	\\
	638.2804	&	0.7	&	E${-}$CH$_3$OH	\\
	638.5259	&	1.0	&	A${-}$CH$_3$OH	\\
	638.8194	&	0.9	&	A${-}$CH$_3$OH	\\
	644.3252	&	0.2	&	NH$_2$	\\
	644.3790	&	0.4	&	SO	\\
	645.2576	&	0.4	&	SO	\\
	645.8756	&	0.4	&	SO	\\
	645.9282	&	0.1	&	CH$_3$OCH$_3	$	\\
	647.0831	&	0.2	&	p${-}$H$_2$CO	\\
	647.6130	&	0.2	&	CH$_3$OCH$_3	$	\\
	648.1956	&	0.2	&	DCO$^+$	\\

	\hline \hline							
	\end{tabular}							
	\begin{flushleft}							
	\end{flushleft}							
								
\end{minipage}								
\end {table}								
\clearpage								
\begin{table}								
\begin{minipage}[t]{0.5\linewidth}								
	\begin{tabular}{lcr}							
				
	\hline \hline							
	Frequency	&	T$_A$	&	Species\\ 			
	   (GHz)	&	(K)	&	\\			
	\hline							
								
	649.5402	&	0.2	&	E${-}$CH$_3$OH	\\
	651.2997	&	0.1	&	SO$_2$	\\
	651.4352	&	0.3	&	NO	\\
	651.6177	&	0.6	&	E${-}$CH$_3$OH	\\
	651.7742	&	0.4	&	NO	\\
	652.0972	&	2.8	&	N$_2$H$^+$	\\
	653.9703	&	0.4	&	p${-}$H$_2$CO	\\
	655.2128	&	0.2	&	o${-}$H$_2$CO	\\
	655.6353	&	0.2	&	p${-}$H$_2$CO	\\
	656.1653	&	0.6	&	o${-}$H$_2$CO	\\
	656.1723	&	0.4	&	E${-}$CH$_3$OH	\\
	656.4663	&	0.5	&	o${-}$H$_2$CO	\\
	658.5544	&	7.7	&	C$^{18}$O	\\
	661.0659	&	30.3	&	$^{13}$CO	\\
	662.2105	&	0.4	&	p${-}$H$_2$CO	\\
	664.8193	&	0.1	&	CH$_3$OCH$_3	$	\\
	665.2485	&	0.2	&	SO$_2$	\\
	665.4440	&	1.3	&	E${-}$CH$_3$OH	\\
	670.3602	&	0.2	&	SO$_2$	\\
	670.4242	&	0.3	&	A${-}$CH$_3$OH	\\
	672.5657	&	0.2	&	SO$_2$	\\
	672.8362	&	0.1	&	A${-}$CH$_3$OH	\\
	672.9037	&	0.3	&	E${-}$CH$_3$OH	\\
	673.4172	&	0.3	&	E${-}$CH$_3$OH	\\
	673.7472	&	1.4	&	A${-}$CH$_3$OH	\\
	674.0107	&	2.4	&	C$^{17}$O	\\
	674.8113	&	1.1	&	o${-}$H$_2$CO	\\
	674.9913$^b$	&	1.4	&	A${-}$CH$_3$OH	\\
		&		&		+ E${-}$CH$_3$OH		\\
	675.0558	&	0.2	&	H$_2$CS	\\
	675.1353	&	0.5	&	E${-}$CH$_3$OH	\\
	675.6138	&	0.5	&	A${-}$CH$_3$OH	\\

	\hline \hline							
	\end{tabular}							
\end{minipage}								
\begin{minipage}[t]{0.5\linewidth}								
	\begin{tabular}{lcr} 							
	\hline \hline							
	Frequency	&	T$_A$	&	Species\\ 			
	   (GHz)	&	(K)	&	\\			
	\hline							
								
		675.7748	&	0.8	&	E${-}$CH$_3$OH	\\
	676.2143	&	0.3	&	A${-}$CH$_3$OH	\\
	676.4618	&	0.1	&	SO$_2$	\\
	676.7523	&	0.3	&	A${-}$CH$_3$OH	\\
	676.8298	&	0.3	&	A${-}$CH$_3$OH	\\
	677.0138	&	0.3	&	E${-}$CH$_3$OH	\\
	677.7113	&	0.4	&	E${-}$CH$_3$OH	\\
	678.2528	&	0.2	&	E${-}$CH$_3$OH	\\
	678.7688	&	0.2	&	Ghost	\\
	678.7864	&	1.6	&	A${-}$CH$_3$OH	\\
	680.0429	&	0.6	&	CN	\\
	680.2599	&	0.8	&	CN	\\
	681.5310	&	-0.1	&	Ghost	\\
	681.9919	&	0.4	&	A${-}$CH$_3$OH	\\
	685.4375	&	1.0	&	CS	\\
	685.5050	&	0.6	&	E${-}$CH$_3$OH	\\
	686.7325	&	0.7	&	A${-}$CH$_3$OH	\\
	687.0220	&	0.1	&	H$_2^{34}$S	\\
	687.1515	&	0.2	&	H$_2^{33}$S	\\
	687.2270	&	0.8	&	A${-}$CH$_3$OH	\\
	687.3040	&	2.3	&	p${-}$H$_2$S	\\
	687.4571	&	0.4	&	SO	\\
	688.2021	&	0.3	&	SO	\\
	688.7286	&	0.3	&	SO	\\
	690.5551	&	0.2	&	H$^{13}$CN	\\
	691.4706	&	64.4	&	CO	\\
	693.8787	&	0.4	&	H$^{13}$CO$^+$	\\
	697.1083	&	0.1	&	Ghost	\\
	697.1448	&	0.2	&	E${-}$CH$_3$OH	\\
	698.5458	&	0.6	&	CCH	\\
	698.6083	&	0.6	&	CCH	\\
	701.3709$^b$	&	1.3	&	o${-}$H$_2$CO	\\
								
	\hline							
	\hline							
	\end{tabular}							
\end{minipage}								
\end {table}

\clearpage								
								
\begin{table}								
\begin{minipage}[t]{0.5\linewidth}								
	\begin{tabular}{lcr} 							
	\hline \hline							
	Frequency	&	T$_A$	&	Species\\ 			
	   (GHz)	&	(K)	&	\\			
	\hline							
		&		&		 + E${-}$CH$_3$OH	\\	
	704.2609	&	0.2	&	Ghost	\\
	704.4134	&	0.2	&	CH$_3$OCH$_3$	\\
	704.9400	&	0.5	&	Ghost	\\
	705.1840	&	0.3	&	E${-}$CH$_3$OH	\\
	705.4000	&	0.2	&	Ghost	\\
	705.9605	&	0.4	&	Ghost	\\
	706.6255	&	0.3	&	Ghost	\\
	707.3180	&	0.4	&	Ghost	\\
	707.7905	&	0.4	&	E${-}$CH$_3$OH	\\
	708.4735$^b$	&	0.5	&	o${-}$H$_2$S 	\\
		&	0.0	&		+ H$_2^{33}$S		\\
	708.7085	&	0.4	&	Ghost	\\
	708.8741	&	3.2	&	HCN	\\
	709.2771	&	0.3	&	Ghost	\\
	713.3422	&	10.0	&	HCO$^+$	\\
	713.9837	&	0.9	&	E${-}$CH$_3$OH	\\
	716.9402	&	0.5	&	p${-}$H$_2$CO	\\
	718.1618	&	0.3	&	A${-}$CH$_3$OH	\\
	718.4378	&	0.9	&	E${-}$CH$_3$OH	\\
	719.6653	&	1.1	&	A${-}$CH$_3$OH	\\
	719.7293	&	1.0	&	Ghost	\\
	719.9548	&	1.5	&	Ghost	\\
	720.0118	&	0.9	&	E${-}$CH$_3$OH	\\
	720.4418	&	1.2	&	A${-}$CH$_3$OH	\\
	721.0118	&	0.2	&	E${-}$CH$_3$OH	\\
	723.0404	&	0.4	&	E${-}$CH$_3$OH	\\
	723.6219	&	0.4	&	A${-}$CH$_3$OH	\\
	724.1224	&	0.7	&	E${-}$CH$_3$OH	\\
	724.3489	&	0.4	&	A${-}$CH$_3$OH	\\
	725.0954	&	0.9	&	Ghost	\\	
	716.9412	&	0.6	&	p${-}$H$_2$CO	\\							
	\hline \hline				
	\end{tabular}							
\end{minipage}								
\begin{minipage}[t]{0.5\linewidth}								
	\begin{tabular}{lcr} 							
\hline \hline							
	\textbf{Band 2b} \\							
	\hline \hline			
							
	Frequency	&	T$_A$	&	Species\\ 			
	   (GHz)	&	(K)	&	\\			
	\hline							
	
	718.1623	&	0.3	&	A${-}$CH$_3$OH	\\
	718.4378	&	1.0	&	E${-}$CH$_3$OH	\\
	719.6668	&	1.2	&	A${-}$CH$_3$OH	\\
	720.4433	&	1.4	&	A${-}$CH$_3$OH	\\
	721.0093	&	0.3	&	E${-}$CH$_3$OH	\\
	723.0429	&	0.5	&	E${-}$CH$_3$OH	\\
	723.2824	&	0.3	&	E${-}$CH$_3$OH	\\
	723.6199	&	0.5	&	A${-}$CH$_3$OH	\\
	724.1219	&	0.9	&	E${-}$CH$_3$OH	\\
	724.3409	&	0.2	&	A${-}$CH$_3$OH	\\
	725.1089	&	0.3	&	HNC	\\
	725.1279	&	0.3	&	A${-}$CH$_3$OH	\\
	725.3169	&	0.2	&	E${-}$CH$_3$OH	\\
	726.0540	&	0.2	&	E${-}$CH$_3$OH	\\
	726.2100	&	0.4	&	p${-}$H$_2$CO	\\
	726.9025	&	0.3	&	E${-}$CH$_3$OH	\\
	728.0535	&	0.3	&	o${-}$H$_2$CO	\\
	728.5860	&	0.3	&	p${-}$H$_2$CO	\\
	728.5960	&	0.2	&	p${-}$H$_2$CO	\\
	728.8650	&	1.5	&	A${-}$CH$_3$OH	\\
	729.2120	&	0.5	&	o${-}$H$_2$CO	\\
	729.7260	&	0.4	&	o${-}$H$_2$CO	\\
	730.5006	&	0.4	&	SO	\\
	730.5206	&	0.4	&	A${-}$CH$_3$OH	\\
	731.1416	&	0.3	&	SO	\\
	731.5981	&	0.3	&	SO	\\
	732.4336	&	0.6	&	A${-}$CH$_3$OH	\\
	732.7761	&	0.2	&	$^{13}$C$^{18}$O	\\
	734.2706	&	0.4	&	H$_2^{34}$S	\\
	734.3286	&	0.8	&	CS	\\						
	734.8952	&	0.6	&	A${-}$CH$_3$OH	\\

	\hline \hline						
	\end{tabular}							
\end{minipage}								
\end{table}								
								
\clearpage								
								
\begin{table}								
\begin{minipage}[t]{0.5\linewidth}								
	\begin{tabular}{lcr} 							
	\hline \hline							
	Frequency	&	T$_A$	&	Species\\ 			
	   (GHz)	&	(K)	&	\\			
	\hline							
	
	735.6762	&	0.6	&	A${-}$CH$_3$OH	\\
	736.0317	&	4.3	&	o${-}$H$_2$S	\\
	737.3407	&	0.4	&	p${-}$H$_2$CO	\\
	737.6272	&	0.2	&	Ghost	\\
	741.2263	&	0.2	&	E${-}$CH$_3$OH	\\
	742.2408	&	0.2	&	H$_2$CS	\\
	745.2119	&	1.8	&	N$_2$H$^+$	\\
	747.3035	&	0.3	&	p${-}$H$_2$S	\\
	749.0735	&	0.8	&	o${-}$H$_2$CO	\\
	751.5560	&	0.4	&	E${-}$CH$_3$OH	\\
	751.6766	&	0.3	&	NO	\\
	752.0321	&	7.9	&	p${-}$H$_2$O	\\
	752.1076	&	0.6	&	E${-}$CH$_3$OH	\\
	752.1366	&	0.4	&	E${-}$CH$_3$OH	\\
	752.1716	&	0.3	&	E${-}$CH$_3$OH	\\
	752.3111	&	0.3	&	E${-}$CH$_3$OH	\\
	753.4161	&	0.2	&	HDO	\\
	753.8671	&	0.1	&	E${-}$CH$_3$OH	\\
	754.2226	&	0.2	&	E${-}$CH$_3$OH	\\
	762.6378	&	0.9	&	E${-}$CH$_3$OH	\\
	763.8823	&	0.3	&	E${-}$CH$_3$OH	\\
	763.9533	&	0.9	&	A${-}$CH$_3$OH	\\
	764.5828	&	-0.5	&	Ghost	\\
	764.8119	&	0.2	&	E${-}$CH$_3$OH	\\
	765.5134	&	0.2	&	E${-}$CH$_3$OH	\\
	765.9404	&	0.2	&	o${-}$H$_2$S	\\
	766.0309	&	0.3	&	E${-}$CH$_3$OH	\\
	766.3974	&	0.4	&	E${-}$CH$_3$OH	\\
	766.6489	&	0.5	&	E${-}$CH$_3$OH	\\
	766.7124	&	1.2	&	A${-}$CH$_3$OH	\\
	766.7624	&	0.8	&	E${-}$CH$_3$OH	\\
	766.8114	&	0.4	&	E${-}$CH$_3$OH	\\							
								
	\hline \hline										
	\end{tabular}							
\end{minipage}								
\begin{minipage}[t]{0.5\linewidth}								
	\begin{tabular}{lcr}							
	\hline \hline							
	Frequency	&	T$_A$	&	Species\\ 			
	   (GHz)	&	(K)	&	\\			
	\hline							
	
	766.9094	&	0.5	&	E${-}$CH$_3$OH	\\
	766.9614	&	0.5	&	E${-}$CH$_3$OH	\\
	766.9844	&	0.5	&	E${-}$CH$_3$OH	\\
	768.2529	&	5.6	&	C$^{18}$O	\\
	768.5404	&	0.2	&	E${-}$CH$_3$OH	\\
	770.8855	&	0.4	&	E${-}$CH$_3$OH	\\
	770.8980	&	0.9	&	o${-}$H$_2$CO	\\
	771.1825	&	27.1	&	$^{13}$CO	\\
	771.5770	&	0.4	&	A${-}$CH$_3$OH	\\
	772.4425	&	0.2	&	A${-}$CH$_3$OH	\\
	772.4545	&	0.6	&	E${-}$CH$_3$OH	\\
	773.2611	&	0.2	&	A${-}$CH$_3$OH	\\
	773.4226	&	0.3	&	A${-}$CH$_3$OH	\\
	773.5136	&	0.3	&	SO	\\
	773.8896	&	0.2	&	E${-}$CH$_3$OH	\\
	773.9481	&	0.2	&	A${-}$CH$_3$OH	\\
	774.0666	&	0.2	&	SO	\\
	774.3331	&	0.3	&	E${-}$CH$_3$OH	\\
	774.4541	&	0.3	&	SO	\\
	775.5996	&	0.4	&	E${-}$CH$_3$OH	\\
	779.0072	&	0.5	&	A${-}$CH$_3$OH	\\
	779.0322	&	0.4	&	E${-}$CH$_3$OH	\\
	779.3822	&	1.3	&	A${-}$CH$_3$OH	\\
	780.5672	&	0.3	&	H$^{13}$CO$^+$	\\
	783.0028	&	0.5	&	A${-}$CH$_3$OH	\\
	783.1993	&	0.6	&	CS	\\
	784.1793	&	0.3	&	A${-}$CH$_3$OH	\\
	785.8058	&	0.3	&	CCH	\\
	785.8679	&	0.2	&	CCH	\\
	786.2829$^b$	&	1.5	&	C$^{17}$O	\\
		&		&		 + p${-}$H$_2$CO		\\
	790.9360	&	0.7	&	Ghost	\\						
	\hline \hline			
	\end{tabular}							
\end{minipage}								
\end{table}								
								
\clearpage								
								
\begin{table}								
\begin{minipage}[t]{0.5\linewidth}								
	\begin{tabular}{lcr} 							
	\hline \hline							
	Frequency	&	T$_A$	&	Species\\ 			
	   (GHz)	&	(K)	&	\\			
	\hline

	793.3410	&	0.4	&	CN	\\
	793.5480	&	0.3	&	CN	\\
	794.5211	&	0.2	&	Ghost	\\
	794.8206	&	0.2	&	Ghost	\\
	797.4306	&	1.7	&	HCN	\\
	798.3106	&	0.2	&	p${-}$H$_2$CO	\\
	\hline \hline							
	\\\\\\\\\\\\\\\\							
		\\\\\\\\\\\\\\\\\\\\\\\\\\\\\\\\\\\

	\end{tabular}							
\end{minipage}								
\begin{minipage}[t]{0.5\linewidth}								
	\begin{tabular}{lcr} 							
	\hline\hline							
	\textbf{Band 3a} \\							
	\hline \hline							
	Frequency	&	T$_A$	&	Species\\ 			
	   (GHz)	&	(K)	&	\\			
	\hline							
	802.2430	&	0.3	&	E${-}$CH$_3$OH	\\
	802.2810	&	0.3	&	o${-}$H$_2$CO	\\
	802.4590	&	7.1	&	HCO$^+$	\\
	803.1130	&	0.3	&	o${-}$H$_2$CO	\\
	806.6501	&	59.4	&	CO	\\
	807.8661	&	0.6	&	A${-}$CH$_3$OH	\\
	809.3431	&	6.8	&	[CI]	\\
	811.4452	&	0.6	&	E${-}$CH$_3$OH	\\
	812.5522	&	0.9	&	A${-}$CH$_3$OH	\\
	813.5442	&	0.2	&	A${-}$CH$_3$OH	\\
	815.0723	&	0.8	&	E${-}$CH$_3$OH	\\
	815.6943	&	0.1	&	HNC	\\
	816.0153	&	0.2	&	E${-}$CH$_3$OH	\\
	816.4953	&	0.2	&	SO	\\
	816.9743	&	0.2	&	SO	\\
	817.3143	&	0.2	&	SO	\\
	818.6674	&	0.2	&	E${-}$CH$_3$OH	\\
	819.4844	&	0.3	&	A${-}$CH$_3$OH	\\
	820.5024	&	0.2	&	A${-}$CH$_3$OH	\\
	820.7644	&	0.4	&	E${-}$CH$_3$OH	\\
	821.4774	&	0.2	&	A${-}$CH$_3$OH	\\
	821.7014	&	0.2	&	A${-}$CH$_3$OH	\\
	821.8694	&	0.2	&	E${-}$CH$_3$OH	\\
	822.5475	&	0.3	&	E${-}$CH$_3$OH	\\
	823.0845	&	0.6	&	o${-}$H$_2$CO	\\
	824.3525	&	0.2	&	E${-}$CH$_3$OH	\\
	824.7255	&	0.2	&	E${-}$CH$_3$OH	\\
	825.2795	&	0.3	&	E${-}$CH$_3$OH	\\
	827.4516	&	0.2	&	A${-}$CH$_3$OH	\\
	829.8906	&	0.7	&	A${-}$CH$_3$OH	\\
	830.3506	&	1.0	&	A${-}$CH$_3$OH	\\								
									
	\hline \hline							
	\end{tabular}							
							
\end{minipage}								
\end{table}								
								
\clearpage								
								
\begin{table}								
\begin{minipage}[t]{0.5\linewidth}								
	\begin{tabular}{lcr} 
	\hline \hline													
	Frequency	&	T$_A$	&	Species\\ 			
	   (GHz)	&	(K)	&	\\			
	\hline							

	831.0487	&	0.3	&	A${-}$CH$_3$OH	\\
	832.0627	&	0.5	&	CS	\\
	832.7567	&	0.4	&	A${-}$CH$_3$OH	\\
	834.6517	&	0.2	&	E${-}$CH$_3$OH	\\
	834.7517	&	0.1	&	E${-}$CH$_3$OH	\\
	834.8397	&	0.2	&	E${-}$CH$_3$OH	\\
	834.9017	&	0.1	&	E${-}$CH$_3$OH	\\
	834.9577	&	0.2	&	E${-}$CH$_3$OH	\\
	835.0037	&	0.1	&	E${-}$CH$_3$OH	\\
	835.1358	&	3.0	&	CH$^+$	\\
	838.3088	&	0.8	&	N$_2$H$^+$	\\
	840.2779	&	0.6	&	o${-}$H$_2$CO	\\
	851.4161	&	0.7	&	A${-}$CH$_3$OH	\\
	851.9151	&	0.3	&	NO	\\
	853.5112	&	0.3	&	E${-}$CH$_3$OH	\\
	855.1542	&	0.3	&	p${-}$H$_2$CO	\\
	857.9613	&	0.7	&	A${-}$CH$_3$OH	\\
	\hline \hline							
	\\\\\\\\\\							
		\\\\\\\\\\\\\\\\	\\\\

	\end{tabular}							
								
\end{minipage}								
\begin{minipage}[t]{0.5\linewidth}								
	\begin{tabular}{lcr} 							
	\hline \hline							
	\textbf{Band 3b} \\							
	\hline \hline							
	Frequency	&	T$_A$	&	Species\\ 			
	   (GHz)	&	(K)	&	\\			
	\hline							
	860.4618	&	0.5	&	E${-}$CH$_3$OH	\\
	863.3639	&	0.6	&	E${-}$CH$_3$OH	\\
	863.4269	&	0.2	&	E${-}$CH$_3$OH	\\
	867.3250	&	0.3	&	A${-}$CH$_3$OH	\\
	869.0370	&	0.5	&	E${-}$CH$_3$OH	\\
	869.9821	&	0.3	&	A${-}$CH$_3$OH	\\
	870.1091	&	0.2	&	E${-}$CH$_3$OH	\\
	870.2811	&	0.2	&	p${-}$H$_2$CO	\\
	873.0361	&	0.2	&	CCH	\\
	873.7811	&	0.1	&	o${-}$H$_2$CO	\\
	875.3692	&	0.2	&	o${-}$H$_2$CO	\\
	876.6462	&	0.1	&	o${-}$H$_2$CO	\\
	877.9232	&	3.4	&	C$^{18}$O	\\
	878.2273	&	0.7	&	A${-}$CH$_3$OH	\\
	879.0153	&	0.2	&	A${-}$CH$_3$OH	\\
	880.9043	&	0.3	&	CS	\\
	881.2703	&	21.6	&	$^{13}$CO	\\
	881.4223	&	0.3	&	A${-}$CH$_3$OH	\\
	881.7833	&	0.8	&	A${-}$CH$_3$OH	\\
	885.9664	&	1.2	&	HCN	\\
	890.1265	&	0.3	&	E${-}$CH$_3$OH	\\
	891.5586	&	5.5	&	HCO$^+$	\\
	893.6387	&	-0.7	&	HDO	\\
	894.6146	&	0.5	&	A${-}$CH$_3$OH	\\
	896.8077	&	0.3	&	o${-}$H$_2$CO	\\
	898.5247	&	0.9	&	C$^{17}$O	\\
	898.9777	&	0.2	&	Ghost	\\
	900.9658	&	0.1	&	E${-}$CH$_3$OH	\\
	902.9388	&	0.5	&	A${-}$CH$_3$OH	\\
	902.9838	&	0.2	&	SO	\\
	905.3959	&	0.2	&	E${-}$CH$_3$OH	\\

	\hline \hline													
	\end{tabular}							
\end{minipage}								
\end {table}								
								
\clearpage								
								
\begin{table}[t]								
\begin{minipage}[t]{0.5\linewidth}								
	\begin{tabular}{lcr}		
		\hline \hline													
	Frequency	&	T$_A$	&	Species\\ 			
	   (GHz)	&	(K)	&	\\			
	\hline							
	905.3959	&	0.2	&	E${-}$CH$_3$OH	\\
	906.5939	&	0.2	&	CN	\\
	906.8029	&	0.2	&	CN	\\
	907.4303	&	0.2	&	NH$_2$	\\
	909.5120	&	0.3	&	o${-}$H$_2$CO	\\
	909.7400	&	0.4	&	E${-}$CH$_3$OH	\\
	910.8110	&	0.2	&	E${-}$CH$_3$OH	\\
	911.6440	&	0.5	&	E${-}$CH$_3$OH	\\
	912.1100	&	0.8	&	E${-}$CH$_3$OH	\\
	916.1771	&	0.5	&	p${-}$H$_2$O	\\
	916.6502	&	0.3	&	E${-}$CH$_3$OH	\\
	917.2702	&	0.4	&	E${-}$CH$_3$OH	\\
	917.4082	&	0.2	&	E${-}$CH$_3$OH	\\
	921.7973	&	59.5	&	CO	\\
	921.9873	&	0.5	&	E${-}$CH$_3$OH	\\
	923.5853	&	0.3	&	p${-}$H$_2$CO	\\
	926.5564	&	0.5	&	A${-}$CH$_3$OH	\\
	926.8944	&	0.3	&	A${-}$CH$_3$OH	\\
	929.7315	&	0.3	&	CS	\\
	930.2035	&	0.3	&	A${-}$CH$_3$OH	\\
	931.3885	&	0.4	&	N$_2$H$^+$	\\
	933.6945	&	0.6	&	A${-}$CH$_3$OH	\\
	937.4826	&	0.4	&	A${-}$CH$_3$OH	\\
	947.4759	&	0.3	&	A${-}$CH$_3$OH	\\
	952.5422	&	-0.3	&	NH$_2$	\\
	952.5740	&	-0.8	&	NH$_2$	\\
	952.6268	&	-0.3	&	NH$_2$	\\
				
	\hline \hline		
	\\\\\\\\	\\

	\end{tabular}							
\end{minipage}								
\begin{minipage}[t]{0.5\linewidth}								
	\begin{tabular}{lcr} 							
	\hline \hline							
	\textbf{Band 4a} \\							
	\hline \hline							
	Frequency	&	T$_A$	&	Species\\ 			
	   (GHz)	&	(K)	&	\\			
	\hline							
	960.4732	&	0.6	&	E${-}$CH$_3$OH	\\
	965.4513	&	0.4	&	E${-}$CH$_3$OH	\\
	974.4895	&	0.7	&	HCN	\\
	974.6895	&	0.3	&	A${-}$CH$_3$OH	\\
	974.8795	&	0.5	&	A${-}$CH$_3$OH	\\
	980.0316	&	0.4	&	A${-}$CH$_3$OH	\\
	980.6376	&	4.3	&	HCO$^+$	\\
	986.1018	&	0.6	&	A${-}$CH$_3$OH	\\
	987.5618	&	2.2	&	C$^{18}$O	\\
	987.9298	&	8.5	&	p${-}$H$_2$O	\\
	991.3259	&	17.6	&	$^{13}$CO	\\
	991.5839	&	0.5	&	A${-}$CH$_3$OH	\\
	993.1019	&	1.7	&	o${-}$H$_2$S	\\
	1002.7782	&	0.8	&	p${-}$H$_2$S	\\
	1006.1242	&	0.4	&	E${-}$CH$_3$OH	\\
	1008.8183	&	0.6	&	E${-}$CH$_3$OH	\\
	1010.7323	&	0.6	&	C$^{17}$O	\\
	1013.5664	&	0.4	&	E${-}$CH$_3$OH	\\
	1023.1986	&	0.6	&	A${-}$CH$_3$OH	\\
	1036.6990	&	1.1	&	SO$_2$	\\
	1036.9070	&	66.9	&	CO	\\
	1039.0150	&	0.5	&	A${-}$CH$_3$OH	\\
	1057.1194	&	0.5	&	E${-}$CH$_3$OH	\\
\hline\hline

		\\\\\\\\\\\\\\\\

	\end{tabular}							
\end{minipage}								
\end {table}								
								
\clearpage								
								
\begin{table}[t]								
\begin{minipage}[t]{0.5\linewidth}								
	\begin{tabular}{lcr}							
		\hline\hline						
	\textbf{Band 4b} \\							
	\hline	\hline						
	Frequency	&	T$_A$	&	Species\\ 			
	   (GHz)	&	(K)	&	\\			
	\hline							
	1057.1239	&	0.5	&	E${-}$CH$_3$OH	\\
	1062.9821	&	1.0	&	HCN	\\
	1069.6982	&	2.8	&	HCO$^+$	\\
	1072.8303	&	1.0	&	o${-}$H$_2$S	\\						
	1092.4668	&	0.4	&	A${-}$CH$_3$OH	\\
	1097.1589	&	1.4	&	C$^{18}$O	\\
	1097.3689	&	3.4	&	o${-}$H$_2$O	\\
	1101.3450	&	12.5	&	$^{13}$CO	\\
	1101.6990	&	-0.7	&	p${-}$H$_2^{18}$O	\\
	1105.3691	&	0.4	&	E${-}$CH$_3$OH	\\
	1113.3472	&	4.0	&	p${-}$H$_2$O	\\							
	\\
		\hline \hline							
	\textbf{Band 5a} \\							
	\hline \hline							
	Frequency	&	T$_A$	&	Species\\ 			
	   (GHz)	&	(K)	&	\\			
	\hline							
	1113.3477	&	3.5	&	p${-}$H$_2$O	\\
	1119.8339	&	0.4	&	A${-}$CH$_3$OH	\\
	1122.9060	&	0.3	&	C$^{17}$O	\\
	1151.4506	&	0.6	&	HCN	\\
	1151.7555	&	0.9	&	A${-}$CH$_3$OH	\\
	1151.9826	&	51.0	&	CO	\\
	1152.9207	&	0.6	&	A${-}$CH$_3$OH	\\
	1153.1267	&	4.4	&	o${-}$H$_2$O	\\
	1153.5527	&	0.6	&	E${-}$CH$_3$OH	\\
	1158.7328	&	1.7	&	HCO$^+$	\\							
	1162.7149$^b$	&	0.6	&	A${-}$CH$_3$OH	\\
		&		&		+ E${-}$CH$_3$OH		\\
	1162.9129	&	4.4	&	o${-}$H$_2$O	\\
	1168.1190	&	0.5	&	A${-}$CH$_3$OH	\\
	1168.4524	&	-1.1&	p${-}$NH$_3$\\		
	1196.0197	&	0.5	&	o${-}$H$_2$S	\\
		
	\hline \hline

	\end{tabular}							
\end{minipage}								
\begin{minipage}[t]{0.5\linewidth}								
	\begin{tabular}{lcr} 	
	\hline \hline							
	Frequency	&	T$_A$	&	Species\\ 			
	   (GHz)	&	(K)	&	\\			
	\hline							
	1206.7279	&	0.7	&	C$^{18}$O	\\
	1207.6540	&	0.5	&	p${-}$H$_2$O	\\
	1211.3260	&	6.8	&	$^{13}$CO	\\
	1214.8529	&	-1.0	&	o${-}$NH$_3$	\\
	1215.2457	&	-1.3	&	p${-}$NH$_3$	\\
	1228.7924	&	1.3	&	p${-}$H$_2$O	\\
	1232.4685	&	-1.2	&	HF	\\\\
								
	\hline \hline							
	\textbf{Band 5b} \\							
	\hline	\hline						
								
	\hline							
	1228.7922	&	1.7	&	p${-}$H$_2$O	\\
	1232.4692	&	-1.2	&	HF	\\
	1239.9125	&	0.4	&	HCN	\\
	1247.7406	&	0.8	&	HCO$^+$	\\
	1267.0091	&	44.6	&	CO	\\
	\\
	
		\hline \hline	
\textbf{Band 6a} \\	
\hline \hline							
	Frequency	&	T$_A$	&	Species\\ 			
	   (GHz)	&	(K)	&	\\			
	\hline								
	1496.9192	&	42.0	&	CO	\\\\
								
	\hline \hline							
	\textbf{Band 6b} \\							
	\hline	\hline						
	1611.7859	&	37.3	&	CO	\\
	1669.9169	&	7.7	&	o${-}$H$_2$O	\\\\
	\hline \hline							
	\textbf{Band 7a} \\							
	\hline	\hline						
	1726.5976	&	26.9	&	CO	\\\\
	\hline \hline							
	\textbf{Band 7b} \\							
	\hline	\hline						
	1841.3367	&	20.1	&	CO	\\
	1900.5261	&	69.5	&	[CII]	\\
								
	\hline \hline	\\						
								
	\end{tabular}							
\end{minipage}								
\end {table}								
								
\clearpage

\begin{table}[t]\caption{Gaussian fits to lines above 5\sig}\label{tab:Gaussian}																
\begin{tabular}{lccccccc}																
																
[CI]	 & 		 & 		 & 		 & 		 & 		 & 		 & 	\\		
\hline\hline																
	 & 	Transition	 & 	Frequency	 & 	T$_A$	 & 	$V_{lsr}$	 & 	$\Delta{V}_{FWHM}$	 & 	$\int{T_{A}}dV$	 & 	$E_u$\\		
	 & 	$^{2s+1}L_{|L+S|}$	 & 	MHz	 & 	(K)	 & 	(km s$^{-1}$)	 & 	(km s$^{-1}$)	 & 	(K km s$^{-1}$)	 & 	(K)\\		
\hline																
	 & 	$^3P_{1}\rightarrow^3P_{0}$	 & 	492161.28	 & 	5.2	 & 	7.6	 & 	4.7	 & 	26.1	 & 	23.62\\
	 & 	$^3P_{2}\rightarrow^3P_{1}$	 & 	809342.91	 & 	6.9	 & 	7.2	 & 	4.2	 & 	30.8	 & 	62.46\\

\hline																
\end{tabular}																
\end{table}

\begin{table}[t]																
\begin{tabular}{lccccccc}																
[CII]	 & 		 & 		 & 		 & 		 & 		 & 		 & 	\\		
\hline\hline																
	 & 	Transition	 & 	Frequency	 & 	T$_A$	 & 	$V_{lsr}$	 & 	$\Delta{V}_{FWHM}$	 & 	$\int{T_{A}}dV$	 & 	$E_u$\\		
	 & 	$^{2s+1}L_{|L+S|}$	 & 	MHz	 & 	(K)	 & 	(km s$^{-1}$)	 & 	(km s$^{-1}$)	 & 	(K km s$^{-1}$)	 & 	(K)\\		
\hline																
	 & 	$^2P_{3/2}\rightarrow^2P_{1/2}$	 & 	1900527.55	 & 	69.1	 & 	8.6	 & 	3.8	 & 	279.8	 & 	91.21\\
\hline

\end{tabular}																
\end{table}

\begin{table}[t]																
\begin{tabular}{lccccccc}																
																
CCH	 & 		 & 		 & 		 & 		 & 		 & 		 & 	\\		
\hline\hline																
	 & 	Transition	 & 	Frequency	 & 	T$_A$	 & 	$V_{lsr}$	 & 	$\Delta{V}_{FWHM}$	 & 	$\int{T_{A}}dV$	 & 	$E_u$\\		
	 & 	$N_{J,F_1}$	 & 	MHz	 & 	(K)	 & 	(km s$^{-1}$)	 & 	(km s$^{-1}$)	 & 	(K km s$^{-1}$)	 & 	(K)\\		
\hline																
	 & 	$6_{6.5,7}\rightarrow5_{5.5,6}$	 & 	523972.07	 & 	2.2	 & 	6.6	 & 	4.5	 & 	10.7	 & 	88.02\\
	 & 	$6_{5.56}\rightarrow5_{4.5,5}$	 & 	524034.40	 & 	1.9	 & 	6.7	 & 	4.0	 & 	8.1	 & 	88.04\\
	 & 	$7_{7.5,8}\rightarrow6_{6.5,7}$	 & 	611267.06	 & 	1.2	 & 	6.8	 & 	5.2	 & 	6.4	 & 	117.36\\
	 & 	$7_{6.5,6}\rightarrow6_{5.5,5}$	 & 	611329.64	 & 	1.0	 & 	7.0	 & 	4.9	 & 	5.0	 & 	117.38\\
	 & 	$8_{8.5,9}\rightarrow7_{7.5,8}$	 & 	698544.47	 & 	0.6	 & 	7.4	 & 	5.9	 & 	3.7	 & 	150.88\\
	 & 	$8_{7.5,7}\rightarrow7_{6.5,6}$	 & 	698607.43	 & 	0.5	 & 	7.0	 & 	5.2	 & 	2.8	 & 	150.91\\
	 & 	$9_{9.5,10}\rightarrow8_{8.5,9}$	 & 	785801.64	 & 	0.3	 & 	7.9	 & 	8.2	 & 	2.3	 & 	188.59\\
	 & 	$9_{8.5,8}\rightarrow8_{7.5,7}$	 & 	785864.19	 & 	0.2	 & 	7.0	 & 	5.8	 & 	1.3	 & 	188.62\\
\hline

\end{tabular}																
\end{table}

\begin{table}[t]																
\begin{tabular}{lccccccc}																
																
CH	 & 		 & 		 & 		 & 		 & 		 & 		 & 	\\		
\hline\hline																
	 & 	Transition	 & 	Frequency	 & 	T$_A$	 & 	$V_{lsr}$	 & 	$\Delta{V}_{FWHM}$	 & 	$\int{T_{A}}dV$	 & 	$E_u$\\		
	 & 	$N_{K,J,F_1}$	 & 	MHz	 & 	(K)	 & 	(km s$^{-1}$)	 & 	(km s$^{-1}$)	 & 	(K km s$^{-1}$)	 & 	(K)\\		
\hline																
	 & 	$1_{1,1.5,1}\rightarrow1_{-1,0.5,1}$	 & 	532721.73	 & 	2.4	 & 	7.1	 & 	4.2	 & 	10.8	 & 	25.73\\
	 & 	$1_{1,1.5,1}\rightarrow1_{-1,0.5,0}$	 & 	532791.46	 & 	1.0	 & 	8.4	 & 	4.8	 & 	4.9	 & 	25.73\\
	 & 	$1_{-1,1.5,2}\rightarrow1_{1,0.5,1}$	 & 	536759.12	 & 	2.2	 & 	8.2	 & 	4.0	 & 	9.2	 & 	25.73\\
	 & 	$1_{-1,1.5,1}\rightarrow1_{1,0.5,1}$	 & 	536779.69	 & 	0.6	 & 	8.3	 & 	3.8	 & 	2.2	 & 	25.76\\
	 & 	$1_{-1,1.5,1}\rightarrow1_{1,0.5,0}$	 & 	536793.54	 & 	1.0	 & 	8.3	 & 	4.1	 & 	4.2	 & 	25.76\\

\hline																
\end{tabular}																
\end{table}																
																
\begin{table}[t]																
\begin{tabular}{lccccccc}																
																
CH$^+$	 & 		 & 		 & 		 & 		 & 		 & 		 & 	\\		
\hline\hline																
	 & 	Transition	 & 	Frequency	 & 	T$_A$	 & 	$V_{lsr}$	 & 	$\Delta{V}_{FWHM}$	 & 	$\int{T_{A}}dV$	 & 	$E_u$\\		
	 & 	J	 & 	MHz	 & 	(K)	 & 	(km s$^{-1}$)	 & 	(km s$^{-1}$)	 & 	(K km s$^{-1}$)	 & 	(K)\\		
\hline																
	 & 	$1\rightarrow0$	 & 	835135.84	 & 	4.0	 & 	8.3	 & 	4.8	 & 	20.4	 & 	40.08\\

\hline																
\end{tabular}																
\end{table}

\begin{table}[t]																
\begin{tabular}{lccccccc}																
A${-}$CH$_3$OH	 & 	 	 & 		 & 		 & 		 & 		 & 		 & 	\\		
\hline\hline																
	 & 	Transition	 & 	Frequency	 & 	T$_A$	 & 	$V_{lsr}$	 & 	$\Delta{V}_{FWHM}$	 & 	$\int{T_{A}}dV$	 & 	$E_u$\\		
	 & 	$J_{+K,\:v_t^\pi}$	 & 	MHz	 & 	(K)	 & 	(km s$^{-1}$)	 & 	(km s$^{-1}$)	 & 	(K km s$^{-1}$)	 & 	(K)\\		
\hline																
	 & 	$3_{2,0^+}\rightarrow3_{1,0^-}$	 & 	480269.83	 & 	0.7	 & 	6.7	 & 	3.7	 & 	2.7	 & 	51.64\\		
	 & 	$2_{2,0^+}\rightarrow2_{1,0^-}$	 & 	481505.55	 & 	0.7	 & 	6.1	 & 	3.3	 & 	2.4	 & 	44.67\\		
	 & 	$10_{0,0^+}\rightarrow9_{0,0^+}$	 & 	483141.47	 & 	1.5	 & 	6.6	 & 	4.3	 & 	6.9	 & 	127.6\\		
	 & 	$10_{2,0^-}\rightarrow9_{2,0^-}$	 & 	483388.95	 & 	0.3	 & 	6.8	 & 	4.9	 & 	1.4	 & 	165.35\\		
	 & 	$10_{3,0^+}\rightarrow9_{3,0^+}$	 & 	483553.40	 & 	0.2	 & 	5.5	 & 	6.4	 & 	1.7	 & 	177.46\\
\end{tabular}																
\end{table}																
\newpage																
\clearpage																
\begin{table}[t]																
\begin{tabular}{lccccccc}																
A${-}$CH$_3$OH	 & 		 & 		 & 		 & 		 & 		 & 		 & 	\\		
\hline\hline																
	 & 	Transition	 & 	Frequency	 & 	T$_A$	 & 	$V_{lsr}$	 & 	$\Delta{V}_{FWHM}$	 & 	$\int{T_{A}}dV$	 & 	$E_u$\\		
	 & 	$J_{+K,\:v_t^\pi}$	 & 	MHz	 & 	(K)	 & 	(km s$^{-1}$)	 & 	(km s$^{-1}$)	 & 	(K km s$^{-1}$)	 & 	(K)\\		
																
\hline																
																
	 & 	$11_{1,0^-}\rightarrow10_{1,0^-}$	 & 	536192.13	 & 	0.6	 & 	6.4	 & 	3.7	 & 	2.4	 & 	169.01\\		
	 & 	$5_{1,0^+}\rightarrow4_{0,0^+}$	 & 	538571.19	 & 	2.1	 & 	6.7	 & 	4.4	 & 	9.7	 & 	49.06\\		
	 & 	$6_{3,0^+}\rightarrow5_{2,0^+}$	 & 	542002.05	 & 	1.2	 & 	6.4	 & 	3.7	 & 	4.7	 & 	98.55\\		
	 & 	$6_{3,0^-}\rightarrow5_{2,0^-}$	 & 	542082.94	 & 	1.2	 & 	6.5	 & 	3.7	 & 	4.7	 & 	98.55\\		
	 & 	$12_{1,0^+}\rightarrow11_{1,0^+}$	 & 	574869.74	 & 	0.3	 & 	6.3	 & 	4.2	 & 	1.4	 & 	193.96\\
\end{tabular}																
\end{table}																
\newpage																
\clearpage																
\begin{table}[t]																
\begin{tabular}{lccccccc}																
A${-}$CH$_3$OH	 & 		 & 		 & 		 & 		 & 		 & 		 & 	\\		
\hline\hline																
	 & 	Transition	 & 	Frequency	 & 	T$_A$	 & 	$V_{lsr}$	 & 	$\Delta{V}_{FWHM}$	 & 	$\int{T_{A}}dV$	 & 	$E_u$\\		
	 & 	$J_{+K,\:v_t^\pi}$	 & 	MHz	 & 	(K)	 & 	(km s$^{-1}$)	 & 	(km s$^{-1}$)	 & 	(K km s$^{-1}$)	 & 	(K)\\		
																
\hline	 & 	$8_{3,0^+}\rightarrow7_{2,0^+}$	 & 	638524.78	 & 	0.9	 & 	6.4	 & 	4.4	 & 	4.3	 & 	133.36\\		
	 & 	$8_{3,0^-}\rightarrow7_{2,0^-}$	 & 	638818.88	 & 	0.9	 & 	6.5	 & 	4.4	 & 	4.0	 & 	133.36\\		
	 & 	$14_{1,0^+}\rightarrow13_{1,0^+}$	 & 	670424.18	 & 	0.2	 & 	6.3	 & 	6.8	 & 	1.6	 & 	256.02\\		
	 & 	$4_{2,0^-}\rightarrow3_{1,0^-}$	 & 	673747.18	 & 	1.3	 & 	6.5	 & 	4.2	 & 	5.8	 & 	60.92\\		
	 & 	$8_{1,0^+}\rightarrow7_{0,0^+}$	 & 	674991.12	 & 	1.3	 & 	6.8	 & 	4.5	 & 	6.2	 & 	97.38\\
\end{tabular}																
\end{table}																
\newpage																
\clearpage																
\begin{table}[t]																
\begin{tabular}{lccccccc}																
A${-}$CH$_3$OH	 & 		 & 		 & 		 & 		 & 		 & 		 & 	\\		
\hline\hline																
	 & 	Transition	 & 	Frequency	 & 	T$_A$	 & 	$V_{lsr}$	 & 	$\Delta{V}_{FWHM}$	 & 	$\int{T_{A}}dV$	 & 	$E_u$\\		
	 & 	$J_{+K,\:v_t^\pi}$	 & 	MHz	 & 	(K)	 & 	(km s$^{-1}$)	 & 	(km s$^{-1}$)	 & 	(K km s$^{-1}$)	 & 	(K)\\		
																
\hline	 & 	$8_{2,0^+}\rightarrow7_{1,0^+}$	 & 	881783.47	 & 	0.7	 & 	6.5	 & 	4.5	 & 	3.6	 & 	121.29\\		
	 & 	$13_{1,0^+}\rightarrow12_{0,0^+}$	 & 	894614.92	 & 	0.4	 & 	6.8	 & 	4.7	 & 	1.9	 & 	223.84\\		
	 & 	$13_{1,0^+}\rightarrow12_{0,0^+}$	 & 	894614.92	 & 	0.4	 & 	6.8	 & 	4.7	 & 	1.9	 & 	223.84\\		
	 & 	$9_{2,0^-}\rightarrow8_{1,0^-}$	 & 	902936.66	 & 	0.4	 & 	6.4	 & 	5.3	 & 	2.5	 & 	142.15\\		
	 & 	$6_{4,0^+}\rightarrow5_{3,0^+}$	 & 	926555.20	 & 	0.4	 & 	7.0	 & 	5.3	 & 	2.5	 & 	129.09\\		
	 & 	$9_{2,0^+}\rightarrow8_{1,0^+}$	 & 	933694.55	 & 	0.6	 & 	6.5	 & 	4.6	 & 	2.7	 & 	142.19\\		
	 & 	$14_{1,0^+}\rightarrow13_{0,0^+}$	 & 	937479.55	 & 	0.3	 & 	6.6	 & 	5.8	 & 	2.0	 & 	256.02\\		
	 & 	$7_{4,0^-}\rightarrow6_{3,0^-}$	 & 	974878.74	 & 	0.4	 & 	6.4	 & 	6.4	 & 	2.9	 & 	145.33\\		
	 & 	$10_{2,0^+}\rightarrow9_{1,0^+}$	 & 	986100.37	 & 	0.5	 & 	6.2	 & 	5.1	 & 	3.0	 & 	165.4\\		
	 & 	$8_{4,0^+}\rightarrow7_{3,0^+}$	 & 	1023196.73	 & 	0.5	 & 	7.2	 & 	4.4	 & 	2.2	 & 	163.9\\

\hline																
\end{tabular}																
\end{table}																
\begin{table}[t]																
\begin{tabular}{lccccccc}																
E${-}$CH$_3$OH	 & 		 & 		 & 		 & 		 & 		 & 		 & 	\\		
\hline\hline																
	 & 	Transition	 & 	Frequency	 & 	T$_A$	 & 	$V_{lsr}$	 & 	$\Delta{V}_{FWHM}$	 & 	$\int{T_{A}}dV$	 & 	$E_u$\\		
	 & 	$J_{\pm K,\: v_t}$	 & 	MHz	 & 	(K)	 & 	(km s$^{-1}$)	 & 	(km s$^{-1}$)	 & 	(K km s$^{-1}$)	 & 	(K)\\		
\hline																
	 & 	$10_{0,0}\rightarrow9_{0,0}$	 & 	482283.13	 & 	0.7	 & 	6.4	 & 	3.8	 & 	2.8	 & 	132.71\\		
	 & 	$10_{-1,0}\rightarrow9_{-1,0}$	 & 	482959.74	 & 	1.4	 & 	6.4	 & 	3.9	 & 	5.7	 & 	125.25\\		
	 & 	$10_{1,0}\rightarrow9_{1,0}$	 & 	483687.10	 & 	0.5	 & 	6.5	 & 	3.9	 & 	2.2	 & 	140.83\\		
	 & 	$10_{2,0}\rightarrow9_{2,0}$	 & 	484023.82	 & 	0.6	 & 	6.6	 & 	4.5	 & 	3.0	 & 	142.08\\		
	 & 	$10_{-2,0}\rightarrow9_{-2,0}$	 & 	484072.13	 & 	0.3	 & 	6.8	 & 	5.0	 & 	1.6	 & 	145.73\\		
	 & 	$7_{0,0}\rightarrow6_{-1,0}$	 & 	495173.84	 & 	1.2	 & 	6.6	 & 	3.9	 & 	4.8	 & 	70.18\\		
	 & 	$7_{1,0}\rightarrow6_{0,0}$	 & 	504294.53	 & 	1.2	 & 	6.4	 & 	3.6	 & 	4.7	 & 	248.24\\		
	 & 	$11_{1,0}\rightarrow10_{2,0}$	 & 	506153.78	 & 	0.2	 & 	6.6	 & 	5.1	 & 	1.0	 & 	166.37\\		
	 & 	$10_{2,0}\rightarrow9_{1,0}$	 & 	509564.94	 & 	0.7	 & 	6.8	 & 	5.2	 & 	3.9	 & 	142.08\\		
	 & 	$2_{-2,0}\rightarrow1_{-1,0}$	 & 	520179.58	 & 	1.1	 & 	6.7	 & 	4.1	 & 	5.0	 & 	24.96\\		
	 & 	$14_{-1,0}\rightarrow13_{0,0}$	 & 	523275.25	 & 	0.3	 & 	6.5	 & 	4.9	 & 	1.4	 & 	241.04\\		
																
\end{tabular}																
\end{table}																
\begin{table}[t]																
\begin{tabular}{lccccccc}																
E${-}$CH$_3$OH	 & 		 & 		 & 		 & 		 & 		 & 		 & 	\\		
\hline\hline																
	 & 	Transition	 & 	Frequency	 & 	T$_A$	 & 	$V_{lsr}$	 & 	$\Delta{V}_{FWHM}$	 & 	$\int{T_{A}}dV$	 & 	$E_u$\\		
	 & 	$J_{\pm K,\: v_t}$	 & 	MHz	 & 	(K)	 & 	(km s$^{-1}$)	 & 	(km s$^{-1}$)	 & 	(K km s$^{-1}$)	 & 	(K)\\		
\hline																
	 & 	$8_{3,0}\rightarrow8_{2,0}$	 & 	530123.52	 & 	0.3	 & 	6.9	 & 	5.4	 & 	1.7	 & 	123.38\\		
	 & 	$11_{0,0}\rightarrow10_{0,0}$	 & 	530184.87	 & 	0.5	 & 	6.6	 & 	4.6	 & 	2.5	 & 	158.15\\		
	 & 	$7_{3,0}\rightarrow7_{2,0}$	 & 	530316.84	 & 	0.3	 & 	6.6	 & 	4.9	 & 	1.8	 & 	104.81\\		
	 & 	$6_{3,0}\rightarrow6_{2,0}$	 & 	530455.52	 & 	0.4	 & 	6.5	 & 	4.1	 & 	1.9	 & 	88.56\\		
	 & 	$5_{3,0}\rightarrow5_{2,0}$	 & 	530549.77	 & 	0.4	 & 	6.7	 & 	4.6	 & 	2.2	 & 	74.63\\
\end{tabular}																
\end{table}																
\newpage																
\clearpage																
\begin{table}[t]																
\begin{tabular}{lccccccc}																
E${-}$CH$_3$OH	 & 		 & 		 & 		 & 		 & 		 & 		 & 	\\		
\hline\hline																
	 & 	Transition	 & 	Frequency	 & 	T$_A$	 & 	$V_{lsr}$	 & 	$\Delta{V}_{FWHM}$	 & 	$\int{T_{A}}dV$	 & 	$E_u$\\		
	 & 	$J_{\pm K,\: v_t}$	 & 	MHz	 & 	(K)	 & 	(km s$^{-1}$)	 & 	(km s$^{-1}$)	 & 	(K km s$^{-1}$)	 & 	(K)\\		
\hline																
	 & 	$13_{-1,0}\rightarrow12_{-1,0}$	 & 	627171.35	 & 	0.5	 & 	6.6	 & 	4.4	 & 	2.3	 & 	208.64\\		
	 & 	$13_{2,0}\rightarrow12_{2,0}$	 & 	629321.48	 & 	0.3	 & 	7.1	 & 	6.0	 & 	1.6	 & 	225.71\\		
	 & 	$13_{-2,0}\rightarrow12_{-2,0}$	 & 	629651.72	 & 	0.2	 & 	7.0	 & 	5.2	 & 	1.1	 & 	229.4\\		
	 & 	$10_{0,0}\rightarrow9_{-1,0}$	 & 	638280.27	 & 	0.7	 & 	6.7	 & 	4.4	 & 	3.1	 & 	132.71\\		
	 & 	$10_{1,0}\rightarrow9_{0,0}$	 & 	651618.53	 & 	0.6	 & 	6.5	 & 	4.6	 & 	2.9	 & 	140.83\\
\end{tabular}																
\end{table}																
\newpage																
\clearpage																
\begin{table}[t]																
\begin{tabular}{lccccccc}																
E${-}$CH$_3$OH	 & 		 & 		 & 		 & 		 & 		 & 		 & 	\\		
\hline\hline																
	 & 	Transition	 & 	Frequency	 & 	T$_A$	 & 	$V_{lsr}$	 & 	$\Delta{V}_{FWHM}$	 & 	$\int{T_{A}}dV$	 & 	$E_u$\\		
	 & 	$J_{\pm K,\: v_t}$	 & 	MHz	 & 	(K)	 & 	(km s$^{-1}$)	 & 	(km s$^{-1}$)	 & 	(K km s$^{-1}$)	 & 	(K)\\		
\hline																
	 & 	$3_{-3,0}\rightarrow3_{-2,0}$	 & 	766983.75	 & 	0.4	 & 	6.4	 & 	3.1	 & 	1.2	 & 	68.74\\		
	 & 	$5_{3,0}\rightarrow4_{2,0}$	 & 	772453.48	 & 	0.5	 & 	7.2	 & 	7.1	 & 	4.0	 & 	74.63\\		
	 & 	$13_{0,0}\rightarrow12_{-1,0}$	 & 	779030.27	 & 	0.4	 & 	7.2	 & 	6.0	 & 	2.2	 & 	215.92\\		
	 & 	$8_{-2,0}\rightarrow7_{-1,0}$	 & 	811444.68	 & 	0.5	 & 	7.0	 & 	6.0	 & 	3.4	 & 	101.6\\		
	 & 	$6_{-4,0}\rightarrow5_{-3,0}$	 & 	815071.72	 & 	0.7	 & 	6.6	 & 	4.3	 & 	3.1	 & 	128.75\\		
	 & 	$6_{3,0}\rightarrow5_{2,0}$	 & 	820764.66	 & 	0.4	 & 	6.2	 & 	4.4	 & 	1.8	 & 	88.56\\		
	 & 	$9_{-2,0}\rightarrow8_{-1,0}$	 & 	860459.44	 & 	0.4	 & 	7.0	 & 	5.5	 & 	2.4	 & 	122.5\\		
	 & 	$7_{-4,0}\rightarrow6_{-3,0}$	 & 	863365.78	 & 	0.5	 & 	6.7	 & 	5.2	 & 	2.9	 & 	145\\		
	 & 	$7_{3,0}\rightarrow6_{2,0}$	 & 	869039.06	 & 	0.5	 & 	6.6	 & 	5.2	 & 	2.6	 & 	104.81\\		
	 & 	$10_{-2,0}\rightarrow9_{-1,0}$	 & 	909738.07	 & 	0.3	 & 	6.9	 & 	4.7	 & 	1.6	 & 	145.73\\		
	 & 	$8_{-4,0}\rightarrow7_{-3,0}$	 & 	911643.69	 & 	0.4	 & 	6.5	 & 	4.5	 & 	2.1	 & 	163.56\\		
	 & 	$3_{-3,0}\rightarrow2_{-2,0}$	 & 	912109.65	 & 	0.7	 & 	6.6	 & 	5.1	 & 	3.5	 & 	68.74\\		
	 & 	$8_{3,0}\rightarrow7_{2,0}$	 & 	917269.03	 & 	0.4	 & 	7.3	 & 	4.9	 & 	1.9	 & 	123.38\\		
	 & 	$4_{-3,0}\rightarrow3_{-2,0}$	 & 	960473.04	 & 	0.6	 & 	6.4	 & 	5.3	 & 	3.1	 & 	78.03\\		
																
\hline																
\end{tabular}																
\end{table}																
																
\begin{table}[t]																
\begin{tabular}{lccccccc}																
																
CN,v=0	 & 		 & 		 & 		 & 		 & 		 & 		 & 	\\		
\hline\hline																
	 & 	Transition	 & 	Frequency	 & 	T$_A$	 & 	$V_{lsr}$	 & 	$\Delta{V}_{FWHM}$	 & 	$\int{T_{A}}dV$	 & 	$E_u$\\		
	 & 	$N_{J,F_1}$	 & 	MHz	 & 	(K)	 & 	(km s$^{-1}$)	 & 	(km s$^{-1}$)	 & 	(K km s$^{-1}$)	 & 	(K)\\		
\hline																
	 & 	$5_{4.5,5.5}\rightarrow4_{3.5,4.5}$	 & 	566729.48	 & 	1.3	 & 	7.3	 & 	5.3	 & 	7.3	 & 	81.59\\
	 & 	$5_{5.5,5.5}\rightarrow4_{4.5,4.5}$	 & 	566946.10	 & 	1.6	 & 	7.6	 & 	5.6	 & 	9.8	 & 	81.64\\
	 & 	$6_{5.5,6.5}\rightarrow5_{4.5,5.5}$	 & 	680046.29	 & 	0.5	 & 	7.5	 & 	5.4	 & 	3.0	 & 	114.23\\
	 & 	$6_{6.5,6.5}\rightarrow5_{5.5,5.5}$	 & 	680263.15	 & 	0.7	 & 	7.5	 & 	5.4	 & 	4.1	 & 	114.29\\
	 & 	$7_{6.5,7.5}\rightarrow6_{5.5,6.5}$	 & 	793337.05	 & 	0.3	 & 	7.3	 & 	6.9	 & 	2.2	 & 	152.38\\
	 & 	$7_{7.5,7.5}\rightarrow6_{6.5,6.5}$	 & 	793552.12	 & 	0.3	 & 	7.6	 & 	6.4	 & 	1.8	 & 	152.38\\

\hline																
\end{tabular}%
\\\\S\textbf{uperscipt ``b'' means it is a blended line and excluded.}																
\end{table}

\newpage																
\clearpage																
\begin{table}           																								
\centering                       																								
\caption{LTE modeling results for species requiring one component fits}  \label{tab:LTE}            																								
																								
%
\\          																								
																		
\end{table}																								

\newpage
\clearpage
\thispagestyle{empty}
\begin{landscape}
\begin{table}
\centering	\caption{LTE modeling results for species requiring two component fits}				\label{tab:LTE2}	
\scalebox{0.70}{
%
																	
}
$^{{\ \rm a}}$  Calculation based on the main isotopologue
\end{table}																		
\end{landscape}																													

\begin{table}
\caption{\textbf{The abundance of species with respect to H$_2$}} \label{tab:wrtH2}
\begin{center}
\newcolumntype{C}[1]{>{\centering}m{#1}}
\end{center}	
\end{table}
\begin{table}													
\centering													
\caption{Column density upper limit  for species in Orion-KL not detected in Orion-S}													
\label{tab:OrionKL}													
\scalebox{0.7}{ 													
%
													
}													
\end{table}													

\begin{table}
\centering
\caption{Column density upper limit  for unmodeled species in Orion-S}
\label{tab:upperS}
%
\end{table}

\newpage
\clearpage

\begin{center}
\includegraphics[angle=0,scale=0.8]{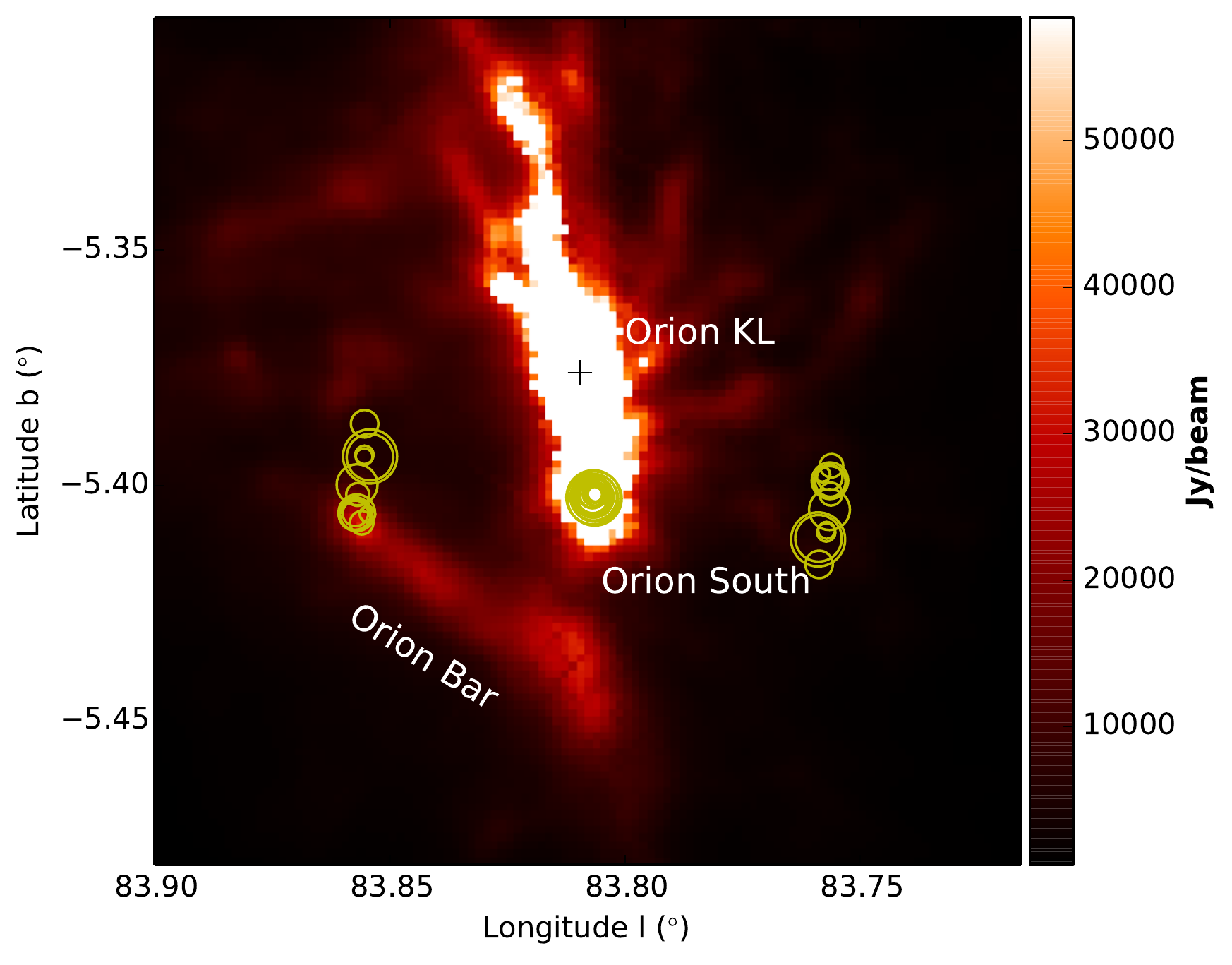}
\figcaption[]{HEXOS Orion-S Observations: The circles indicate the beam positions with diameters corresponding to the FWHM at the center frequency of the HIFI bands. Circles near the center of the image (at l $\sim$ 83.81$^\circ$) indicate the position of the spectral scan observations of Orion-S.  Circles to the left and to the right (at l $\sim$ 83.86$^\circ$ and 83.76$^\circ$ respectively) indicate the off position observations. The background shows the \textit{Herschel}/SPIRE 250 $\mu m$ dust emission in the Orion-KL region (white regions indicate saturated pixels).  From the location of the beam circles near the Orion Bar, it is apparent that some of the observations see emission in at least one reference beam. \label{Orion_KL_dust}}
\end{center}

\begin{center}
\includegraphics[angle=0,scale=0.8]{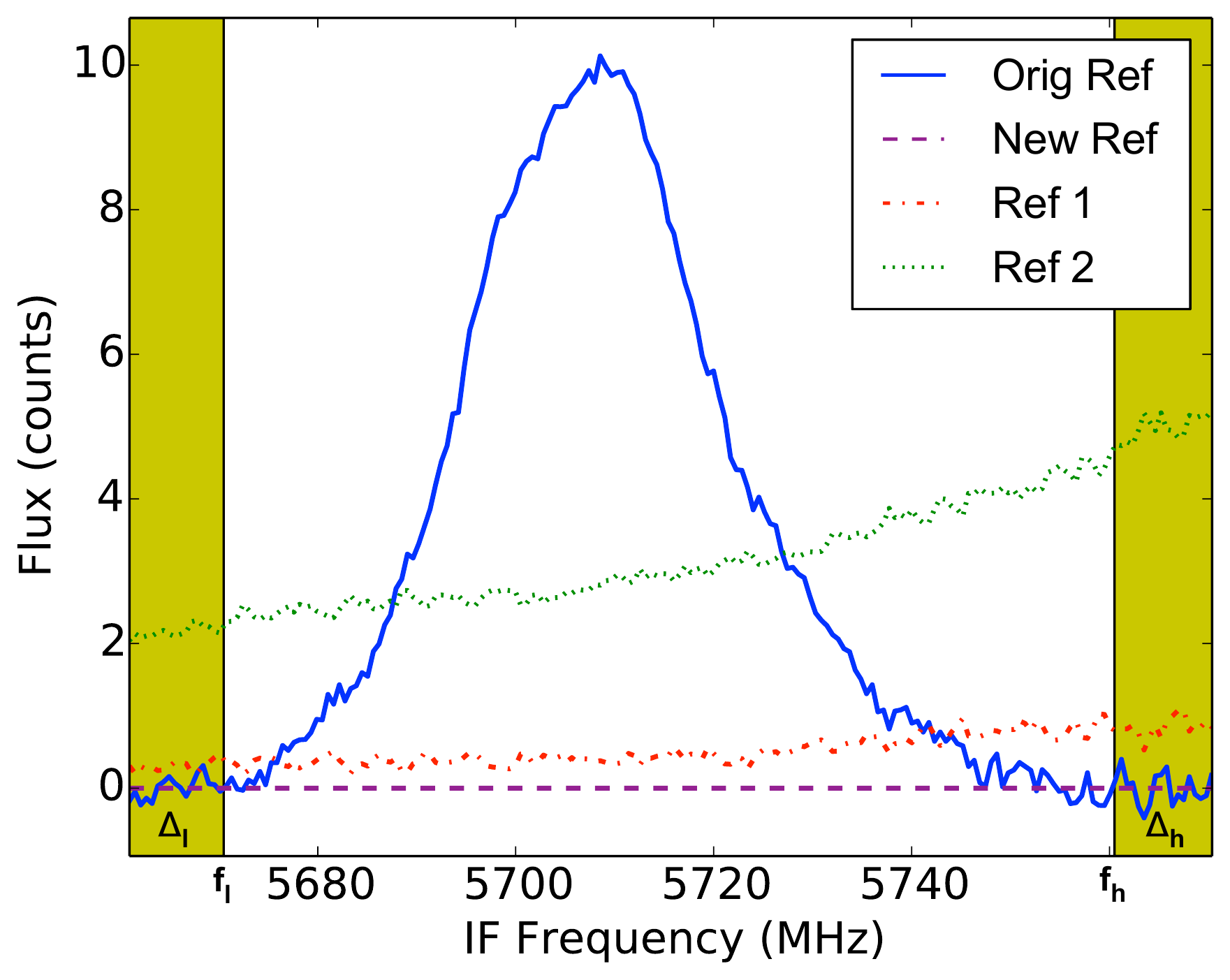}
\figcaption[]{Removing the reference beam emission in the HEXOS Orion-S spectral scan: The blue solid line shows the emission in a reference beam. The red dashed and the green dotted lines show the references for an earlier scan and a later scan with slightly changed LO-settings, but covering the same IF interval. Averaging and scaling these reference scans using just the frequency ranges marked with green background results in the new reference spectrum, which replaces the old reference spectrum only in the frequency range shown (the total spectrum is still 4 GHz wide or $\sim$1 GHz for each of the 4 HIFI WBS sub-bands). For display purposes we subtracted the new reference from all scans causing it to appear as a straight line. \label{hexos_os_contsm}}
\end{center}

\begin{center}
\includegraphics[angle=0,scale=0.6]{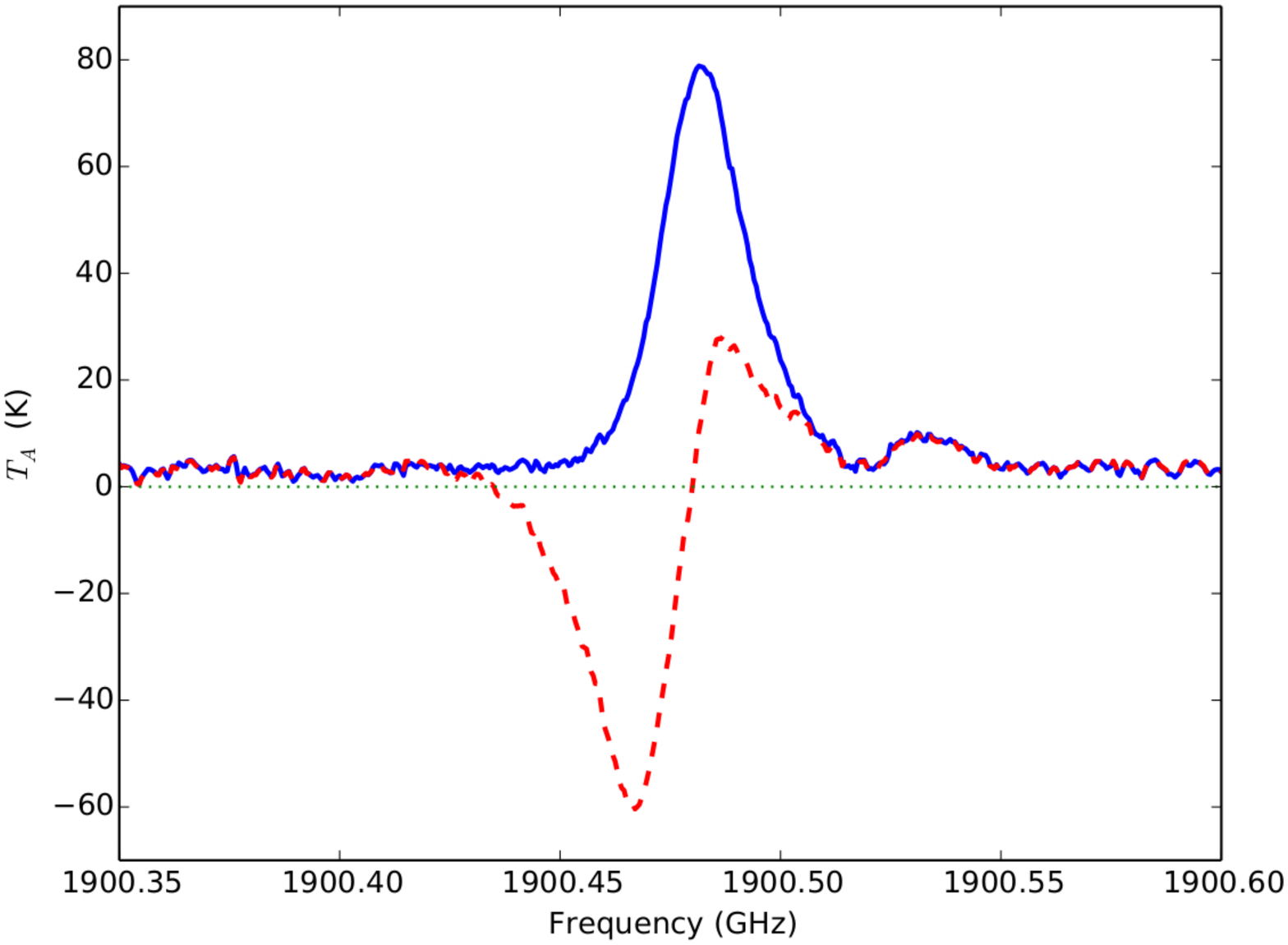}
\figcaption[]{Example of the repaired [CII] line at 1900.526 GHz. The red dashed line shows the original result with emission in both reference beams and the blue solid line shows the result with the emission in the reference beams removed.  \label{hexos_os_final}}
\end{center}

\begin{center}
\includegraphics[scale=1.2]{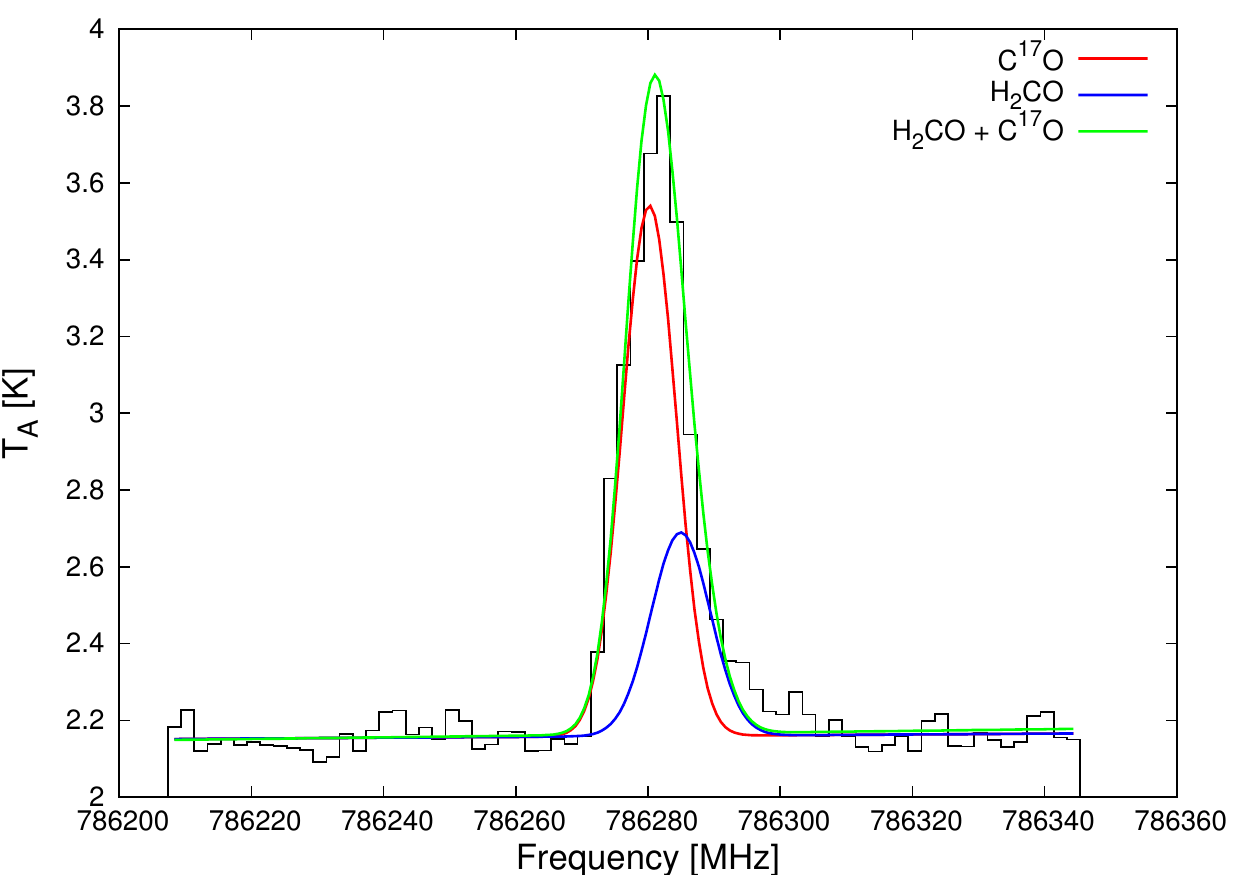}
\figcaption[]{Example of a blended line at 786.3 GHz, produced by C$^{17}$O and H$_2$CO. The red line shows the LTE modeled synthetic spectrum for C$^{17}$O and the blue line shows that for  H\2CO (Table \ref{tab:LTE}  and \ref{tab:LTE2}).  The green line is the superposition of these two components.  Data are shown by the black histogram.
\label{fig:Blended}}
\end{center}

\begin{figure}
\centering
\includegraphics[scale=0.7]{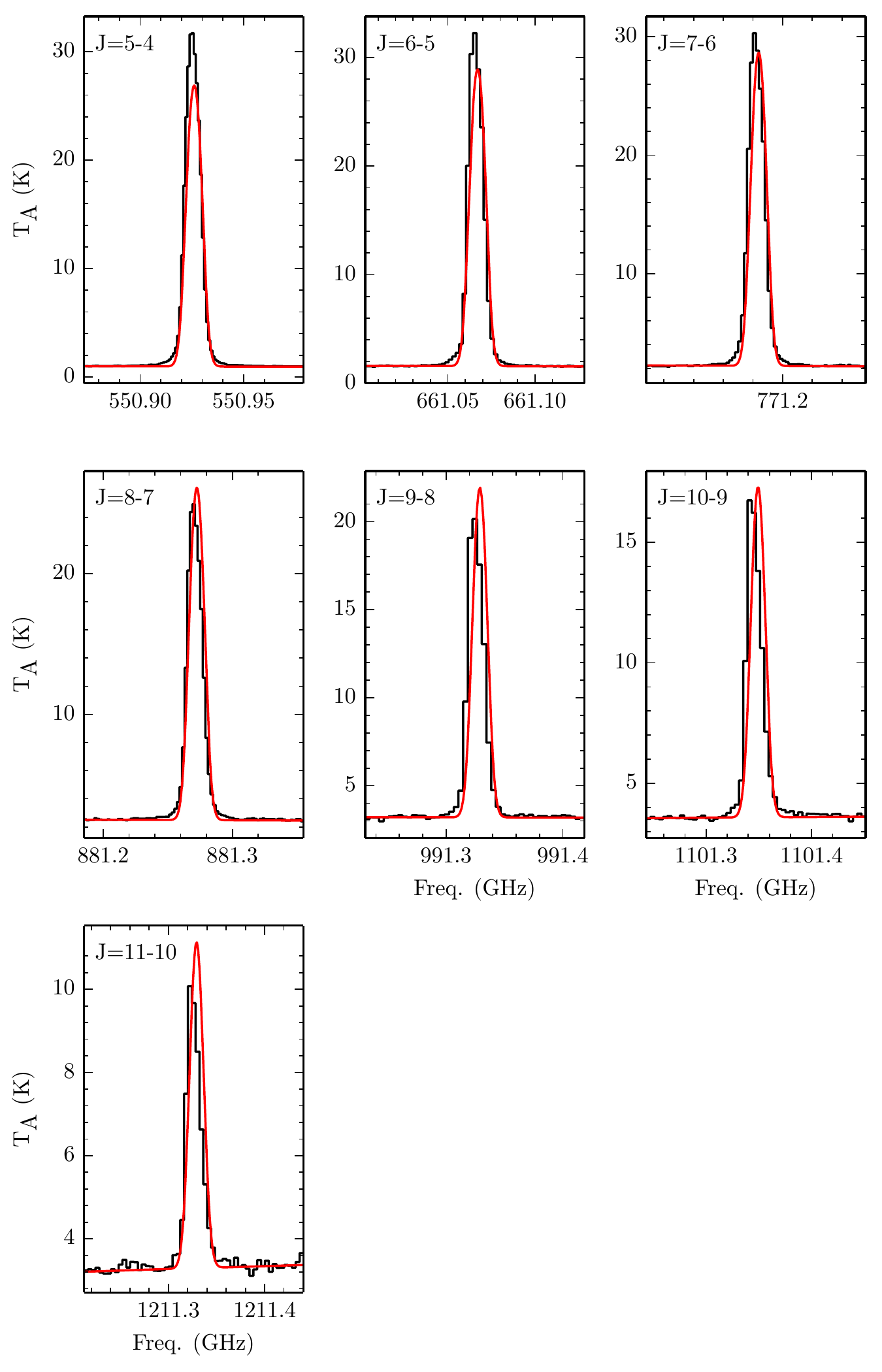}
\caption{One component LTE modeling for $^{13}$CO. Black histogram shows the data. Resulting model spectra are shown in red. LTE model parameters are provided in Table \ref{tab:LTE}.} 
\label{C13O}
\end{figure}

\end{document}